\documentclass{aa}
\usepackage[varg]{txfonts}
\usepackage{graphicx}

\usepackage{natbib}
\bibpunct{(}{)}{;}{a}{}{,}
\usepackage[colorlinks=true,citecolor=blue]{hyperref}
\usepackage{threeparttablex}
\usepackage{longtable}
\usepackage{lscape}
\usepackage{bm}
\usepackage{verbatim}
\usepackage{makecell}
\usepackage{caption}
\usepackage{subcaption}
\usepackage{multirow}
\usepackage{booktabs}
\usepackage{hyperref}

\usepackage{colortbl}
\usepackage{xcolor}

\def\teff{$T_{\rm eff}$}
\def\logg{$\log\,g$}

\newcommand{\kader}[2]{\begin{center}\fbox{\parbox{#1}{#2}}\end{center}}

\begin{document}

\title{Asteroseismic modelling
  of fast rotators and its opportunities for astrophysics}
\author{Conny Aerts\inst{1,2,3} \and Andrew Tkachenko\inst{1}}

\institute{Institute of Astronomy, KU Leuven, Celestijnenlaan 200D,
  B-3001 Leuven, Belgium
  \\ \email{conny.aerts@kuleuven.be; andrew.tkachenko@kuleuven.be}
\and
Department of Astrophysics, IMAPP, Radboud University Nijmegen, PO Box 9010,
6500 GL, Nijmegen, The Netherlands
\and 
Max Planck Institute for Astronomy, Koenigstuhl 17, 69117, Heidelberg, Germany
}

\date{Received XXX / Accepted XXX}

\abstract{
  Rotation matters for the life of a star.  It causes a
  multitude of dynamical phenomena in the stellar interior during a
  star's evolution, and its effects accumulate until the star dies.
  All stars rotate at some level, but most of those born with a mass 
  higher than 1.3 times the mass of the Sun rotate rapidly during more than
  90\% of their nuclear lifetime.  Internal rotation guides the
  angular momentum and chemical element transport throughout the
  stellar interior. These transport processes change over time as the
  star evolves.  The cumulative effects of stellar rotation and its
  induced transport processes determine the helium content of the core
  by the time it exhausts its hydrogen isotopes. The amount of helium
  at that stage also guides the heavy element yields by the end of the
  star's life.

  \hspace{0.5cm} A proper theory of stellar evolution and any 
  realistic models for the chemical enrichment of galaxies must be
  based on observational calibrations of stellar rotation and of the
  induced transport processes. In the last few years, asteroseismology
  offers such calibrations for single and binary stars.  We review
  the current status of asteroseismic modelling of rotating stars for
  different stellar mass regimes in an accessible way for the
  non-expert.  While doing so, we describe exciting opportunities
  sparked by asteroseismology for various domains in astrophysics,
  touching upon topics such as exoplanetary science, galactic
  structure and evolution, and gravitational wave physics to mention
  just a few. Along
  the way we provide ample sneak-previews for future `industrialised'
  applications of asteroseismology to slow and rapid rotators from
  the exploitation of combined {\it Kepler}, Transiting Exoplanet Survey
  Satellite (TESS), PLAnetary Transits and Oscillations of stars
  (PLATO),
  {\it Gaia}, and ground-based
  spectroscopic and multi-colour photometric surveys.

  \hspace{0.5cm} We end the review with a list of takeaway
  messages and achievements of asteroseismology that are of
  relevance for many fields of astrophysics.}

\keywords{Asteroseismology -- Waves -- Binary Stars -- Stars: Rotation
  -- Stars: Interiors -- Stars: evolution -- Stars: oscillations
  (including pulsations)-- Stars: magnetic field -- Gravitational
  Waves -- Convection -- Hydrodynamics -- Methods: data analysis --
  Methods: statistical -- Surveys -- Sun: interior -- Sun:
  helioseismology -- Binaries: eclipsing -- Stars: subdwarfs -- Stars:
  supergiants -- (Galaxy:) open clusters and associations: general
  -- Stars: black holes --
  Stars: neutron -- Binaries (including multiple): close -- Blue
  stragglers -- Stars: emission-line, Be }

\titlerunning{Asteroseismology of fast rotators and  opportunities for astrophysics}
\authorrunning{Conny Aerts \& Andrew Tkachenko}
\maketitle

\section{Asteroseismology: A fountain of
  opportunities for astrophysics}

Asteroseismology is the science of probing stellar interiors by
modelling detected oscillation modes of stars \citep[see][for a
  comprehensive monograph]{Aerts2010}. It has been an established
research field within stellar astrophysics for more than a decade. Rapid progress in numerous successful applications of this
technique has occurred due to major advances in the observational
aspects brought by space telescopes.  {This is mainly thanks to
  space missions} dedicated to the collection of uninterrupted
high-cadence high-precision photometric light curves, such as the past satellites
Microvariability and Oscillations of STars
\citep[MOST,][]{Walker2003},
Convection, Rotation, and planetary Transits \citep[CoRoT,][]{Auvergne2009}, and {\it
  Kepler\/} \citep{Koch2010},  and the currently operational
Transiting Exoplanet Survey
  Satellite telescope
\citep[TESS,][]{Ricker2015}. The oscillation frequencies deduced
from such time series data are independent of stellar models as they
are derived directly from the Fourier transforms of the light
curves. The mode frequencies are typically one to several orders of
magnitude more precise than the classical observables used to evaluate
stellar structure and evolution models in Hertzsprung-Russell diagram (HRD) or
colour-magnitude diagram (CMD)  \citep[see Table\,1 in
][]{Aerts2019-ARAA}.
Moreover, oscillation mode frequencies are
influenced by the physical and chemical conditions of the matter in
the part of the stellar interior where the modes have their dominant
probing power. It is then no surprise that the exploitation of seismic
observables determined directly by the deep interior of stars brings a
revolution for astrophysics \citep{Aerts2021-RMP}.  Now that the field
of asteroseismology is mature and well established, we enter its
`industrial revolution' thanks to the large surveys brought by TESS
and the PLAnetary Transits and Oscillations of stars
mission currently under construction \citep[PLATO,][]{Rauer2024}.

Figure\,1 represents a fountain of opportunities fed by
asteroseismology. It provides a non-exhaustive preview of the themes touched
upon in this review. Asteroseismology currently plays an
important role in the progress of each of these topics of modern
astrophysics.  This fountain highlights the major capabilities of
asteroseismology by uncovering the detailed properties of stars, all the
way from their central core to their surface. Thanks to this unique
capability, asteroseismology brings novel and crucial ways to improve
stellar structure and evolution theory, impacting all research in
astrophysics that relies on stellar models. For each of the topics shown
in Fig.\,1, we highlight the role of asteroseismic input, with
emphasis on the improvements achieved or to be expected in the not-too-distant future.

\begin{figure*}[t!]
  \phantom{a}\hspace{-1cm}
  {\resizebox{19.cm}{!}{\includegraphics{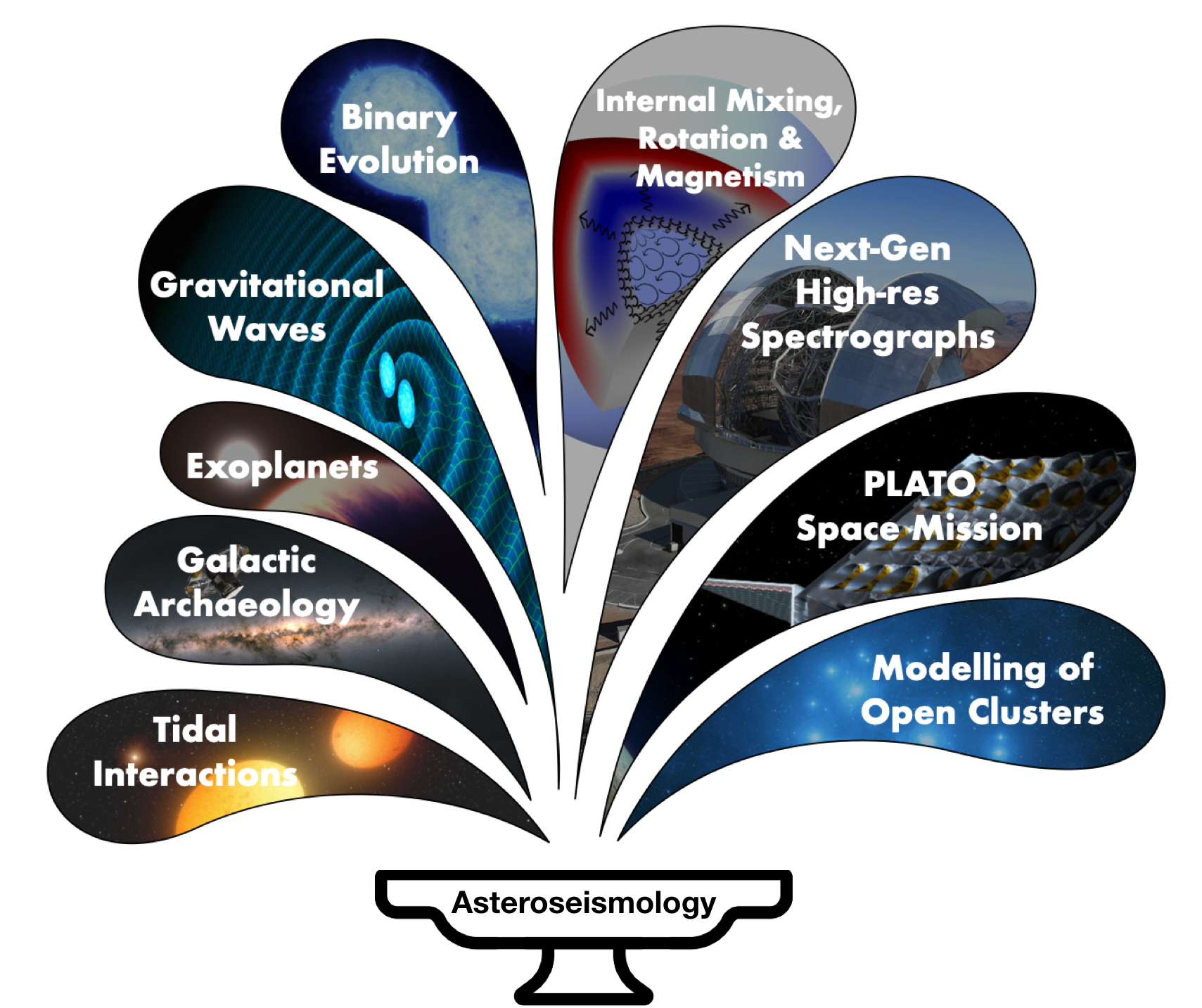}}}
  \caption{\label{Fountain}
    Asteroseismology  provides a fountain of opportunities for
    astrophysics.
    Some research areas benefitting from it 
    are indicated, notably exoplanetary science,
  {archaeological} studies of the Milky Way and Magellanic Clouds,
  stellar evolution theory of single and binary stars, stellar
  populations and clusters, technology development and instrument
  calibration, and gravitational wave and theoretical physics, among others.
{Each of these}
  topics is discussed in this review, highlighting the breadth and
  impact of asteroseismology. The figure
  was designed by Conny Aerts and implemented by Clio Gielen.}
\end{figure*}

The extensive general and observational papers by
\citet{Aerts2021-RMP} and \citet{Kurtz2022}, respectively, already
offer comprehensive reviews on the history, beginnings, methodology,
and numerous applications of asteroseismology in the current space
era, written for a broad readership. Here we offer new insights and
more recent applications for the general A\&A reader that have not
yet been covered or are too recent to have been included in these two comprehensive
reviews. In particular, we focus this review on applications of
asteroseismology to rapidly rotating stars, and discuss why and how
they differ from the {asteroseismology} of the Sun-like stars.
{Such stars are} slow
rotators during most of their life because they are
subject to magnetic braking during their initial life phases.

The character of an oscillation mode is mainly determined by its
dominant restoring force(s).  {Examples are the pressure gradient
  for pressure (p)~modes, also known as acoustic modes; the buoyancy
  force of Archimedes for gravity (g)~modes; and the Coriolis force for
  inertial modes.  In the absence of rotation, the family of
  spheroidal modes is categorised into three branches: the
  high-frequency p~modes and the low-frequency g~modes are separated
  by the fundamental (f) mode, which corresponds to surface
  gravity waves. This simple picture becomes more complex in the
  presence of rotation.}  We come back to the different types of modes, which
probe different layers of the star, in Sect.\,3.  In this review we
  highlight  the modelling results based on identified oscillation
modes probing the deep internal physics of observed rotating stars. A
key motivation for this emphasis is the major impact offered by the
interpretation of  gravito-inertial oscillation modes.
{These modes allow us} to improve the theory of stellar evolution in the presence
of rapid rotation. Gravito-inertial asteroseismology only came into
being a few years ago, and has not been covered extensively in previous
review papers, while it is of major importance for calibrating stellar evolution
and chemical yield computations
\citep[e.g.][]{Hirschi2004,Hirschi2005,Kaiser2020,Brinkman2024}.

Rotation has a major impact during the longest nuclear life phase of
all the single and binary stars of intermediate and high mass
\citep{Maeder2009}.  In that sense, the internal rotation during that
phase is an important driver for the chemical evolution of galaxies
\citep{Karakas2014,Kobayashi2020}. It should be kept in mind that the
effects of rotation, notably the transport and mixing phenomena it
induces, are cumulative throughout the evolution of the star.
Rotationally induced internal mixing and angular momentum transport in
stars born with a convective core and a radiative envelope leave
strong fingerprints during the initial $\sim$90\% of their life, that
is, while they are burning hydrogen into helium in their rotating
core. 

Throughout this  main-sequence phase, the cumulative
effect of rotationally induced or affected transport processes may
lead to the production of far more helium compared to the case where
rotation is not active or only mildly so.  The efficiency of the heavy
element production beyond the main sequence is largely determined by
the amount of helium available as fuel in the final $\sim$\!10\% of
stellar life \citep{Kippenhahn2013,SalarisCassisi2017}. During that
short and complicated end phase of stellar evolution, stars are slowed
down tremendously as they expand and lose angular momentum due to
winds, outflows, and explosions (for those born with a mass above
about eight solar masses).
{In this way, stars enrich
  their surrounding interstellar medium near the end of their life.} 
To assess the amount and
kind of processed nuclear material expelled by dying stars,
it is essential to measure the internal rotation, mixing, and
the masses of the growing helium and carbon-oxygen cores during the
stages of central hydrogen and helium burning.
Asteroseismology is able to measure these helium and carbon-oxygen cores as
relics of the accumulated effect of the internal rotation and mixing
throughout the two longest nuclear life phases of stars. 

Non-radial oscillations in fast rotators provide critical and new
calibrations of the internal rotation and magnetic fields to stellar
evolution theory. Aside from this basic role of asteroseismology, it also
offers guides to improve three-dimensional (3D) magnetohydrodynamical
(MHD) simulations of core-collapse supernovae
\citep[e.g.][]{Hammer2010,Wongwathanarat2015,Lentz2015,Summa2016,Mueller2017,Ott2018,OConner2018,Andresen2019,Burrows2019,Varma2023}
and to calibrate binary population synthesis studies. These two
  topics, among others, guide
predictions for gravitational wave emission from merging neutron star
and black hole binaries
\citep[e.g.][]{deMink2015,deMink2016,Marchant2016,Belczynski2016,Farr2017,Belczynski2020,Laplace2020,Marchant2020,Marchant2021,Landry2021,Schneider2021,Schneider2023,Mezzacappa2023,Wong2023,Agrrawal2023,Jiang2023,Cheng2023}.
No such studies available today have been done with
asteroseismically calibrated internal rotation and magnetic field
profiles of massive stars. Major future endeavours are therefore to be
anticipated to study the bridge between asteroseismology of massive close
binaries and multi-messenger astronomy. Asteroseismology is an
essential approach to bring a solid observational foundation to the
interpretation of gravitational wave detections in terms of the
progenitors of their merging compact
binaries (see Fig.\,\ref{Fountain}).

\section{Existing asteroseismology reviews and the aspect of binarity}

Aside from the recent reviews by \citet{Aerts2021-RMP} and
\citet{Kurtz2022} mentioned already, several earlier asteroseismology
review papers exist. These works were published during the past decade and are
dedicated to particular types of pulsators, following the tremendous
and rapid progress delivered by the 4 yr light curves assembled by
the nominal NASA {\it Kepler\/} space mission \citep{Koch2010} and its
refurbished version K2 \citep{Howell2014}.  These previous manuscripts
offer comprehensive overviews of the impact of asteroseismology on
astrophysics for specific types of stars, notably slow rotators. We
do not repeat their content here, but do take the opportunity to
present some of the latest asteroseismology results for slow rotators
not yet covered in the available reviews.  In doing so we touch upon
a few topics that stand out because of their unique potential for
impact in future astrophysics research, notably the internal magnetism of
stars. 

\begin{figure}[h!]
  \centering
\rotatebox{0}{\resizebox{8.5cm}{!}{\includegraphics{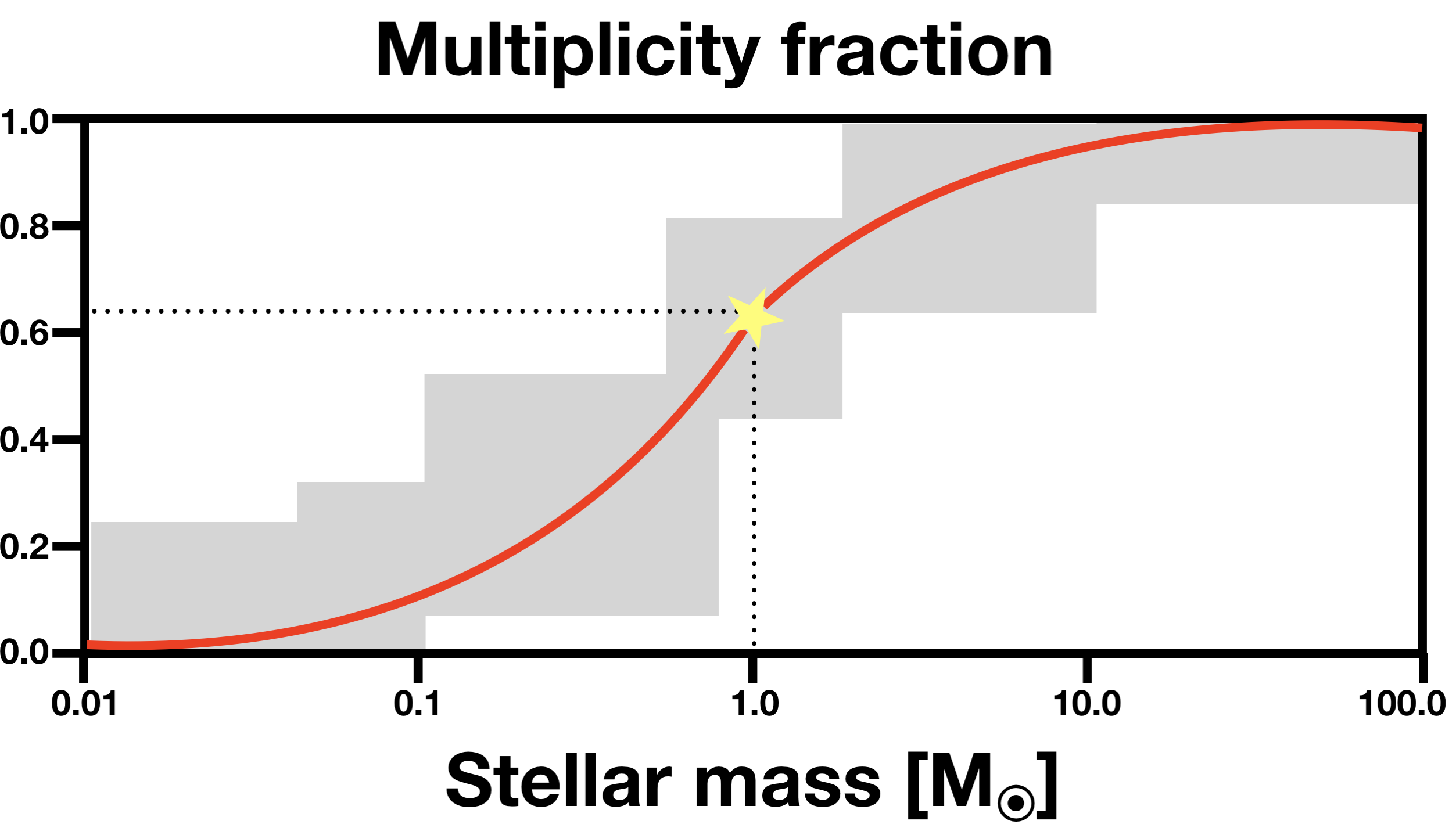}}} 
\caption{\label{BinaryFraction}Sketch representing the multiplicity
  fraction for stars and brown dwarfs (red line), with the
  uncertainties summarising 
  numerous observational studies in the literature (grey boxes).
  More than half of
  the solar-like stars situated near the yellow star symbol are expected to
  have a companion. The figure was inspired by \citet{Moe2017}.}
\end{figure}

To date, the playground of asteroseismology (i.e.\ {the range in  birth masses of stars with identified
non-radial oscillation modes) is from about 0.75\,M$_\odot$
for the K5V star $\varepsilon\,$Indi\,A \citep{Lundkvist2024}
to as high as about 25\,M$_\odot$ for the O9V $\beta\,$Cep star HD\,46202
\citep{Briquet2011}.}
However, by far the most efforts in asteroseismic modelling have been
concentrated on low-mass stars, from their birth as dwarfs through
their evolved stages as sub-giants and red giants, or as sub-dwarfs
due to loss of their hydrogen envelope, and all the way to their
end-of-life as cooling compact remnant white dwarfs \citep[see the
  reviews by][]{ChaplinMiglio2013,HekkerJCD2017,Corsico2019,GarciaBallot2019,Giammichele2022}.
Multiple reasons for this focus on stars born with
a low mass (defined here as stars with a mass in the range
0.7\,M$_\odot\leq M \leq 1.2$\,M$_\odot$)
are {relevant}. Low-mass stars have the following characteristics: 
\begin{enumerate}
\item
  They constitute the dominant population in our Milky Way galaxy, allowing
   the study of its history, {archaeology,} and current structure.
  {This is currently being done from} the multitude of {\it Gaia\/} data
  \citep{Prusti2016,Vallenari2023}
  and ongoing or near-future dedicated all-sky survey
  high-resolution spectroscopy
  \citep{Kollmeier2017,Pinsonneault2018,Bensby2019,Jin2023};
\item
  They reveal high-frequency solar-like oscillations excited stochastically
  by the turbulent convection in their envelopes throughout almost
  their entire nuclear burning life. Such oscillations obey scaling
  relations, which were already {summarised} by \citet{Kjeldsen1995} long
  before the era of space asteroseismology. These scaling relations
  make it easy to deduce the mass, radius, and age from radial,
  dipole, and quadrupole modes, with high precisions of a few
  percent. This is achieved by comparing the observed properties of
  detected low-degree
  global oscillation modes with those of the Sun
  \citep[see e.g.][for an extensive review]{GarciaBallot2019}.
  It has the major advantage that
  one can use the heritage from helioseismology
  \citep{JCD2002,JCD2021} and transfer it to asteroseismic
  applications to solar-like oscillations;
\item
  They cover the most important parameter space of exoplanet host stars,
  including those with multiple Earth-like rocky planets in the
  habitable zone;
\item
  They have a relatively modest binary and multiplicity fraction compared
  to stars of higher mass (see Fig.\,\ref{BinaryFraction}); 
\item    
  They are all slow rotators from early in their life
  as they are subject to efficient magnetic 
braking during their core hydrogen burning phase. As we   detail in
the next sections, this implies that the
impact of rotation on the modelling of their oscillations is in the
easiest regime of the frequency domain. We come back to this important
aspect throughout this review.
\end{enumerate}
None of these five aspects is valid for moderate to fast rotators of
intermediate to high mass. We define stars born with 1.3\,M$_\odot\leq
M \leq 8$\,M$_\odot$ as intermediate-mass stars and those with $M >
8$\,M$_\odot$ as high-mass stars.  For the pulsators in these two mass
regimes we do not have a main-sequence calibrator like the Sun, whose
{helioseismology} \citep{JCD2002} has been essential to guide similar
applications to low-mass stars.

Though a minority compared to their low-mass colleagues, stars
born with $M\geq 1.3$\,M$_\odot$ deliver a major fraction of the
energy and heavy elements to galaxies and to the Universe as a
whole. For this reason, and given the fast progress of
asteroseismology for this under-represented category of fast rotators,
we dedicate most of this review to asteroseismic applications of stars
born with intermediate or high mass. In the next section we clarify
what we mean by moderate and fast rotation.

Figure\,\ref{BinaryFraction} shows the binary fraction for stars and
brown dwarfs. This figure makes it clear that we cannot ignore stellar
multiplicity in a review focusing on the  asteroseismology of intermediate-
and high-mass stars. In particular, close binarity dominates the
evolution of the most massive stars \citep{Sana2012,Sana2013,Sana2014}. While
this was firmly established observationally   a decade ago, it is recognised that
binary evolution theory is still affected significantly by the major
uncertainties occurring in the theory of single-star evolution
\citep{MarchantBodensteiner2024}. Hence, this must be given the
highest priority, and asteroseismology is the optimal way to do so
\citep{Aerts2021-RMP}.  We   touch upon various aspects of
binarity and multiplicity for asteroseismic studies in Sect.\,5 and
Sect.\,6.  The class of sub-dwarfs is discussed briefly in Sect.\,7. It
deserves special attention in this respect as the binary fraction is
high \citep{Han2003,Hu2008,Vos2017,Vos2019}.  In fact, all sub-dwarfs
originate from binary interactions because the seemingly single class
members can only be understood in terms of a common envelope phase
resulting in a merger \citep{Heber2009}.  Due to the efficient loss of
their envelope, the core-helium burning sub-dwarfs are of low mass in
their current phase of evolution (about 0.5\,M$_\odot$). Those
with oscillations reveal high-frequency modes with periods much
shorter than their rotation and/or orbital period \citep[see][for a
  review]{LynasGray2021}.  We also briefly touch upon merger
seismology for higher mass objects, offering a way to hunt for such stars
asteroseismically. Section\,8 is dedicated to future outlooks, putting
emphasis on {\it Gaia\/}'s role in asteroseismology.

Rather than trying to be exhaustive in reviewing the progress of
asteroseismology, this paper takes the approach of offering the
non-expert reader highlights from the recent literature based on
prominent case studies. This way, we illustrate what can be
achieved with current and future data and theoretical developments.
The plan of the paper is as follows. We provide an {accessible}
introduction to asteroseismology in Sect.\,3, touching upon the
wave equations representing stellar oscillations and the accompanying
frequency regimes. Section\,4 covers recent applications of
asteroseismology of single stars, starting with the Sun and
gradually increasing the level of
complexity in the mode computations
as the ratio of the oscillation mode periods to the
rotation period increases. In Sect.\,5 we highlight how 
model-independent dynamical masses and radii help asteroseismology,
before moving on to tidal asteroseismology in Sect.\,6 and sub-dwarf
and merger seismology in Sect.\,7.
We end the paper with an extensive outlook for the bright future of asteroseismic applications
to stars of intermediate and high mass from combined {\it Gaia\/} data, survey
spectroscopy, and ongoing TESS or future PLATO space photometry.
{There is a pertinent need for new theory and better modelling
tools} in order to interpret the oscillations of
fast rotators and close binaries up to
current measurement precisions.

\section{Asteroseismology's unique probing power}

This review is focused on applications of asteroseismology to rotating
stars, while omitting jargon so as to make it {easily} accessible.
A major focus is put on future prospects and
opportunities (see Fig.\,\ref{Fountain}). Hence, we   only
introduce concepts and equations that are absolutely necessary to keep
the text self-contained. We refer to the monographs by
\citet{Unno1989} and \citet{Smeyers2010} for full derivations of the
oscillation equations. {Further, we point} to \citet{Aerts2010} and
\citet{BasuChaplin2017} for extensive descriptions of the analysis tools
and the principles of stellar modelling in the modern era of
space asteroseismology.

Here it is enough to know that non-radial oscillations lead to
periodic deviations from the equilibrium structure of the star. Each
mode moves the fluid elements away from the position they would have
without any seismic activity. Each mode displaces the fluid elements
with a periodicity corresponding to the mode's frequency and its
nodes. Since this 
perturbs a three-dimensional (3D)
body, three numbers are required to describe the geometry of each
mode. Two of these are called `wavenumbers'
and stand for the number
and position of nodal lines at the stellar surface with respect to the
symmetry axis of the oscillations. In the case of a spherical star in
equilibrium that gets perturbed by the modes, these wavenumbers are
usually denoted as $(l,m)$, which characterise the nodal lines of a
spherical harmonic function $Y_l^m(\theta,\phi)$ used to describe the
angular part of the displacement vector due to the mode.
{Assignment of the third number, $n$, is connected with the
  number of nodes in the radial component of the eigenfunction. It is
  more involved than the cases of $l$ and $m$ because it depends on
  the dominant restoring force of the oscillations
  \citep{Aerts2010}. For the simplest case of radial ($l=0$) oscillations, the
  pressure gradient is the restoring force, and hence these are
  acoustic eigenmodes (or p~modes) of the star. The radial mode with the $n$-th
  lowest eigenfrequency has $n$ nodes in its displacement
  vector. Since the centre of the star is a node, $n-1$ nodes occur
  between the centre and the surface of the star and one speaks
  of the $(n-1)$-th overtone, while the radial mode having only a node
  in the star's centre is called the fundamental mode. The
  classifcation of non-radial modes ($l\neq 0$) in terms of their
  overtone is considerably more complex because additional restoring
  forces, such as buoyancy, come into play.  We refer to
  \citet{Takata2012} for a detailed mathematical description of mode
  classification in terms of the overtone $n$, denoted here as p$_n$ and
  g$_n$ for p and g~modes of radial order $n$, respectively.}
{We end this part on the mode numbers by pointing out that the white dwarf
  community has historically denoted the radial order as $k$ instead
  of $n$.}

\subsection{Stellar evolution models in equilibrium}

Modern asteroseismic modelling applications require numerical
solutions of the relevant oscillation equations described in the next
sub-section.  The modelling hence relies on realistic unperturbed
background stellar models.  Any fluid element inside a rigidly
rotating star is characterised by its position vector $\bm{r}$ with
respect to the star's centre and the time coordinate $t$ measured
since the  birth of the star (called the zero age main sequence or ZAMS
defining time zero). The fluid elements  have to fulfil the
equations of stellar structure at any time during the star's evolution
\citep[see the monograph on models of rotating stars by][for the
  derivations and solutions of these equations]{Maeder2009}.

Since we wish to introduce and discuss the dynamical
properties of stellar oscillations, we limit ourselves to a
description of the equation of motion, and we focus on a rigidly
rotating star for now.
Expressed in a frame of reference co-rotating with the star's rotation
vector $\bm{\Omega} = \Omega\, \bm{e}_z$, where $\bm{e}_z$ is the unit
vector along the rotation axis, the  equation of motion reads
\begin{equation}
\renewcommand{\arraystretch}{2.8}
\begin{array}{l}
  \label{motion}
  \displaystyle{
    \frac{\partial \bm{v}}{\partial
      t}+\left(\bm{v}\cdot\nabla\right)\,\bm{v}+
    \underbrace{2\,\bm{\Omega}\,\times\,\bm{v}}_{\rm
      Coriolis}\,+\underbrace{\,\bm{\Omega}\,\times\,\left(\bm{\Omega}\,\times\,\bm{r}\right)}_{\rm
      centrifugal}}\ =\ \\
\displaystyle{
  \ \ \ \ \ \ \ \ \ \ \ \ \ \ \ \ \ \ \ \ \ \ \ \ \ \ \ \ \ \ 
  \underbrace{-\ \frac{\nabla
      p}{\rho}}_{\rm pressure}\ \underbrace{-\ \nabla\Phi}_{\rm
    gravity}\ +\ \bm{a}_{\rm extra}^{\rm internal}\ +\ \bm{a}_{\rm
    extra}^{\rm external}\, .} 
\end{array}
\end{equation}
In this equation, $\rho$ is the density, $p$ is the pressure, $\bm{v}$
is the velocity vector, and $\Phi$ is the gravitational
potential fulfilling the equation $\nabla^2\Phi=4\pi G\rho$ with $G$
the gravitational constant. Aside from the two accelerations due to
the pressure and gravitational forces, the two terms
$\bm{a}_{\rm   extra}$ represent extra accelerations due to the joint effect of
any active forces not spelled out explicitly, one term for forces active
inside the star and the other one for forces imposed by external sources. An
example of $\bm{a}_{\rm extra}^{\rm internal}$ is the Lorentz force
caused by an internal magnetic field or accelerations due to radiative
forces resulting in a dust-driven or line-driven stellar
wind. Additionally, extra accelerations $\bm{a}_{\rm extra}^{\rm  external}$ may be
caused by external forces such as magnetic fields and/or tides due to one or more companions.

For an unperturbed background model of a single star not subject to
any extra forces aside from those due to pressure gradients, gravity, and rotation,
the equilibrium state at time $t$ is described by
$\bm{v}=\bm{0}$. In that case, it follows from Eq.\,\eqref{motion}
that the star is an oblate spheroid flattened due to its centrifugal force
\citep[see][for an extensive discussion]{EspinozaLara2013}.  Most
of the current stellar evolution codes simplify the oblateness of stars
caused by rotation and describe the gaseous spheroids by using only
one spatial coordinate, for instance the distance $r$ from a fluid element to
the centre of the star.  Thus, any fluid element is characterised by 
two coordinates, denoted here as  $(r,t)$ with $t$ the star's age.

Almost all the asteroseismic applications discussed in this review are based on
such one-dimensional (1D) background models evolving with time, assuming
either rigid rotation with constant frequency $\Omega$ or shellular
rotation \citep{Zahn1992} for which the rotation frequency only
depends on $r$ and not on latitude $\theta$ or longitude
$\phi$. Shellular rotation is denoted here as 
$\Omega(r)$ and results from the assumption of strong
horizontal turbulence, forcing a constant rotation rate along isobars.
We discuss prospects for asteroseismic applications relying on more
realistic yet more complex 2D or 3D background models in the last section.

\subsection{The wave equations describing the dynamics of stellar
  oscillations in a rotating star}

Each oscillation mode causes the fluid elements in the star to become displaced from
their equilibrium position according to the Lagrangian vector
$\bm{\xi}  (\bm{r},t)$.
In general, the oscillation equations for a rotating star are
of the form
\begin{equation}\label{nonlinear}
  \ddot{\vec{\xi}} + 
  2\,\bm{\Omega}\times \dot{\vec{\xi}} + \bm{O}^{(1)}(\vec{\xi}) =
\bm{O}^{(2)} (\bm{\xi},\bm{\xi})\ + \bm{O}^{(3)}
(\bm{\xi},\bm{\xi},\bm{\xi}) + \ldots\, ,
\end{equation}
where a superscript dot stands for a time derivative and the operators
$\bm{O}^{(i)}$ group all acting forces such that their application
represents a collection of terms of $i$-th order in
$\bm{\xi}$ for $i=1, 2, 3, \ldots$.
In order to solve Eq.\,\eqref{nonlinear}, one can take various
approximations and approaches, depending on the importance of the Coriolis, centrifugal,
magnetic, and tidal forces. The validity of approximations depends
strongly on the mode frequency regimes with respect to the rotation
frequency, whether or not we have to take into account deformation due
to the centrifugal force, and whether we are dealing with rigid or non-rigid rotation.
We briefly discuss some of the options, but refer to the references for details.

Asteroseismology is most often applied in a linear framework, where it
is assumed that any perturbation of the background equilibrium model
stemming from an oscillation mode is sufficiently small to ignore
the effects of order higher than one in the
displacement $\bm{\xi}$ in Eq.\,\eqref{nonlinear}. This means that we
can ignore all terms on the right-hand side in Eq.\,\eqref{nonlinear}.
{In this case, the equation of motion in Eq.\,(\ref{motion}) due to the
mode $\bm{\xi}$ is the much simpler version of Eq.\,\eqref{nonlinear},
namely
\begin{equation}\label{SL}
  \ddot{\vec{\xi}} + 
  2\,\bm{\Omega}\times \dot{\vec{\xi}} + \bm{O}^{(1)} (\vec{\xi}) =
\bm{a}_{\rm extra}^{\rm  external}\, ,
\end{equation}
where the operator $\bm{O}^{(1)}$ contains all acting internal forces.
If we further write the Lagrangian displacement experienced by a fluid
element inside a non-rotating non-magnetic single spherical
star at position $\bm{r}$ and time $t$ due to a periodic
linear eigenmode  with frequency  $\omega$ as
\begin{equation}
  \label{ksit}
    \vec{\xi}(\bm{r},t) = \vec{\xi}(\bm{r})\,\exp\,({-{\rm i} \omega
      t)}\,  ,
\end{equation}
the wave equation simplifies to
\begin{equation}\label{simplest-wave}
\omega^2 {\vec{\xi}} = \bm{O^{(1)}}(\vec{\xi}) ,
\end{equation}
with $\bm{O}^{(1)}$ a linear function determined by the equilibrium
values and  first-order perturbations of the density,
pressure, and gravitational potential}
\citep[e.g.][Chapter\,3]{Aerts2010}. This is the simplest version of
the wave equation to solve for a family of linear spheroidal
non-radial oscillation modes. We highlight some of the latest findings
of helio- and asteroseismology in this simplest approximation in
Sect.\,4.1.

Versions of non-linear theory of non-radial oscillations based on
Eq.\,\eqref{nonlinear} have also been developed, up to second
\citep{Dziembowski1982,BuchlerGoupil1984}, third
\citep{VanHoolstSmeyers1993,Buchler1995,Mourabit2023}, or fourth
\citep{VanHoolst1994} order in $\bm{\xi}$, while ignoring the rotation
of the star or by considering it to cause only a small perturbation. However, applications of asteroseismic modelling based on 
higher-order (in $\bm{\xi}$) theory are  scarce. They
occur for oscillation modes in white dwarfs \citep{Zong2016a},
sub-dwarfs \citep{Zong2016b}, red giants
\citep{Weinberg2019,Weinberg2021}, and a $\delta\,$Sct star
\citep{Mourabit2023}, all of which treat rotation as a small
perturbation for the computation of the displacement vectors.
Modelling of tides in close binaries may also require non-linear
non-radial oscillation theory to solve Eq.\,\eqref{nonlinear} when
linear tides do not provide a sufficiently accurate description, as
theorised by \citet{Weinberg2012,Ogilvie2014}.

A different aspect of simplifying Eq.\,\eqref{motion} and
Eq.\,\eqref{nonlinear} deals with  the treatment of rotation, notably the
importance of the Coriolis and centrifugal forces and how they compare
to the Lorentz force. The rotational effect in the equation of motion
leads to terms up to $\Omega^2$.  The Coriolis force creates vorticity.
It may become the dominant restoring force instead of the
pressure force or gravity. Hence, one encounters additional families of
modes, which do not exist in non-rotating stars. A well-known example
from both geophysics and astrophysics involves the family of toroidal
modes. For a pedagogical derivation and discussion of the various
types of mode families in rotating stars, we refer to
\citet{Townsend2003}.

The strategy used to compute oscillation modes in a rotating star depends
entirely on how the mode periods compare to the rotation period. If
these two are of comparable order,  the rotation cannot be treated as a
small effect to compute $\bm{\xi}$ (see also the next section).  If
the rotation happens on a much longer timescale than that of the
oscillations,   the rotational effects can be treated in a perturbative
approach.
{This is typically a good strategy when the rotation period is at
  least ten times longer than the mode periods.
In such a case,} perturbation theory for proper computation of
$\bm{\xi}$ can be developed up to any order in
$\varepsilon\equiv\Omega/\omega$, where $\varepsilon$ is  
a small expansion parameter for computing the displacement in the form
$\bm{\xi} = \bm{\xi}_0 + \varepsilon\,\bm{\xi}_1 +
\varepsilon^2\,\bm{\xi}_2 + \ldots\,$. We note that this methodology
only makes sense if $\varepsilon$ is sufficiently small as an
expansion parameter, and hence  the modes should have far shorter periods
in the frame of reference co-rotating with the star compared to the
rotation period. Within such an approach, one must then also decide
whether the deformation of the star, which is $\propto\Omega^2$, can
be ignored or not. If it must be taken into account, the seismic
modelling requires a way to incorporate the deformation, either at the
level of the equilibrium structure or at the level of the expression
for $\bm{\xi}$, or both.
Often one considers
{the simplest} deformation
only at the level of the equilibrium structure, where the
{contribution to the potential is approximated by its spherically
  symmetric component}
\citep[see Eq.\,(30) in][]{Aerts2021-RMP}.
Whatever the choice of how to deal with the
deformation, one also needs to specify whether and up to what order the
terms due to the Coriolis and centrifugal forces couple to each other
when computing $\bm{\xi}$.

A variety of first-, second-, and third-order
perturbative non-radial pulsation theories for the computation of
spheroidal and toroidal families of mode solutions is available in the
literature, relying on the assumption that all the terms within the
operator $\bm{O^{(1)}}$ as well as the term $2\,\bm{\Omega}\times
\dot{\vec{\xi}}$ in Eq.\,\eqref{nonlinear} cause only small deviations
from the solutions to Eq.\,\eqref{simplest-wave} represented by
$\bm{\xi}_0$.  We note that the computation of $\bm{\xi} = \bm{\xi}_0
+ \varepsilon\,\bm{\xi}_1 + \varepsilon^2\,\bm{\xi}_2 + \ldots\,$ from
perturbation theory to treat the rotation may involve mode coupling,
particularly among spheroidal and toroidal modes, but this is a
different matter than adopting non-linear oscillation theory by relying
on the right-hand side of Eq.\,\eqref{nonlinear}.  The
series $\bm{\xi} = \bm{\xi}_0 + \varepsilon\,\bm{\xi}_1 +
\varepsilon^2\,\bm{\xi}_2 + \ldots\,$ in perturbation theory used to
compute linear modes of rotating stars can be truncated after taking
sufficient terms in numerical computations, depending on the
envisioned precision required for the modelling application.
Perturbative approaches for asteroseismic modelling adhering to the
requirement $\Omega<\!<\!\omega$ were developed theoretically
 long before the modern era of space asteroseismology
\citep[e.g.][{each of which 
  containing a lot of technical
  details}]{Ledoux1951,Saio1981,GoughThompson1990,Dziembowski1992,DziembowskiGoode1996,Soufi1998,Daszynska2002,Karami2008}.

In summary, for each particular asteroseismic modelling application, proper
balancing between the acting forces in Eq.\,\eqref{motion} and
Eq.\,\eqref{nonlinear} is necessary in order to decide upon the best
theoretical formalism and the most suitable numerical approach to
calculate the oscillation modes. The choice between a perturbative or
non-perturbative approach to calculate the modes of equilibrium models
depends entirely on the frequency regime of the detected modes under
investigation, notably the value of $\varepsilon$ \citep[see
  e.g.][]{Ballot2010,Ballot2013}. This is the major
reason why we organise the applications discussed in the following sections
according to the regimes of the frequencies of the detected
oscillations.

\subsection{Oscillation mode frequency regimes}

\begin{figure*}[t!]
\begin{center}
  {\resizebox{18.cm}{!}{\includegraphics{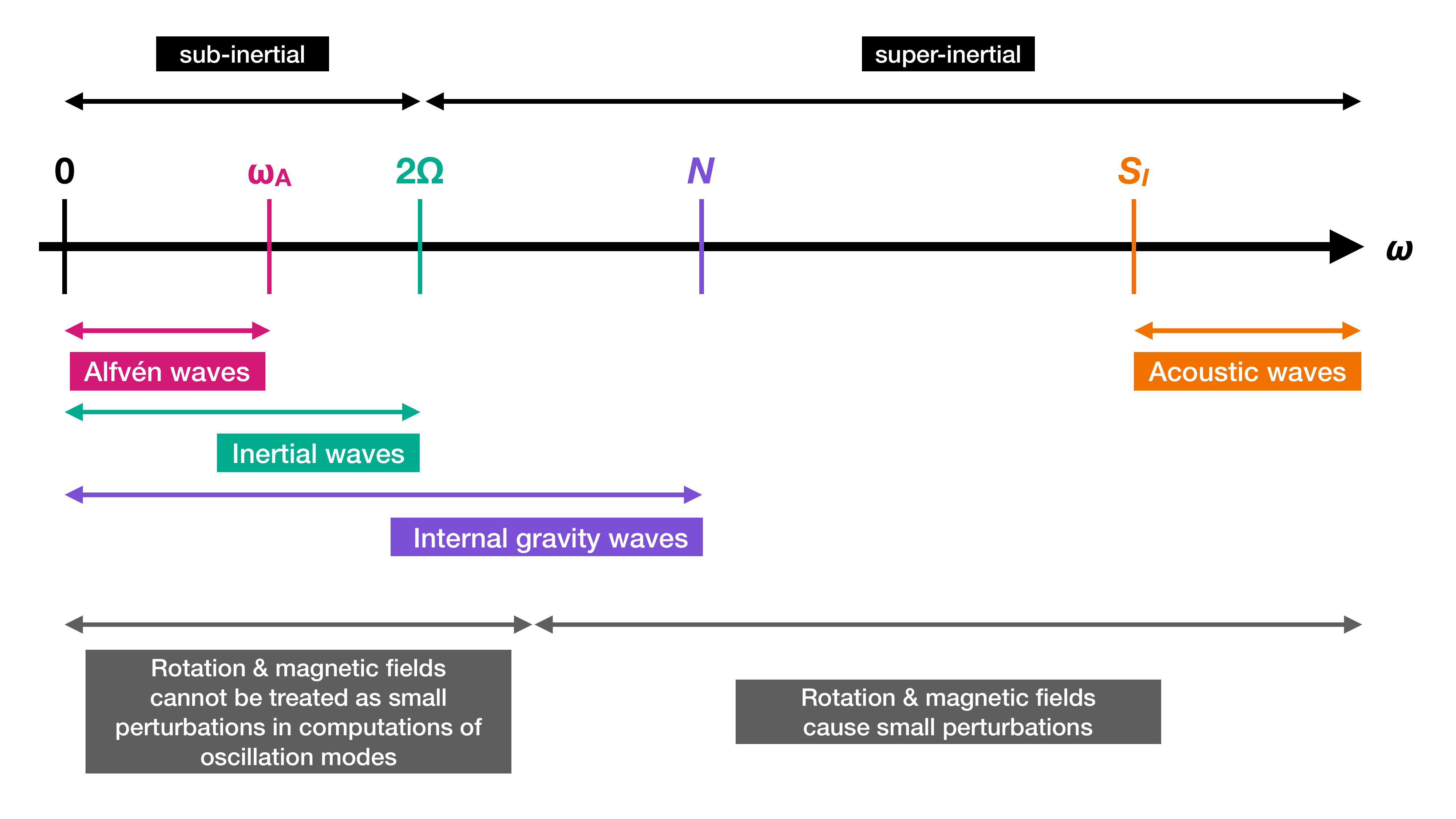}}}\vspace{-0.5cm}
\end{center}
  \caption{\label{Frequency-Axis} Axis showing various frequencies
    of relevance inside a rotating star,
    notably the Lamb frequency for a wave
    of degree $l$ denoted as $S_l$, the Brunt-V\"ais\"al\"a frequency
    $N$, the rotation frequency $\Omega$, and the frequency due to
    Alfv\'en waves $\omega_{\rm A}$. The relevant frequency regimes
    for different types of waves are indicated, as well as the regimes
    where particular forces can (or cannot) be treated perturbatively in
    the computation of oscillation modes. Waves in the sub-inertial
    regime have spin parameter $s\equiv 2\Omega/\omega>1$, while waves
    in the super-inertial regime are characterised by $s<1$. The
    indicated frequencies change throughout the evolution of the star.
    The figure was inspired by the scheme in \citet{MathisDeBrye2011} and
    is adapted from the version in \citet{Aerts2019-ARAA}, by
    Clio Gielen.}
\end{figure*}
  
All stars rotate, even if many do so slowly, by which we mean that
their rotation velocity is only a small fraction (up to 10\%) of
their critical break-up velocity. The seemingly simple concept of
break-up velocity, defined as the velocity of fluid elements at the
stellar equator high enough to overcome the gravitational attraction,
is non-trivial. The centrifugal acceleration due to fast
rotation can be accompanied by extra outward forces, making it easier
to overcome the inward force of gravity. An example is the
case of a radiation-driven wind \citep{KudritzkiPuls2000}
or the tidal pull by a companion \citep{Fuller2019},
helping matter to escape more easily
from the star compared to the case without such helpful forces.

In this review we focus on intermediate- and high-mass stars with
sufficient detected and identified  non-radial oscillation
modes in terms of their angular mode dependence to scrutinise stellar
equilibrium models. This means that the identified modes allow 
quantitative measurements of the
internal rotation profile and possibly the  internal
magnetic field. With a few exceptions,
asteroseismic modelling based on sufficient modes to achieve this in
such types of stars has only begun in the space asteroseismology era.
At the time of writing, stars fulfilling these requirements have a
mass typically below 25\,M$_\odot$ for the core-hydrogen burning
phase, and lower for the hydrogen-shell or core-helium burning
phases. We   therefore ignore the effect of a dynamical wind and
rely on a static outer boundary condition to compute the oscillation
modes via Eq.\,\eqref{nonlinear}.  We treat the case of tidal forces
due to a close companion in Sect.\,6.

We use the definition of the Keplerian angular
critical frequency given by
\begin{equation}
  \Omega_{\rm crit}^{\rm Kep} \equiv\,\sqrt{\frac{GM}{R_{\rm eq}^3}}\,  ,
\end{equation}
  where $M$ is the mass and $R_{\rm eq}$ the equatorial radius of the
  star, in comparison with the star's oscillation frequencies in a
  co-rotating frame.  The equatorial radius of a star is most often
  unknown, unless the star has a well-resolved interferometric image 
{such as the fast-rotating stars Altair
  \citep{Domiciano2005,Monnier2007,Bouchaud2020}, Rasalhague
  \citep{Monnier2010}, Vega \citep{Monnier2012}, or Archenar
  \citep{Domiciano2014}.}
  Lack of knowledge of $R_{\rm eq}$ is one of
  the reasons why the critical rotation rate adopted in calculations
  often corresponds to the Roche critical frequency, defined as
\begin{equation}
  \Omega_{\rm crit}^{\rm Roche} \equiv\,\sqrt{\frac{8}{27}
    \frac{GM}{R_{\rm pole}^3}}\ , 
\end{equation}
with $R_{\rm pole}$ the star's polar radius. These two expressions are
not the same because the polar radius only equals $2R_{\rm eq}/3$ when
the actual critical rotation velocity of the star is reached (which is
never the case as the star would no longer exist). For a more
elaborate discussion on the relationship between $\Omega_{\rm
  crit}^{\rm Roche}$ and $\Omega_{\rm crit}^{\rm Kep}$, we refer to
\citet{Rieutord2016}.  In citing the results below, we   explicitly
mention the definition adopted by the authors for the critical
rotation frequency whenever available.

Here  we define a star as a slow rotator if its rotation
frequency is less than 10\% of its Keplerian critical frequency, a
moderate rotator if it has a rotation frequency between 10\% and 70\%
of its Keplerian critical frequency, and a fast rotator if its
rotation frequency is above 70\% of the Keplerian critical frequency.
This may differ substantially from the
{definitions adopted by spectroscopists, who usually only consider
  $v\sin\,i$ to decide whether a star is a fast or a slow rotator.  As an
  illustration of this, we computed $\Omega_{\rm crit}^{\rm Roche}$
  and $\Omega_{\rm crit}^{\rm Kep}$ for the 12\,M$_\odot$ $\beta\,$Cep
  star HD\,192575, which has a measured
  $v\sin\,i\simeq\,27\,$km\,s$^{-1}$ and whose oscillation spectrum is
  shown in Fig.\,\ref{Siemen} (discussed further in the text).  Based on
  the high-resolution spectroscopic estimate of $v\sin\,i$, one would
  be tempted to classify this star as a slow rotator because most
  B-type stars have 5 to 15 times higher $v\sin\,i$.  However,
  asteroseismologists are not hampered by the unknown $\sin\,i$ factor
  as they measure $\Omega$ directly from the oscillation
  frequencies \citep[see][for a sample of B stars]{Pedersen2022b}.
For HD\,192575 this leads to an equatorial rotation velocity between
  75 and 100\,km\,s$^{-1}$ and an inclination angle between $10^\circ$
  and $30^\circ$ \citep{Burssens2023}.
  Moreover, what matters for asteroseismic modelling is
  the ratio of twice the rotation frequency and the oscillation mode
  frequency in a frame of reference co-rotating with the star,
  called the mode spin parameter and defined as $s\equiv 2\Omega/\omega$.
From the asteroseismic modelling of this star by \citet{Burssens2023},
we find that the star's rotation frequency
at the position of the $\mu$-gradient zone adjacent to the receding
convective core has $\Omega/\Omega_{\rm crit}^{\rm Roche}=32\pm 5\%$
and $\Omega/\Omega_{\rm crit}^{\rm Kep}=17\pm 3\%$, while its
radiative envelope near the surface rotates at $\Omega/\Omega_{\rm
  crit}^{\rm Roche}=28\pm 4\%$ and $\Omega/\Omega_{\rm crit}^{\rm
  Kep}=15\pm 2\%$. The spin parameters of the dipole and
quadrupole modes range from about 5\% to 20\%.} Hence, we classify this
star as a moderate rotator because second-order rotational effects should
not be neglected to achieve an optimal interpretation of its observed
rotationally split mode frequencies shown in Fig.\,\ref{Siemen}.  We
come back to the unique asteroseismic study of this star as a key
example of low-order p- and g-mode asteroseismology of a supernova
progenitor in Sect.\,4.2.

Aside from the mode frequency $\omega$ and the star's rotation
frequency $\Omega$ (assuming rigid rotation for now), several more
frequencies connected with the internal dynamical properties of a
star are of importance. The cavity of the p~modes has a characteristic acoustic frequency 
for each of the modes, known as the Lamb frequency $S_l$ and defined as
\begin{equation}
  \displaystyle{S_l^2(r) = \frac{l ( l +1)\ c^2(r)}{r^2}} ,
\end{equation} 
with $c(r)$ the local sound speed. Pressure modes of degree $l$ are
only propagative in the region where $\omega > S_l$. Gravity modes, on
the other hand, are only propagative if their frequency is below the
Brunt-V\"ais\"al\"a frequency $N$, defined as
\begin{equation}
\displaystyle{
N^2(r) = g(r) \left( \frac{1}{\Gamma_1\cdot p(r)} \frac{{\rm d} p(r)}{{\rm d} r} 
- \frac{1}{\rho (r) } \frac{{\rm d} \rho (r)}{{\rm d} r }\right)} \;  ,
\end{equation}
with $\Gamma_1$ the first adiabatic exponent:
\begin{equation}
  \label{Gamma1}
\Gamma_1 = { \left( \frac{\partial \ln p}{\partial \ln \rho} \right) }_{\rm ad}\; .
\end{equation}
Finally, some layers in stars may be subject to a magnetic field,
whose origin, evolution, and properties vary from star to star. Such layers
experience plasma waves restored by the Lorentz force. These waves are
characterised by their  Alfv\'en frequency, $\omega_{\rm A}$,
corresponding to a wave speed
\begin{equation}
v_{A}=\bm{B}\cdot\bm{k}/\sqrt{\mu_0\,\rho} , 
\end{equation}
  with $\bm{B}$ the
magnetic field, $\bm{k}$ the wave vector, and $\mu_0$ the permeability.

Figure\,\ref{Frequency-Axis} assembles all the introduced frequencies
into a single axis representing a classification of the types of waves
corresponding to their dominant restoring force.  The dynamical
properties of the resonant eigenmodes depend on the relationship between
their frequencies $\omega$ and the Lamb, Brunt-V\"ais\"al\"a, rotation, and Alfv\'en
frequencies.  The types of approximations that can be made to compute
the eigenmode frequencies $\omega$ via Eq.\,\eqref{SL} for $\bm{a}_{\rm
  extra}^{\rm external}=\bm{0}$ are also indicated in the figure. In particular,
given that {almost} all stars rotate
{with periods shorter than a decade}, we make a distinction of their waves in
the sub-inertial and super-inertial regimes, defined as those having
spin parameter $s\equiv 2\Omega/\omega$ above or below 1.
Throughout the life of a star, the frequencies indicated in
Fig.\,\ref{Frequency-Axis} change appreciably. We refer to
the extensive Table\,A.1 in \citet{Aerts2010} for
typical values of mode periods for all the different types of
pulsators, along with typical ranges of the dominant mode amplitudes,
luminosities, and effective temperatures corresponding to their
excitation mechanism and evolutionary stage as indicated by their
position in the HRD discussed in Fig.\,1 of \citet{Aerts2021-RMP}.

\begin{figure*}
\begin{center}
{\resizebox{17cm}{!}{\includegraphics{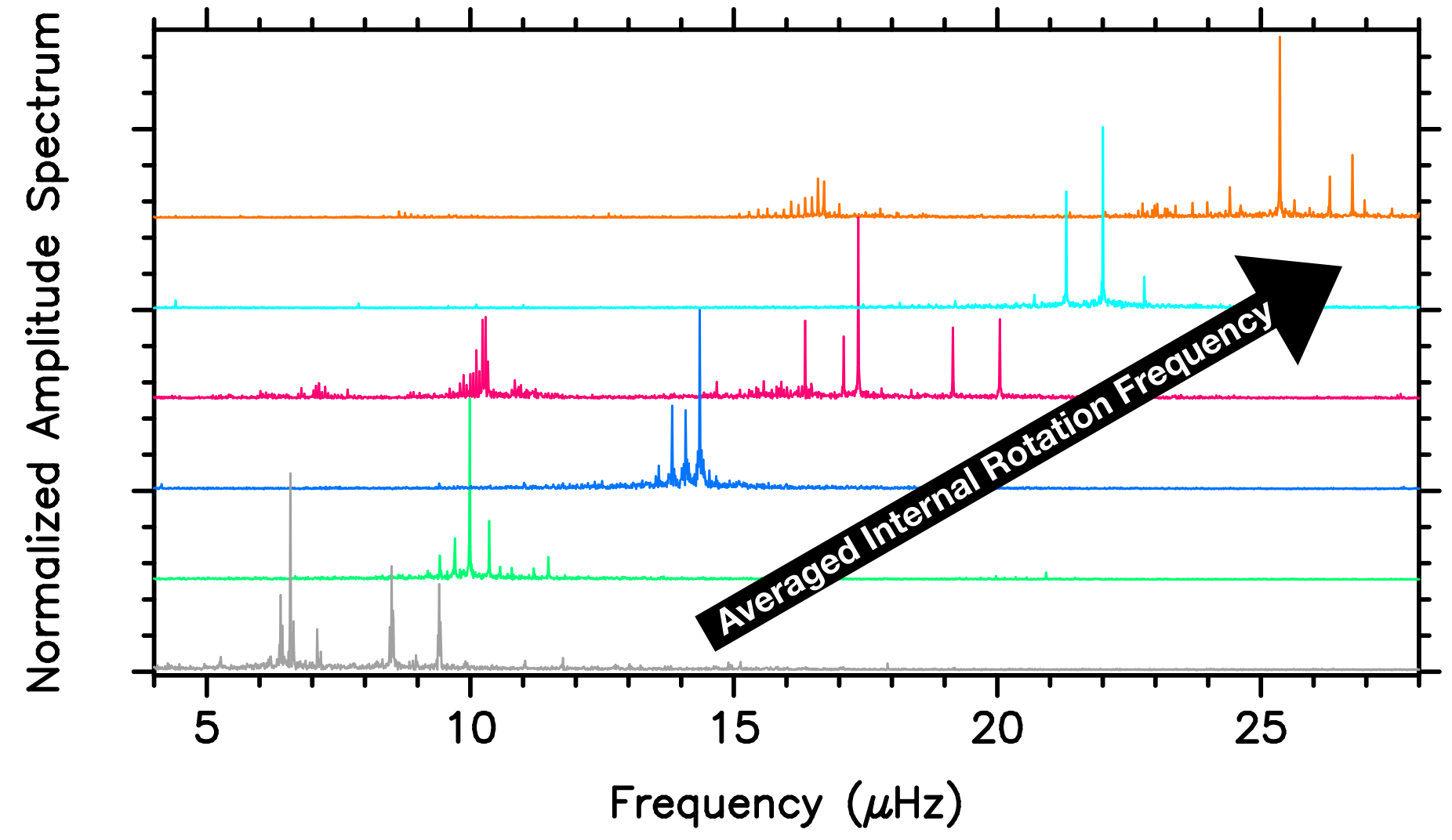}}}\vspace{-0.5cm}
\end{center}
\caption{\label{FT-gmodes} Gallery of amplitude spectra deduced  from 4 yr {\it Kepler\/}
light curves for six dipole prograde g-mode pulsators of intermediate mass.
The graph illustrates typical frequency shifts in an inertial frame of
reference connected with $(l,m)=(1,+1)$ modes in such stars.
The lower curve is for the only  slow-rotating SPB
star with detected rotationally split
triplets, KIC\,10526294. This star's
19 dipole mode triplets were discovered by \citet{Papics2014} and 
subjected to rotation inversion, following the forward asteroseismic
modelling of its zonal dipole modes by \citet{Moravveji2015}. Its 
internal rotation profile deduced from mode inversions 
led to a rotation frequency of  0.16\,$\mu$Hz \citep{Triana2015}.
The upper curve represents the data for the $\gamma\,$Dor
star KIC\,9210943, rotating essentially rigidly, with
frequency 19.73\,$\mu$Hz near its convective core
and 19.70\,$\mu$Hz at its surface
\citep{VanReeth2018}.}
\end{figure*}

We emphasise
here that the frequencies of gravity modes or internal
gravity waves of a moderate or fast rotator detected by an observer in the
inertial frame of reference characterised by $(r,\theta,\phi,t)$ are
appreciably different from the frequencies computed for a reference frame
co-rotating at frequency $\Omega$ with the star characterised by
$(r',\theta',\phi')$.  This is 
due to purely geometrical reasons, even without taking into account
the Coriolis or centrifugal forces. We can easily understand this
based on the simplest case of a slow rotator, for which the geometry
of an eigenmode can be approximated by a spherical harmonic,
$Y_l^m(\theta',\phi')$. In terms of time dependence, the
transformation from the co-rotating to an inertial reference frame
follows from
\begin{equation}
  \label{transfo}
(r',\theta',\phi')\,=\,(r,\theta,\phi - \Omega\,t)\, ,
\end{equation}
such that a frequency in the co-rotating frame, $\omega_{\rm corot}$
will be detected as $\omega_{\rm inertial} = \omega_{\rm corot} +
m\,\Omega$ by an observer.  While this geometrical shift with
$m\,\Omega$ may be relatively small for high-frequency modes of a slow
rotator, the effect is large for low-frequency modes of a fast
rotator. As an example, let us consider typical low-degree prograde
($m > 0$) or retrograde ($m < 0$) modes of slowly pulsating B (SPB) or
$\gamma\,$Doradus ($\gamma\,$Dor) stars. These {modes} typically have
frequencies of about 5 to $20\ \mu$Hz in a co-rotating frame of
reference. These mode frequencies get shifted by $m\,\Omega$ in
an observer's inertial frame, as illustrated in Fig.\,\ref{FT-gmodes}.
Following the measured rotation frequencies $\Omega$ for these two
classes of pulsators \citep[cf.\ Fig.\,6 in][]{Aerts2021-RMP}, this
geometrically based shift can reach values up to 30$\ \mu$Hz and
$-30\ \mu$Hz for prograde and retrograde dipole ($l=1$) modes,
respectively. Moreover, the frequency shifts increase dramatically as
the mode degree increases, given that $m \in [-l,l]$.  In addition,
the Coriolis force induces extra frequency shifts, lifting the
degeneracy with respect to zonal ($m=0$) mode frequencies in the
co-rotating frame. These shifts must also be taken into
account, in addition to the geometrical shifts. It is then clear that one
cannot make meaningful comparisons between observed frequencies as in
Fig.\,\ref{FT-gmodes} and their theoretical predictions in a
co-rotating frame without applying the proper frequency shifts between
the reference frames. This is an essential point of attention for
asteroseismic modelling of moderate and fast rotators.  Such modelling
requires the identification of the azimuthal orders $m$ in order to
interpret the collective effect of all the numerous detected
oscillations in observations of rotating stars whose modes and waves
cannot be treated perturbatively with respect to the Coriolis force
(see Fig.\,\ref{Frequency-Axis}).

In the following sections we discuss some applications of asteroseismology,
with an emphasis on moderate to fast rotators. However, we begin with
slow rotators for ease of understanding and to focus on some recent
highlights for such stars. We discuss applications for various types of stars,
giving descriptions of case studies with the aim to encourage
future applications.

\section{Applications of asteroseismic modelling to single stars}

In order to understand the  asteroseismology of fast rotators, it is
convenient to first recall how it works for slow rotators.  After a
brief description of the basic principles of asteroseismic modelling,
we recall some of the challenges involved in the probing of the Sun's
internal sound speed from its acoustic modes. Subsequently, we provide
some recent topical updates of asteroseismology applied to slowly
rotating low-mass stars before moving on to faster
rotators. Throughout this section, Fig.\,\ref{Frequency-Axis} will be
our guide for the applications. We start on the right with
high-frequency acoustic oscillations and will gradually shift to the
left along the frequency axis.

\subsection{Principles of asteroseismic modelling of slow rotators}

The linear free oscillation modes of a non-magnetic single star whose
oblateness caused by the  centrifugal force can be ignored are
obtained by solving the simplified version of Eq.\,\eqref{SL}, that is,
by setting $\bm{a}_{\rm extra}^{\rm external} = \vec{0}$ and ignoring
all forces aside from  gravity and the pressure and Coriolis
forces. The wave equation then becomes
\begin{equation}
  \label{simplest}
  - \,\omega^2 \vec{\xi} - 2\,{\rm i}\, \omega\,\bm{\Omega}\times \vec{\xi}
  + \bm{O}^{(1)}(\vec{\xi}) = 0\, ,
\end{equation}
where $\omega$ is the frequency in the
co-rotating frame for simplicity of notation. The full expression of
the linear operator $\bm{O^{(1)}}$ is omitted here for brevity, as we   do for
all theoretical expressions for the eigenmodes in the rest of this
review. It can be found in Eq.\,3.340 of \citet{Aerts2010}. 
For modes in the super-inertial frequency regime, one can treat the Coriolis
force in Eq.\,\eqref{simplest} perturbatively, while it has to be taken
into account in full for moderate and fast rotators (see Fig.\,\ref{Frequency-Axis}).
In the applications spelled out below, we   gradually  upgrade
towards higher complexity caused by rotation, focusing on pedagogy
rather than completeness in quoting the literature.

Simplifying maximally to zeroth order in $\Omega$ (i.e. no
rotation) implies the full separability of Eq.\,\eqref{simplest} in terms
of spherical coordinates and time as it reduces to
Eq.\,\eqref{simplest-wave}. We thus find a family of spheroidal
oscillation modes with frequencies $\omega_{nl}$. A degeneracy with
respect to the azimuthal order $m$ occurs, and the time-independent
part for the Lagrangian displacement, $\vec{\xi}(\bm{r})$, can be
written in terms of spherical harmonics, $Y_l^m$.  {The
  expression for $\bm{\xi} (r,\theta,\phi,t)$ is derived in full
  detail and spelled out explicitly in standard books treating
  non-radial oscillations of stars (see \citet{Unno1989} and
  \citealt[][Chapter\,3, Eq.\,3.132]{Aerts2010}).}

Forward asteroseismic modelling of slow rotators to estimate their
basic stellar parameters such as mass, radius, core mass or envelope
mass, and age is usually done by matching the observed frequencies,
$\omega_{nl}$, of
identified zonal ($m=0$) modes of degree $l$ and radial order $n$, with those predicted from grids of 1D stellar evolution
models.  Such grid modelling is minimally a 4D optimisation problem
as any stellar evolution code  requires the mass, initial chemical
composition (any combination of the mass fractions of hydrogen $X$,
helium $Y$, or the metals $Z$), and age as input in order to compute
the stellar structure for that moment in the star's
evolution. However, the problem to solve is in practice of much higher
dimension, as numerous free parameters occur in the codes due to
limitations in our knowledge of the input physics and/or
simplifications of inherently 3D physical macroscopic processes into
1D prescriptions. While this 3D-to-1D simplification is fine for gravity,
thermodynamics, and the microphysics (e.g. nuclear reactions,
equation-of-state, atomic diffusion), it gives bad
approximations for macroscopic transport processes due to rotation,
magnetism, and tides. Hence, for computations with codes assuming the
stellar structure models to be described in 1D, one is forced to
introduce free parameters summarising the effects of these 3D
macroscopic phenomena.  One thus rapidly ends up in a high-dimensional
parameter space for the fitting of the oscillation frequencies; there are currently 
ample opportunities  to attack the regression problem with
machine-learning tools
\citep[cf.][]{Bellinger2016,Hendriks2019,Bellinger2019,Bellinger2020,Angelou2020,Hon2020}.

\subsubsection{A few updates on solar modelling}

\begin{figure*}[t!]
\includegraphics[width = 18cm]{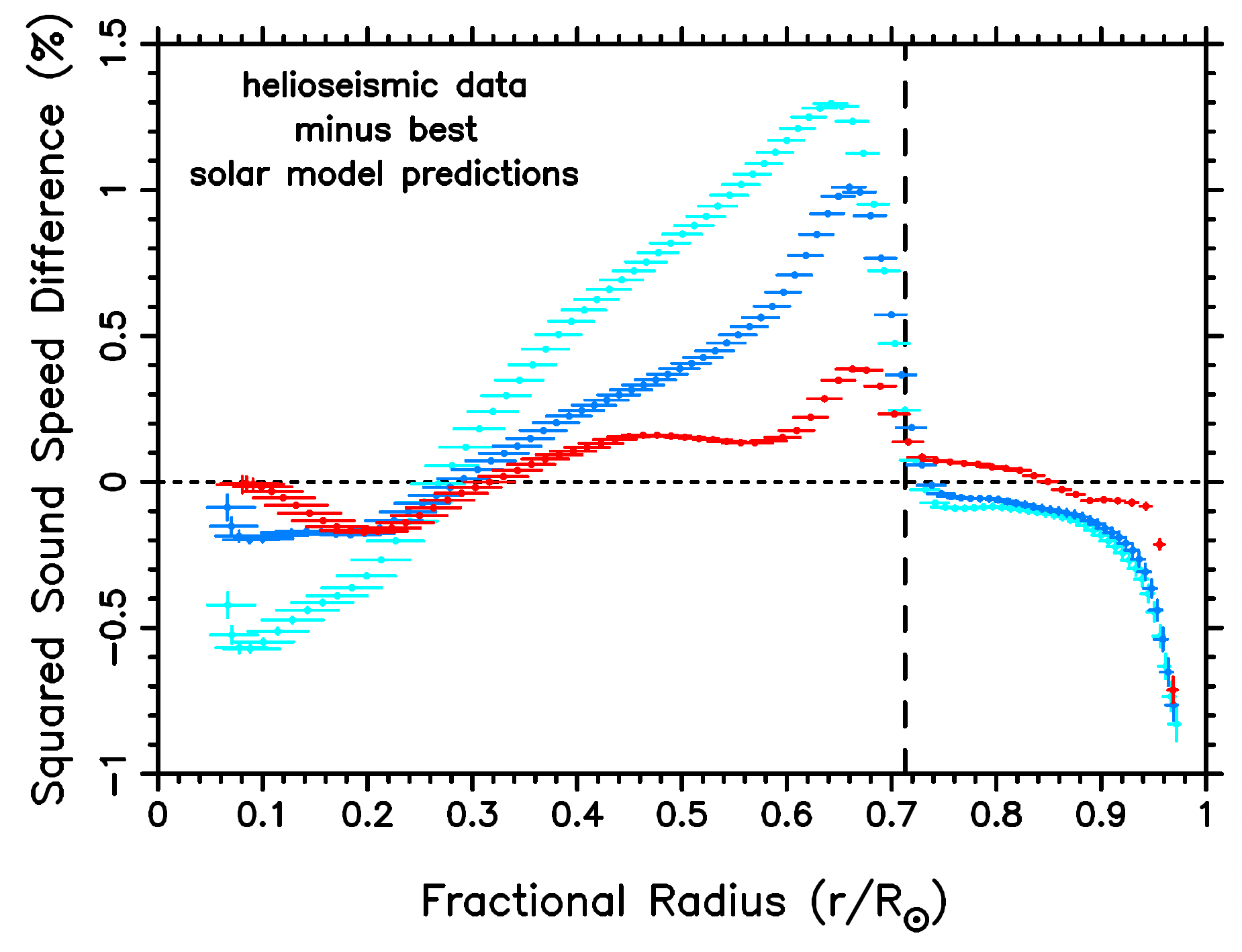}
\caption{\label{sun} Squared sound speed differences between
  helioseismic observations and predictions based on three models of
  the Sun from oscillation mode structure inversions. 
  The errors along
  the $y$-axis are often smaller than the symbol size. The numerous
  small regions indicated by the horizontal bars along the $x$-axis
  represent the intervals in fractional radius over which the
  localised mode kernel averaging was done for the inversions.
{Models delivering perfect 
  agreement with the solar oscillations would coincide with 
the dotted horizontal line. The vertical
    dashed line indicates the helioseismic derivation of the
    base of the solar convection zone.}
  The
  agreement between the observations and the best model predictions
  has degraded since the era of Model S \citep{JCD1996}, despite
  immense improvements in the description of the physics in terms of
  the equation of state, opacities, and transport of chemical elements
  in the solar interior, as explained in \citet{JCD2021}. This
  downgrade is due to the adopted solar surface abundances measured by
  \citet{Asplund2021}, who relied on 3D non-local thermodynamic equilibrium (NLTE) atmosphere models and
  their line predictions leading to a ratio of metals to hydrogen of
  $Z/X = 0.0187$ based on the solar calibration carried out in
  \citet{Buldgen2024}, while LTE abundances from 1D atmosphere models
  from \citet{GN1993} were used for the older red curve from
  Model\,S. The blue curve results from averaged 3D NLTE-based
  abundances by \citet{Magg2022} leading to $Z/X = 0.0225$, and was
  computed by Dr.\ Ga\"el Buldgen relying on the physical descriptions
  adopted in \citet{Buldgen2023} and \citet{Buldgen2024}. The figure
  was produced from data kindly made available by
  J.\ Christensen-Dalsgaard (red curve) and G.\ Buldgen (cyan and blue
  curves).}
\end{figure*}

The dimensionality of the regression problem involved in stellar
modelling can be reduced appreciably if proper calibrations for the
input physics or model-independent ranges for the parameters are
available. We discuss this latter aspect for intermediate- and
high-mass stars in Sect.\,5.
For slow-rotating low-mass stars, one often
freezes the input physics, notably the mixing length parameter used in
time-independent convection theory, to parameter values calibrated
from helioseismology. While doing so, it must be kept in mind that the
models may include extra forces in the term $\bm{a}_{\rm extra}^{\rm
  internal}$ in Eq.\,\eqref{motion} that may be operational in the stars
of the application, while being absent or less important for the Sun.
We illustrate here that, even for the star we know best,
helioseismology is still an active yet specialised sub-field of
asteroseismology, for which the slow solar rotation matters in
improving the dynamics in the Sun.

{In terms of physical
  ingredients, the progress in solar models compared to the famous Model S
  \citep[see][for details]{JCD1996} has been great \citep{JCD2021}. However,
  the physical description for two regions inside the Sun, notably the
  base of the convective envelope and the surface layers, require
  improvements.  The properties of the internal rotation of the Sun as
  summarised by \citet[][Fig.\,7]{Thompson2003} and updated by
  \citet{JCD-Thompson2007} reveal a decrease in the surface rotation
  frequency from about 450\,nHz at the solar pole (zero latitude) to
  about 330\,nHz at latitude $75^\circ$. The level of latitudinal
  differentiality decreases from the solar surface to the base of the
  convection zone situated at a fractional radius of
  $r/R_\odot=0.713$. The transition region between the differentially
  rotating convective envelope and the quasi-rigidly rotating (at about
  440\,nHz) radiative interior is called the tachocline.}

{Figure\,\ref{sun} illustrates the deviations between the squared
  sound speed profile of the true Sun delivered by its thousands of
  identified acoustic modes and those based on some of the best models.
The agreement between the 1D solar models and
  helioseismic data is better than 1.5\%. However, the remaining
  discrepancies between theory and observations are highly significant
  due to the extreme precision of the observations.}
Aside from the ongoing controversy between the `old' and `new' Sun in
terms of its surface abundance determinations
\citep{Buldgen2023,Buldgen2024}, the deviations
{between the Sun and the current solar models in the tachocline}
are likely linked to the interplay of processes on
microscopic and macroscopic scales.
{This is a question of} limited knowledge of
opacities \citep{Bailey2015}, of mixing processes at the base of the
convection zone \citep{Baraffe2022}, and of angular momentum transport
\citep{Eggenberger2022a}, {among other aspects.}

All 1D models of the Sun and of low-mass
stars also suffer from the what is known as the `surface effect' {(see the
  lack of points between $r/R_\odot=0.97$ and 1 in Fig.\,\ref{sun}).}
This represents an offset between measured frequencies of identified
modes and those predicted by the best 1D solar models of the order of a few
$\mu$Hz, which is much larger than observational uncertainties.  The
offsets are mode dependent and increase for increasing radial order of
the acoustic modes (higher mode frequencies).
Improving envelope models of the Sun and of low-mass stars requires a
full 3D description incorporating the interplay between acoustic
oscillations and 3D time-dependent convection \citep{Mosumgaard2020},
keeping in mind the non-adiabaticity of the gas \citep{Houdek2019}.

Novel 3D simulations of the whole convective solar envelope, including
shear-driven magnetic buoyancy due to rotation in the tachocline,
provide new insights into the local mixing it induces
\citep{Matilsky2022,Duguid2023}.  This illustrates that rotation
matters, even for slow rotators such as the Sun. In addition,
thanks to long-term monitoring from space by the Heliospheric and
Magnetic Imager on board NASA’s Solar Dynamics Observatory
\citep[SDO,][]{Pesnell2012}, a novelty was introduced into
helioseismology in 2018. Aside from the well-known spheroidal
high-frequency acoustic modes occurring in the
high-frequency regime to the right in Fig.\,\ref{Frequency-Axis} and
used in Fig.\,\ref{sun}, toroidal inertial modes restored by the
Coriolis force with frequencies below twice the solar rotational
frequency were discovered in the SDO data by \citet{Loptien2018}.
They are related to equatorial Rossby modes as well as high-latitude
inertial modes \citep{Gizon2021}.  These recently discovered
retrograde modes open up inertial-mode
helioseismology, an exciting new way to probe the dynamics of the
rotating tachocline and solar envelope \citep{Bekko2022}. We come back
to Rossby modes in Sect.\,4.4, where we point out that such
modes were already discovered in numerous rapidly rotating stars of
intermediate mass years prior to those found in the Sun.

{To conclude, despite the challenges shown in Fig.\,\ref{sun} the
precision of the solar models at the level of 1.5\% or better is a
remarkable achievement of helioseismology.  Additional 
limitations for models of fast rotators will be discussed as of Sect.\,4.3.}

\subsubsection{Updates on forward modelling of slow rotators}

For the distant stars, we cannot rely on high-degree (e.g.\ $l>4$)
modes from space photometry, as their effects cancel out when
integrating the variability across the visible hemisphere in the line
of sight. Forward modelling in such cases relies on observables in
addition to the mode signals \citep{Gent2022}, notably spectroscopic
and astrometric input to delineate the grids of 1D equilibrium models
used for the regression in the modelling of the stellar
interior.  Spectroscopic effective temperatures and
gravities, as well as luminosities from {\it Gaia\/} parallaxes
\citep{Prusti2016} have become available for large ensembles of
low-mass dwarfs and red giants in the Milky Way.
{These inferred stellar properties are}
used in tandem
with their seismic observables from CoRoT, {\it Kepler\/} and TESS for
galactic studies
\citep[see Fig.\,\ref{Fountain};][]{Anders2014,Chiappini2015,Pinsonneault2018,Zinn2019,Zinn2020,Claytor2020,Mackereth2021,Gruntblatt2021,Zinn2022}.
We note the large difference in relative precision between
such classical quantities and the seismic observables connected
with identified modes \citep[factors 10 to 1000, see Table\,1 in
]{Aerts2019-ARAA}.

The matching between observed and theoretically predicted frequencies
of identified modes is mostly done in a Bayesian framework, where the
error estimation is often tackled numerically from a Markov chain
Monte Carlo approach
\citep{Appourchaux2009,Gruberbauer2012,Gruberbauer2013,Chaplin2014,SilvaAguirre2017}.
Deducing stellar parameters with high precision for large ensembles is
one major asset of asteroseismology
\citep[e.g.][]{Stello2011,Hekker2011,Huber2011,Huber2012,Hekker2013,Davies2016,Brogaard2018,Farnir2021,Brogaard2022,TandaLi2022,TandaLi2023}
as input for other studies in astrophysics
(see Fig.\,\ref{Fountain}), notably galactic {archaeology}
\citep[][etc.]{Miglio2009,Miglio2013,Stello2013,Ness2016,Stello2017,Anders2017a,Anders2017b,SilvaAguirre2018,Hekker2019,Ness2019,Miglio2021,Ness2022,TandaLi2022,Hon2022,Anders2023},
old open clusters
\citep{Brogaard2012,Miglio2012,Miglio2016,Brogaard2021}, and
exoplanetary research where age-dating and star--planet interactions
are crucial aspects
\citep{Huber2013,Lebreton2014,SilvaAguirre2015,Huber2019,Chontos2021,Huber2022}.
However, the ages of the stars deduced from any method, including
asteroseismology, remain dependent on the input physics used to
compute the stellar evolution models and their isochrones, no matter
how precisely the mass and initial chemical composition of the stars
have been derived. For this reason, stellar ages are at best precise,
but not necessarily accurate.

The unknown level of internal mixing
due to element transport causes a major systematic uncertainty for the
age-dating of stars \citep{SalarisCassisi2017}, notably for {those} of
intermediate or high mass born with a convective core.  As an example,
the age-dating of red giants from non-rotating isochrones while ignoring the
fast rotation of their progenitors on the main sequence {may lead
  to systematic age uncertainties in the red giant phases of up to
  20\% \citep{Fritzewski2024b}}. {However,} this effect is often ignored when
providing age estimates {and their uncertainties for}
red giants.  More generally, the cumulative
effect of convective core overshooting during the main-sequence phase,
which may partially be caused by rotational mixing and possibly be
inhibited somewhat by a magnetic field, is among the most important
unknown factors in age-dating, for all levels of internal rotation
rates and masses
\citep[e.g.][]{Deheuvels2015,Bellinger2019,Mombarg2019,Pedersen2021,Johnston2021,Noll2021,Noll2023}.

Developing a calibrated theory of transport processes
\citep{Mathis2013} for all stellar masses and across all evolutionary
phases is thus a major  aim of asteroseismology. This is
currently the only feasible method to offer a proper value for the
numerous free parameters occurring in 1D implementations of these
processes because their parameters change on an evolutionary
timescale.  Major progress has been achieved to improve internal
angular momentum transport, guided by the overarching measured
properties of internal rotation across stellar evolution
\citep{Aerts2019-ARAA}.  Simulations based on new theoretical
ingredients, involving magnetic fields and/or internal gravity waves
are successful in explaining the
{internal rotation rates deduced from}
high-precision space
asteroseismology
\citep{Rogers2013,Fuller2014,Rogers2015,Eggenberger2017,Eggenberger2019,Fuller2019,Rathish2020,Takahashi2021,Eggenberger2022b,Moyano2023}.
Nevertheless, more work is needed when it comes to the understanding
of asteroseismically determined levels of internal mixing for various
ensembles of pulsators
{\citep{RogersMcElwaine2017,Deal2018,Deal2020,Mombarg2020,Pedersen2021,Mombarg2022,Varghese2023}.
Asteroseismically measured values of envelope mixing assuming
a diffusive coefficient for the element transport} in the stellar envelope
range from 1\,cm$^2$\,s$^{-1}$ to
10$^6$\,cm\,s$^{-1}$ \citep[][Table\,1]{Aerts2021-RMP}. Both the
level and functional form of internal shear mixing is the dominant unknown
ingredient affecting the ages and convective core masses of stars
across stellar evolution
\citep{VanGrootel2010a,VanGrootel2010b,Giammichele2018,Charpinet2019a,Tkachenko2020,Johnston2021,Pedersen2021,Pedersen2022a}.

Future age-dating of stars with 10\% accuracy instead of precision
from asteroseismically calibrated
transport processes for exoplanet host stars is a challenging aim in
the core science programme of the ESA PLATO space mission
\citep{Rauer2024}. PLATO also offers an extensive
Complementary Science programme  open to the worldwide community offering
the study of internal rotation, magnetism, and mixing across the
entire HRD (see Fig.\,\ref{Fountain}).

Making progress in our understanding of internal mixing due to element
transport requires a good knowledge of ${\rm d}\Omega(r)/{\rm d}r$.
This brings us back to the quest to deduce the internal rotation
profile of stars for different phases of their evolution. Taking the
Coriolis force in Eq.\,\eqref{simplest} into account lifts the
degeneracy of the mode frequencies with respect to $m$ and gives rise
to resolved rotationally split multiplets if the duration of the time
series data is sufficiently long compared to the average rotation
period of the star.
Ignoring terms in $\Omega^2$ in the equation of motion and
adopting a perturbative first-order approach in
$\Omega$ for shellular rotation leads to observed frequencies
{in an intertial frame of reference}
according to Ledoux splitting \citep{Ledoux1951} given by
\begin{equation}
\omega_{nlm}  = \omega_{nl}  + \ m\ (1-C_{nl}) \int_0^R K_{nl} (r) 
\Omega (r) {\rm d} r  \; , 
\label{LedouxSplitting}
\end{equation}
where $K_{nl} (r)$ are the  rotational kernels, which can be
computed from the identified modes and the equilibrium structure of
a 1D stellar model \citep[][Eq.\,45]{Aerts2021-RMP}.
{This expression includes the geometrical shift introduced in
  Eq.\,(\ref{transfo}), as well as the Ledoux constant $C_{nl}$ caused
  by   the Coriolis force \citep{Ledoux1951}.
  Hence, following Eq.\,(\ref{LedouxSplitting}),}
the rotation
frequency $\Omega(r)$ throughout the stellar interior gives rise to
frequency multiplets $\omega_{nlm}$ with $2l+1$ components in the
line of sight, provided that all the modes in the multiplet are
excited to observable amplitude.   We note that the latter circumstance is
a good assumption for solar-like oscillation modes excited
stochastically by envelope convection, but that this is not necessarily
the case for heat-driven oscillations, as shown in this era of
high-precision space asteroseismology \citep{Aerts2021-RMP}. Modes of particular $m$ may not be excited intrinsically by the heat
mechanism, but may be pumped up in mode energy from non-linear mode
interactions involving the rotation frequency. We give some examples
of this below,  as  measured in fast rotators.

Equation\,\eqref{LedouxSplitting} reveals that treating the Coriolis
force perturbatively up to first-order creates symmetrical Ledoux
splittings for fixed $n$ and $l$, given that $m$ ranges from
$m=-l,\ldots,0,\ldots,+l$.  Moreover, in the limit of high-order p and
g~modes, one can show that $C_{nl}\simeq 0$ and $C_{nl} \simeq
1/[l(l+1)]$, respectively \citep[see][for a summary of derivations of
these approximations]{Aerts2010}.  Hence, for such modes,
the measured rotational splittings provide a direct estimate
of the local rotational profile $\Omega (r)$ averaged by the mode
energy, since $K_{nl} (r)$ is determined by the square of the mode
displacement $\bm{\xi}$ \citep[see  Eq.\,(45) in][]{Aerts2021-RMP}.

The Ledoux rotational splitting in Eq.\,\eqref{LedouxSplitting} is a
good approximation for the  p~modes of low-mass dwarfs, for the p~modes of sub-giants and red giants, for the p and g~modes of
sub-dwarfs, and for the g~modes of white dwarfs. For most of these
pulsators, the oscillation periods range from several minutes to a few
hours and are small fractions of the rotation periods ranging from
about a day for white dwarfs \citep{Hermes2017} to several months for
red giants \citep{Mosser2012,Gehan2018}.  Aside from a few exceptions,
the spin parameters of the modes of all these types of pulsators in the co-rotating frame,
$2\Omega/\omega_{nlm}$, put them far into the super-inertial regime
(completely to the right in Fig.\,\ref{Frequency-Axis}). Almost all of
these pulsators are hence slow rotators in our definition, with their
forward modelling being done from their $m=0$ mode frequencies and the
subsequent derivation of $\Omega (r)$ from Eq.\,\eqref{LedouxSplitting}
relying on the best forward seismic model.  With the notable exception
of the high-order p~modes in the Sun-like star $\eta\,$Bootis modelled
with inclusion of the centrifugal deformation by \citet{Suarez2010},
summaries of asteroseismic modelling results for the slow rotators can
be found in \citet{GarciaBallot2019}, \citet{HekkerJCD2017},
\citet{LynasGray2021}, and \citet{Giammichele2022} for low-mass
dwarfs, red giants, sub-dwarfs, and white dwarfs,
respectively. Homogeneous analyses for the sub-giant phase were
somewhat lacking after the {\it Kepler\/} mission finished, but are
now also well underway \citep{Ong2021,Noll2021}, in anticipation
of TESS light curves for this evolutionary phase.

\subsubsection{Detection of internal rotation and core magnetism in red giants}

The CoRoT space mission revealed that red giants are non-radial pulsators
\citep{DeRidder2009}. It was a lucky circumstance that the mission
programme could not avoid red giants in the observing fields
dedicated to exoplanet research. The mode periods roughly range from
half an hour to half a day, while the rotation periods of red giants
typically range from ten days to hundreds of days.  As discussed in the
previous section, this implies that rotation can be treated
perturbatively following the Ledoux approximation and that modelling can be
done from the zonal modes.  The CoRoT discovery of non-radial
oscillations in red giants allowed the application of age-dating for various
areas of the Milky Way \citep{Hekker2009,Miglio2009};   major
advances in galactic archaeology
\citep[e.g.][]{Miglio2013,Chiappini2015,Montalban2021} is one of the
important spin-offs of asteroseismology, as indicated in
Fig.\,\ref{Fountain}.

Following the theoretical predictions from \citet{Dupret2009}
triggered by CoRoT, dipole mixed modes were discovered in
{CoRoT data of a sub-giant \citep{Deheuvels2010} and}
in {\it   Kepler\/} data of a red giant \citep{Beck2011}.
 {Subsequently, such modes were found} in a whole
sample of red giants by \citet{Bedding2011}. The latter breakthrough study
led to the important capacity to discriminate between stars on the red
giant branch (RGB) and in the first or secondary red clump, even
though such stars share the same surface properties. This
discriminating capacity was also nicely illustrated for red giants in
open clusters \citep{Stello2011} and relies on the fact that the mixed
modes have g-mode character in the deep interior of the star, yet are
of acoustic nature in the envelope. Figure\,\ref{Frequency-Axis}, helps us understand the nature of these
modes: the p- and g-mode cavities delineated by the
$S_l$ and $N$ symbols come closer to each other as a star evolves from
being a dwarf to a red giant. As shown by \citet{Dupret2009}, this
narrows the zone between $N(r)$ and $S_1(r)$ in such a way that the
waves can tunnel through both cavities without becoming fully
evanescent in the transition zone, thus creating dipole mixed modes
that probe the entire star.

Core rotation frequencies were detected from the splitting
of mixed dipole modes in a few {\it Kepler\/} red giants after two
years of uninterrupted photometric monitoring \citep{Beck2012}.
The positive effect of the longer duration of the {\it Kepler\/} light
curves beyond two years implied major progress in   seismic
precision \citep{Hekker2012}.  {It led} to a revolution in the
understanding of angular momentum transport from large samples of
evolved stars initiated by \citet{Mosser2012}, with refined analyses
from rotation inversions for a few RGB pulsators by
\citet{Deheuvels2012} and red clump giants by \citet{Deheuvels2015}.
Following the initial breakthroughs, the internal rotation of evolved
stars became an industrialised observational science of major
importance \citep[e.g.][]{Gehan2018,Li2024}.

Meanwhile, space asteroseismology  delivered the internal rotation for several thousands of
evolved low- or intermediate-mass stars, notably sub-giants and their
successors climbing up the red giant branch, as well as giants in the
first and secondary red clumps, sub-dwarfs, and white dwarfs. A summary
of the rotational properties of about 1200 slow-rotating field stars
is available in the review paper by \citet{Aerts2019-ARAA} and is not
repeated here. Figure\,4 in that paper, along with the updated figures
with measured internal rotation rates
until the end of 2019 in \citet{Aerts2021-RMP},
place the internal rotational properties of
low- and intermediate-mass stars in a global evolutionary
picture. This revealed that the angular momentum of the core of helium
burning red giants is in agreement with the angular momentum of white
dwarfs.

The major new insight from asteroseismology regarding internal
rotation across stellar evolution 
is that the level of radial-differential rotation of 
single dwarfs of low and intermediate mass is low during 
their longest phases of evolution, when
they fuse hydrogen into helium and helium into carbon and oxygen.
As discussed in \citet{Aerts2019-ARAA}, who summarised all
results obtained until mid-2018, any explanation of this
asteroseismic picture of stellar evolution requires more efficient
transport of angular momentum during these phases than anticipated
prior to {\it Kepler\/} space asteroseismology. Said differently,
much stronger coupling between the convective core and envelope occurs
during the two central burning stages of intermediate-mass stars,
while strong radial-differential rotation with a factor up to 20 for the
core-to-envelope rotation rates occurs during the RGB \citep{Li2024}.  
{New theories to address this too low angular momentum transport
  in models of stars with a convective core have  been developed.
One way to increase the
angular momentum transport while such stars evolve is to include
internal gravity waves in the models. The occurrence of such waves was
already proposed as an efficient way to transport angular momentum in low-mass
stars like the Sun by \citet{Charbonnel2005,Rogers2006} and in high-mass stars by
\citet{Rogers2013}.}

{
Another way to achieve higher angular momentum
transport in models is by including magnetic effects.}
The existence of internal magnetic fields in red giants was already
inferred by \citet{Fuller2015} as an interpretation for depressed
dipole mixed modes found in {\it Kepler\/} data for a small fraction
of red giants.
{The accompanying study by
\citet{Cantiello2016} showed that a 
magnetic greenhouse effect
may be
operational} in red giants, turning g~modes
trapped in the core into Alfv\'en waves. The mode damping requires internal
magnetic field strengths above about $10^5$\,G, as confirmed by
the theoretical predictions in \citet{Loi2017}.
A test of this greenhouse 
scenario for earlier phases of stellar evolution was developed by
\citet{Stello2016}, namely about half of the F dwarfs should have
a strong internal magnetic field during their main-sequence phase. So far,
it has not been possible to test this hypothesis from a representative 
asteroseismic population of F dwarfs, but internal magnetic field
strengths of the 37 best modelled {\it Kepler\/} $\gamma\,$Dor
pulsators inferred by \citet{Aerts2021-GIW} are in agreement with the
scenario by \citet{Fuller2015} keeping in mind the minimal field
strengths deduced by \citet{Cantiello2016}.

Although the interpretation by \citet{Fuller2015} was contested by
\citet{Mosser2017}, the presence of internal magnetic fields in
{a fraction of}
red giants is no longer a hypothesis. An observational breakthrough
was achieved by \citet{GangLi2022} and by \citet{Deheuvels2023}, who
found evidence of the presence of such fields inside the cores
of  respectively 3 and 11  red giants from 4 yr {\it Kepler\/}
light curves.  \citet{GangLi2022} detected the core magnetism from the
occurrence of asymmetrical rotational splittings in the triplets of
mixed dipole modes. They interpreted this asymmetry as being due to
the small effects induced by the Lorentz and Coriolis forces on the
modes.  In their approximation, the terms caused by
$2\,\bm{\Omega}\,\times\,\bm{v}$ and the magnetic force
in $\bm{a}_{\rm extra}^{\rm internal}$ are considered small
compared to those due to the gas pressure {gradient} and gravity in
Eq.\,\eqref{motion}, as confirmed by their numerical computations.  
Elegant analytical and numerical work to interpret the observational
findings in the adopted approximation was offered by \citet[][see the
  Supplementary Material]{GangLi2022} and by
\citet{MathisBugnet2023} for various magnetic field topologies,
following earlier theoretical {studies} on the effect of magnetism on modes
of red giants by
\citet{Loi2018,GomesLopes2020,Loi2020a,Loi2020b,Loi2020c,Loi2021,Bugnet2021,Bugnet2022}.
 {
\citet{Deheuvels2023}, on the other hand, inferred the presence of a
core magnetic field in 11 RGB stars from the effect of the Lorentz
force on the frequencies of zonal dipole modes.
These 11 red giants do not reveal asymmetries in their dipole mode
splittings. Rather, the Lorentz force affects the
dipole zonal mode frequencies of consecutive
radial order. The effect depends on the strength of the magnetic
field and allows us to deduce lower limits on the core field strengths.
As such, \citet{Deheuvels2023} found fields of 40\,kG to 610\,kG for
these 11  stars. One of them, KIC\,6975038,
has depressed dipole modes, lending further support to
the original interpretation of a magnetic greenhouse effect
by \citet{Fuller2015}.}

Meanwhile, \citet{GangLi2023} detected a core magnetic field from
asymmetrical splittings in  13 of the $\sim\! 1200$ red giants
 {with detected dipole mixed mode triplets found in the sample of
  8000 compiled by \citet{JieYu2018} and \citet{Gehan2018}.
The way the authors selected these 1200 RGB stars disfavours such stars
with depressed dipole modes, as well as stars in an advanced stage of
evolution along the RGB.}
The average core magnetic field strengths of these
13 stars range from 20 to 150\,kG, representing a range from 5\%
to 30\% of the  critical field strength above which
magneto-gravity waves are dissipated in the core rather than being
propagative \citep{Fuller2015,RuiFuller2023}.  The detected fields
were found to have a variety of horizontal field geometries.  {The
  core magnetic fields detected in less than 2\% of the {\it Kepler\/}
  red giants with dipole mixed modes by \citet{GangLi2023} and
  \citet{Deheuvels2023} are important to improve stellar evolution
  models.  The low fraction found in the large homogeneous database of
  the {\it Kepler\/} 4 yr light curves does not necessarily imply the
  absence of a field for the majority of stars as the fields may
  remain undetected for large parts of the evolution. The
  field strengths were found to decrease as the cores of the stars
  shrink along their evolutionary path. The magnetic red giants have
  core rotation properties fully in line with those of the thousands
  of red giants without detected magnetic signatures.  Both findings
  suggest that the magnetic fields do not cause much extra transport
  of angular momentum compared to the case of a non-detectable
  magnetic field. The sample of red giants with detected core fields
  is currently too small to make strong general conclusions. The effects
  of the core fields on internal mixing, if any, remain to be studied
  given the recent character of these discoveries. This may help to
  assess whether the magnetic field has any effect on element transport as
  the red giants evolve.}


\subsection{Asteroseismology of supernova progenitors: Low-order p and
g~modes in $\beta\,$Cep stars}

Moving somewhat to the left on the frequency axis in
Fig.\,\ref{Frequency-Axis}, we find moderate rotators among the
$\beta\,$Cep stars, such as HD\,192575 already discussed above.  The
$\beta\,$Cep stars are dwarfs or (super)giants of spectral type O or early B
with masses roughly in the range 8 to 25\,M$_\odot$. They are
hence precursors of supernovae.  This class of pulsators has always
played a special role in asteroseismology, and this is no different in
this review.  As early as 1902, \citet{Frost1902} discovered
radial-velocity variations for the prototype of the class, the star
$\beta\,$Cephei. At that time, the only explanation was that the star
is a spectroscopic binary. Although the variability of tens of similar
stars was misinterpreted in terms of orbital motion for about half a
century, a breakthrough was achieved thanks to the three periodicities
found in the radial-velocity time series for the class member
$\beta\,$CMa \citep{Struve1950,VanHoof1953}. These properties in fact
led \citet{Ledoux1951} to write his breakthrough paper on non-radial
oscillation modes in a rotating star, introducing the concept of
rotationally split multiplets due to the joint effect of the Coriolis
force and the projection in the line of sight for an observer, as discussed
in Sect.\,3.3.  Even though we know today that Ledoux's interpretation
of the variability of $\beta\,$CMa was wrong \citep[e.g.][for a modern
  asteroseismic interpretation]{Mazumdar2006}, it led him to derive a
general description of the observed characteristics of non-radial
oscillations in a rotating star, which is the one still in use today
for slow rotators with modes in the super-inertial regime.

\begin{figure*}
{\resizebox{18.cm}{!}{\includegraphics{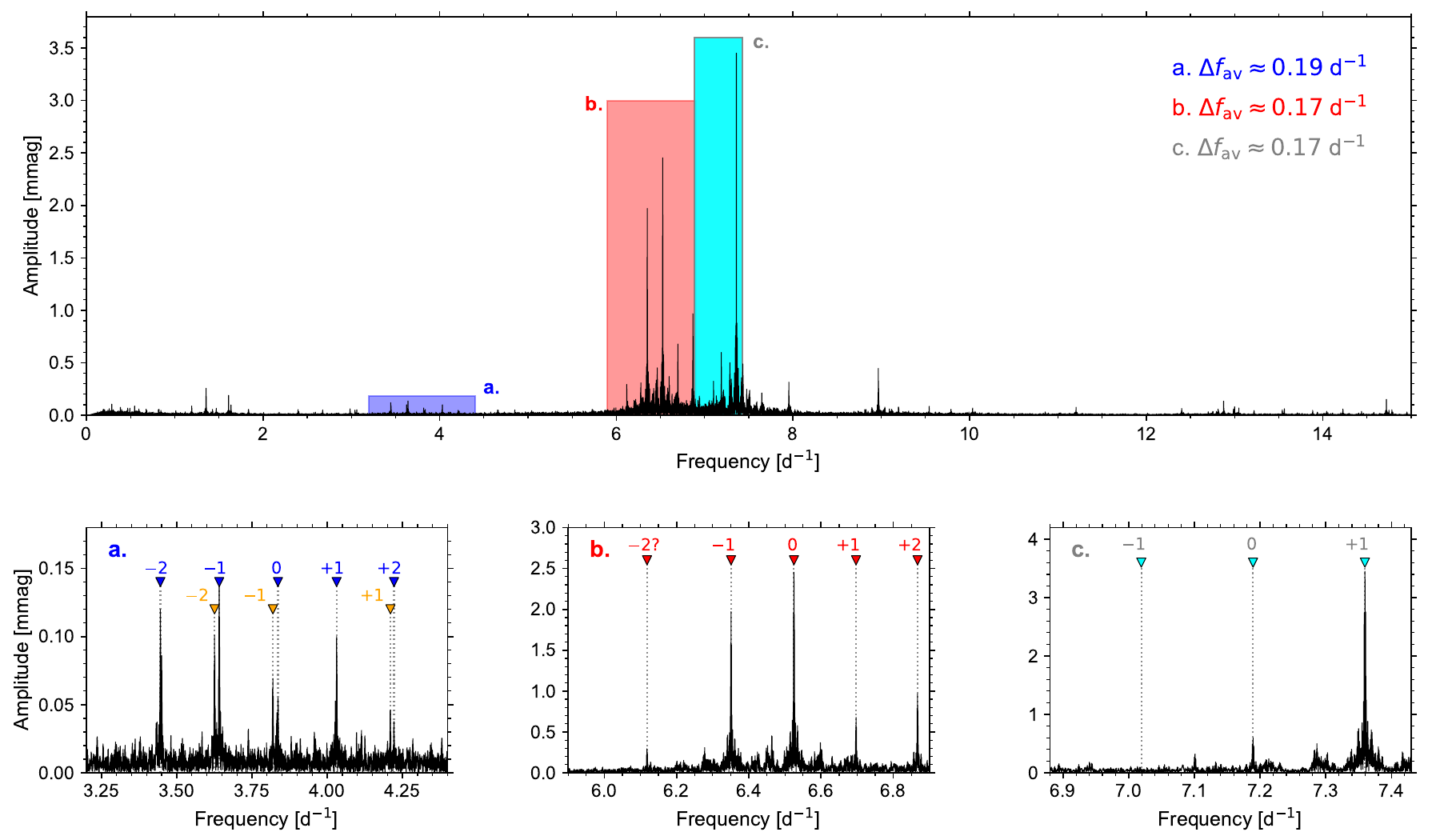}}}
\caption{\label{Siemen}  
  The Fourier transform of the TESS merged
  Cycle\,2 and 4 light curve of the $\beta\,$Cep star HD\,192575 (top). The
  three lower panels are zoomed-in versions of the coloured and labelled
  frequency ranges, whose averaged splittings are indicated in the
  upper panel.  They consist of a merged quintuplet and quadruplet undergoing an
  avoided crossing (left), an isolated (possibly incomplete)
  quintuplet (middle), and a
  triplet whose prograde mode has the highest amplitude in the entire
  oscillation spectrum (right). The identifications for the azimuthal order $m$ of
  these multiplet components are indicated above each peak. The figure was
  made by Siemen Burssens.}
\end{figure*}

Rotational splitting
following Eq.\,\eqref{LedouxSplitting}
applied to 21 years of
multi-colour photometric monitoring led to the first asteroseismic
detection of non-rigid rotation in a star other than the Sun
\citep[the $\beta\,$Cep star HD\,129929][]{Aerts2003,Dupret2004}.  In
terms of modes, the $\beta\,$Cep stars reveal low-order p and g~modes
with periodicities of several hours, while their rotation periods cover
the range from more than 100\,d down to a considerable fraction of
their critical rotation period. None rotate close to critical
\citep{Stankov2005}, however, so following our definition they are
slow to moderate rotators. 

Following \citet{Goupil2000}, we assess the relative importance of
second-order effects due to the Coriolis force, which scales as
$(\Omega/\omega)^2$, and the effects due
to the centrifugal force, which is proportional to
$(\Omega^2 R^3)/{\rm G}M$. Thus, the relative role of the centrifugal 
versus Coriolis forces in any second-order theory in $\Omega$ behaves as
\begin{equation}
  \label{Coriolis-Centrifugal}
  \displaystyle{
    \frac{{\rm centrifugal}}{{\rm 2nd-order\ Coriolis}}
    \ \sim
    \frac{\omega^2\,R^3}{{\rm G} M}\ .
  }
  \end{equation}
This ratio shows that the Coriolis force dominates over the
centrifugal force for high-order g~modes in many of the brightest SPB
pulsators for which this ratio has been measured, and is typically
between 0.01 and 0.1 \citep[see Fig.\,17 in][]{DeCatAerts2002}. One
can therefore ignore the oblateness caused by the centrifugal force
for such moderately rotating pulsators, but   cannot ignore the
Coriolis force in their asteroseismic modelling \citep{Aerts2021-GIW}.
\citet{Ballot2010} unravelled the domains of validity of a
perturbative approach for the rotation
to compute modes in the super-inertial regime
and concluded that the
second-order approach still gives satisfactory results for equatorial
velocities up to $\sim$100\,km\,s$^{-1}$ for $\gamma\,$Dor stars and
up to $\sim$150\,km\,s$^{-1}$ for SPB stars. A third-order
{perturbative}
approach increases the domains of validity by a few tens of
km\,s$^{-1}$. However, we have known since the era of space asteroseismology 
that most of these pulsators have their modes in the sub-inertial
regime \citep[see Fig.\,\ref{Frequency-Axis} and][]{Aerts2021-GIW}
where perturbative approaches for the rotation are not valid. We treat
such high-order g-mode pulsators by adopting     the traditional
approximation of rotation (TAR) in Sect.\,4.4,
 {where we explain  that the TAR is a mathematically
  elegant approximation that ignores the horizontal component of the
  rotation vector.}

The ratios from Eq.\,\eqref{Coriolis-Centrifugal} for typical p~modes in
$\beta\,$Cep or $\delta\,$Sct stars are roughly between 1 and 10
\citep{Goupil2000}.  This explains why the super-inertial modes in slowly
to moderately rotating $\beta\,$Cep and $\delta\,$Sct stars can
be described by a second-order perturbative theory.
\citet{Ballot2010} evaluated the validity domain
for these p-mode pulsators and concluded that this approximation is
fine for equatorial rotation velocities up to about
70\,km\,s$^{-1}$. They noted  that adding third-order terms hardly
improves this because p~modes are only weakly sensitive to the
Coriolis force.  For the cases where low-order linear oscillation
modes in moderately
rotating $\delta\,$Sct or $\beta\,$Cep stars can still be described perturbatively,
we refer to \citet{Soufi1998}, \citet{Karami2008}, and
\citet{Guo2024} for appropriate mode expressions.

Rotationally split p and g~modes should in principle be easy to
recognise for the slower rotators in the class and help identify the
modes. However, despite the hundreds of OB pulsators found in space
photometry
\citep{Aerts2006a,Balona2011,Balona2016,Balona2019,Pedersen2019,Burssens2019,Burssens2020,Balona2022},
the frequency spectra turn out to contain so many frequencies that
the modes are hard to identify from the single-passband photometric
variations. In particular, zonal modes are hard to find. 
As a result, forward modelling based on the fitting of individual
identified modes has been done
for only a few of these supernova progenitors from space photometry
\citep{Aerts2006b,Handler2009,Briquet2011,Aerts2011,Daszynska2017,Handler2019}.
Given their mass range and extensive convective cores, the forward
modelling problem to fit the zonal modes is at least five-dimensional
because convective core overshooting cannot be ignored and is an
uncalibrated mixing phenomenon requiring at least one fit parameter.

Some of the most detailed forward models for this class result from
earlier months-long ground-based multi-site campaigns for several
bright class members
\citep{Pamyatnykh2004,Briquet2007,Dziembowski2008,Briquet2012}.  A
detailed comparison among the forward models resulting for the three
pulsators HD\,129929, $\theta\,$Oph, and V2052\,Oph, which have very
similar dominant modes including an $l=2$, g$_1$ quintuplet, a radial
mode, and an $l=1$, p$_1$ triplet, revealed that the magnetic star
V2052\,Oph has lower core overshooting than the two   non-magnetic
pulsators. This hinted at inhibition of mixing by a magnetic field.

To date, internal rotation profiles assuming shellular rotation have been
determined for five $\beta\,$Cep stars. This had led to ratios of the
near-core and envelope rotation rates between 1 and 5
\citep[see Table\,1 and Fig.\,5 in][for a summary]{Burssens2023},
where the rigid rotation case occurs in the only close binary pulsator
among this small sample.  However, the methods adopted to derive the
internal rotation are diverse.
{For the extensively studied star $\nu\,$Eri,
  \citet{Dziembowski2008} adopted
  the first-order Ledoux approximation according to
  Eq.\,\eqref{LedouxSplitting} and used the
  fundamental
  radial mode and the two $l=1$ triplets g$_1$ and p$_1$ to
  obtain a core-to-envelop rotation ratio of about 5.   
  \citet{Suarez2009}, however,  performed
a more detailed analysis based on a second-order
perturbative approach, including  the 
$l=1,$~p$_2$ triplet. They
found a factor $\sim\!2.6$ lower envelope than near-core rotation
from the three observed triplets, while also being able to explain
most (but not all) of the measured triplet asymmetries.}
The approach by \citet{Suarez2009} is a
promising route to understand angular momentum redistribution and
chemical mixing due to radial-differential rotation and its induced
turbulence in moderately rotating $\beta\,$Cep stars.

The $\beta\,$Cep class member with the richest set of multiplets so
far is the $\sim\!12\,$M$_\odot$ star HD\,192575. This pulsator was
discovered from TESS data covering the northern continuous
viewing zone (N-CVZ) of the satellite. Its oscillation spectrum based on
available TESS data covering about three years (with a one-year gap in the
middle) is shown in Fig.\,\ref{Siemen}. This is an updated oscillation
spectrum compared to the one in the discovery paper by
\citet{Burssens2023}, who used only the first 352\,d of TESS data.
Figure\,\ref{Siemen} reveals a complete identification for the
$l=2$ quintuplet (in blue), rather than an incomplete one in
\citet{Burssens2023}. This changes the avoided crossing with the other
quintuplet indicated  in Fig.\,\ref{Siemen} (in orange) and leads to a
somewhat more evolved star (without affecting the other parameters
from the forward modelling). Moreover, the left component of the
quintuplet in panel b\ (labelled  `-2?') is now significant and
may point to an asymmetry in that $l=2$ multiplet or to a frequency
unrelated to the quintuplet.
Rotation inversion of the securely identified
multiplets of this star leads to a level of radial-differential rotation of about 1.5
between the core and envelope (Vanlaer et al. in prep.),
in line with the
$\mu$-gradient shear layer solution found by \citet{Burssens2023} and
the result from 2D stellar evolution modelling by \citet{Mombarg2023}. The
asymmetries in the splittings of the quintuplets of 
this star stand in sharp contrast to the symmetrical triplets.
The detected asymmetries point to the presence of a core
magnetic field
in this supernova progenitor (Vanlaer et al. in
prep.). This is of vast interest because it would deliver an
asteroseismic calibration to 3D simulations of core-collapse
supernova explosions as an important spin-off from the asteroseismology of
massive stars, in line with
Fig.\,\ref{Fountain} already discussed in Sect.\,1.

\citet{Fritzewski2025} brought the first ensemble asteroseismology for
a population of 164 $\beta\,$Cep stars discovered from {\it Gaia\/}
DR3 by \citet{DeRidder2023}. The nature of these pulsators was 
confirmed from their light curves extracted from the TESS full
frame images (FFIs) by \citet{HeyAerts2024}.
We come back to their mode identification in
Sect.\,8.1, but point out here that this study
increased the sample of $\beta\,$Cep stars
with a measurement of internal core-to-surface rotation substantially.
The level of radial-differential
rotation for 48 of these newly discovered pulsators ranges from 1 to 2,
except for a few stars that have a value up to 4.

Only a few of the numerous modes active in the fastest rotators among
the $\beta\,$Cep stars have been unambiguously identified so far. Much
remains to be done on that front, as is the case for the p~modes in
early-type Be pulsators \citep[such as the CoRoT target
  HD\,49330,][]{Huat2009}.  This is not surprising, given the
findings of \citet{Daszynska2002} who studied mode coupling among
linear p~modes in models of rapidly rotating $\beta\,$Cep stars up to
second order in $\Omega$.  Their conclusion is essentially that mode
identification is only possible if the inclination angle of the
rotation axis is well measured,  a situation hardly encountered in
practice, although modern interferometry gives a promising
perspective \citep{Domiciano2014,Bouchaud2020}.  Lack of mode
identification is currently the  major challenge to advance
asteroseismology of fast-rotating p-mode pulsators of spectral types O
to B3, including the pulsating early-type OBe stars whose mode
frequencies occur in a similar regime as to those of the $\beta\,$Cep stars
\citep{Neiner2009-CoAst}.  We anticipate major progress from
asteroseismology of $\beta\,$Cep stars and p-mode Be stars from
applying second- or third-order perturbative methods in $\Omega$ as in
\citet{Suarez2009}. Updated modelling from such an approach to all the
stars with already identified modes, and even more so from future 2D
or 3D stellar evolution models, is an opportunity we come back to at
the end of this review.


\subsection{Asteroseismic inferences on transport processes from 
modes of rotating $\delta\,$Sct stars}

We now move on to the case of slowly, moderately, and rapidly
rotating $\delta\,$Sct stars. An extensive catalogue of these pulsators
was already available prior to the space era
\citep{Rodriguez2001}.  With their masses between roughly 1.5 and
2.5\,M$_\odot$, the majority of stars in this class have the most
complex internal structures among dwarfs, with a fast-rotating convective core
and thin convective outer envelope requiring time-dependent
convection theory in a non-adiabatic environment for a proper
mode description \citep{Dupret2005a,Dupret2005b,Antoci2014}.

\citet{Bowman2016-Springer} gave an extensive summary of the
observational properties of $\delta\,$Sct stars as evaluated by the
end of the nominal {\it Kepler\/} mission, highlighting that their
oscillation spectra are characterised by strong amplitude modulation
and moderate to fast rotation. This observational behaviour is readily
understood in terms of the theory of non-linear mode coupling by
\citet{Buchler1997} and \citet{Goupil1998}. With the TESS mission
ongoing, the class of $\delta\,$Sct stars keeps increasing with
thousands of members \citep{Antoci2019} with well characterised global
properties deduced from the {\it Gaia\/} space data \citep{Murphy2019}.  Many
of the {\it Kepler\/} $\delta\,$Sct stars actually turned out to be
hybrid $\gamma\,$Dor--$\delta\,$Sct pulsators experiencing both
high-order g~modes and low-order p~modes
\citep{Grigahcene2010,Uytterhoeven2011,Bowman2016,GangLi2020}.
These are extremely interesting asteroseismic targets as the
combination of high-order g~modes and low-order p~modes offers probing
power throughout the entire stellar interior, notably the rotational
properties \citep{Audenaert2022}.

The Ledoux approximation in Eq.\,\eqref{LedouxSplitting} is still valid
for a small fraction of the $\delta\,$Sct pulsators. Slow rotation
notably occurs among several of the $\gamma\,$Dor--$\delta\,$Sct hybrid
pulsators.  The first-order Ledoux approximation was readily applied to
the amplitude spectrum of the slow-rotating hybrid pulsator
KIC\,11145123 by \citet{Kurtz2014}. This delivered the first
$\gamma\,$Dor--$\delta\,$Sct star with a determination of internal
rotation. This star reveals nearly uniform rotation with a period of
about 100\,d and a slightly faster surface rotation compared to core rotation.  Models
with atomic diffusion, including radiative levitation, offer an
excellent explanation of the g~modes and surface abundances for this
hybrid pulsator \citep{Mombarg2020}.  Nevertheless, it has been
suggested that the star's `weird' and slow rotation, along with its
surface abundances deviating from solar values, may be due to a merger
event \citep{Takada-Hidai2017}.  While such a scenario should be no
surprise (given Fig.\,\ref{BinaryFraction}), another such slowly
rotating hybrid $\gamma\,$Dor--$\delta\,$Sct star, KIC\,9244992, has
solar-like surface abundances and very similar seismic properties.  Its rotation
near the surface deduced from p-mode splittings amounts to 66\,d,
which is slightly slower than the rotation period of 64\,d in the
near-core region measured from its g-mode splittings \citep{Saio2015}.
Both pulsators
have a mass of about 1.5\,M$_\odot$ and are at an advanced stage of
main-sequence evolution, with a central hydrogen mass fraction $X_c$
of about 10\%.

Two additional similar cases of slow-rotating hybrid
$\gamma\,$Dor--$\delta\,$Sct pulsators were discovered in the 15.3\,d
eccentric orbit ($e=0.45$) binary KIC\,10080943 by
\citet{Keen2015,Schmid2015}. Their orbital separation is wide enough
to perform asteroseismic modelling assuming  the stars to be single yet
coeval, as done by \citet{SchmidAerts2016}. This delivered an age of
1.08\,Gyr with a 1.67\,M$_\odot$ primary and a 1.60\,M$_\odot$
secondary having $X_c=22\%$ and 29\%, respectively. The two stars have
similar small convective core overshooting and envelope mixing.
The primary's and secondary's near-core rotation periods are 7.16\,d
and 11.05\,d, respectively. This delivers spin parameters for the
detected and identified dipole g~modes between 0.20 and 0.26 for the
primary, and between 0.14 and 0.24 for the secondary. These values
place them well into the super-inertial regime
(see Fig.\,\ref{Frequency-Axis}).  For both stars, the Coriolis force can
still be treated perturbatively and the linear modes can be modelled
from first-order Ledoux splitting \citep{SchmidAerts2016}.  The envelope
rotation deduced from rotational splitting of p~modes resulted in
about 7.46\,d and 8.20\,d for the primary and secondary, respectively.
This gives core-to-envelope levels of 0.96 and 1.35, respectively.
Thus, the radial-differential rotation of the secondary again reveals
a slower core than envelope, just as for KIC\,11145123. The
centrifugal and tidal forces active in this binary result in only
small perturbations, up to 1.7\% for the former and below 0.04\% for
the latter.  Thus, the centrifugal force has a stronger
influence on the oscillation modes than the tides.

The core-to-envelope rotation rates of these four
$\gamma\,$Dor--$\delta\,$Sct stars were found to be fully in line with
angular momentum transport due to internal gravity waves, an
interpretation reached from hydrodynamical simulations with
appropriate levels of core-to-envelope rotation by \citet{Rogers2015}.  However,
these stars are not representative of the  $\delta\,$Sct
class as a whole.  The large majority of these pulsators
rotate so fast that even second-order perturbation theories in $\Omega/\omega$ 
are too limited according to evaluations of
Eq.\,\eqref{Coriolis-Centrifugal} for typical class members.  For this
reason, a more elaborate third-order perturbation
formalism taking into account the Coriolis force and the
centrifugal flattening with its coupling with the Coriolis force, was
developed by \citet{Soufi1998}. For the full expressions of the
operators for this third-order perturbative theory in $\Omega$ we refer
to that paper and the corrections proposed by \citet{Karami2008}.
To date, this third-order theory has hardly been applied to
unravel the internal physics of fast-rotating $\delta\,$Sct
stars. This is due to lack of proper mode identification for a
sufficient number of modes, just as for the fast-rotating $\beta\,$Cep
and early-type Be pulsators. While \citet{Bedding2020} managed
to overcome the hurdle of mode identification for a sample of 60 very
young $\delta\,$Sct stars from regular patterns delivering global
parameters of these stars from their large frequency spacing, no
rotational splitting was found for this sample. Pressure-mode
frequency spacings alone cannot give tight constraints on transport
processes to improve the input physics of the models.

Most $\delta\,$Sct stars do offer the potential to calibrate transport
processes. However, this requires more theoretical work on non-linear
mode coupling in order to interpret their complex frequency spectra,
which include numerous combination frequencies in addition to a
variety of frequency structures \citep{Bowman2016}. So far this
complexity in the observations has prevented unambiguous mode
identifications except partially for a few slow rotators
 {
\citep{Zima2006,Lenz2008,Chen2017a,Chen2017b,Sun2023}} or overtone radial pulsators
\citep{Murphy2020b,Daszynska2023}.   More than a decade ago,
\citet{Breger2012} nicely illustrated that the low-frequency g~modes
and rotational signals are physically connected to the high-frequency
p~modes in the fast-rotating hybrid $\delta\,$Sct--$\gamma\,$Dor
pulsator KIC\,8054146. This {\it Kepler\/} target is a prototypical
case of a non-linearly coupled mode pulsator according to
Eq.\,\eqref{nonlinear} in the presence of fast rotation ($v\sin i=300\pm
20$\,km\,s$^{-1}$).  Predictions for linear oscillation modes applied
to non-rotating evolutionary models following the simplified
Eq.\,\eqref{SL} as computed by \citet{Murphy2023} are likely insufficient
to identify the modes correctly.  For hybrid
$\delta\,$Sct--$\gamma\,$Dor stars, non-linear asteroseismology by
means of coupled mode equations following Eq.\,\eqref{nonlinear}, as
originally developed by \citet{Buchler1997} and \citet{Goupil1998}, offers a way
forward, provided that the theoretical formalisms used are generalised
to include the Coriolis force to fit the low-frequency g~modes as for
SPB pulsators \citep[discussed in the next
  section;][]{Lee2012,Lee2022,VanBeeck2024}.

For the fast-rotating pure p-mode pulsators among the $\delta\,$Sct
stars (e.g.\ Altair \citep{Bouchaud2020}), the challenge is even
greater because the mode identification cannot rely on period spacing
patterns as for the hybrid pulsators. High-degree p~modes also
occur among the high-amplitude modes, as \citet{LeDizes2021} and
\citet{Rieutord2023,Rieutord2024} have shown from combined MOST and
TESS space photometry
and high-resolution spectroscopy. High-degree mode asteroseismology at
the level of fitting the frequencies of individually identified modes
to calibrate transport processes has yet to be developed. For stars
such as Altair, this must be done by using future 2D or 3D stellar
evolution models, for example those delivered by the ESTER code for stars
of higher mass \citep{Rieutord2016,Mombarg2023} coupled to 2D or 3D
pulsation codes such as TOP \citep{Reese2006,Reese2021} or ACOR
\citep{Ouazzani2012,Ouazzani2015}.

{\citet{Reese2009a} have shown that the acoustic p~modes
  occurring in spherical stars are turned into island
  modes for rotationally deformed stars in 2D pulsation
  computations. As shown by \citet{Reese2009b}, these island
  modes can potentially still be identified in observations, provided
  that another family of modes, namely chaotic modes, are absent in
  the frequency spectra.
  Moreover, \citet{Reese2021} have further shown
  from 2D adiabatic pulsation computations in the asymptotic regime of
  high radial order that island modes
  still show a  pseudo large frequency separation. The typical
  differences between actual numerically computed frequencies and
  those deduced from asymptotic approximations range from 5\% to
  12\%. This relatively modest frequency difference
  should in principle safeguard the opportunity to identify
  island modes from pseudo large frequency separations predicted to
  occur in deformed p-mode pulsators.  This is definitely a promising way forward to advance
  p-mode asteroseismology of rapidly rotating $\delta\,$Sct,
  $\beta\,$Cep, and some of the Be pulsators. Currently these theoretical 2D
  pulsation predictions for p~modes in deformed fast rotators
  have hardly been turned into practical applications, with the
  notable exception of Altair in \citet{Rieutord2024}.

\subsection{Gravito-inertial asteroseismology: High-order
g~modes in moderate rotators of intermediate mass}

Moving to the left in Fig.\,\ref{Frequency-Axis} we end up with dwarfs
undergoing slow resonant internal gravity waves with periodicities on
the order of days.
Gravity-mode asteroseismology was kick-started from data
assembled by the CoRoT space telescope \citep{Auvergne2009}, with the
discovery of periodic deviations from a constant g-mode period spacing
in the B3V star HD\,50230 \citep{Degroote2010}.  This pulsator's g
modes behave according to the theoretical predictions for non-rotating
pulsators made before its discovery in the pedagogical paper by
\citet{Miglio2008}.  Bringing this theory and the CoRoT data together
allowed   an estimation of the mixing near its convective core.
This SPB star is a prototype of the sub-class of slow rotators among
g-mode pulsators of intermediate mass, including  the SPB star
KIC\,10526294 whose amplitude spectrum is  shown in the bottom
part of Fig.\,\ref{FT-gmodes} and the $\delta\,$Sct--$\gamma\,$Dor
hybrids KIC\,11145123 and KIC\,9244992 discussed above. The core
boundary mixing of such slow-rotating g-mode
pulsators have since been deduced in several studies
\citep{TaoWu2018,TaoWu2019,TaoWu2020,Mombarg2020,Pedersen2021}.

However, the majority of B and F g-mode pulsators are
moderate to fast rotators whose high-order g-mode periods are of
similar order
{to}
their rotation period (cf.\ blue to orange stars
indicated
{in} Fig.\,\ref{FT-gmodes}). Most of the identified modes in
the 63 well-studied prototypical samples of $\gamma\,$Dor and SPB
stars assembled by \citet{Aerts2021-GIW} indeed have frequencies in
the co-rotating reference frame well below 2$\Omega$. Thus, these modes are in
the sub-inertial regime. In this case it is no longer appropriate to
treat the Coriolis force perturbatively in Eq.\,\eqref{nonlinear}, not even
for moderate rotators (see Fig.\,\ref{Frequency-Axis}). On the other
hand, the centrifugal force is less important than the Coriolis
force to describe such waves (see Eq.\,\eqref{Coriolis-Centrifugal}).
When the perturbation {to} the
gravitational potential can be ignored (the Cowling approximation), one
can make the oscillation equations separable in $(r,\theta,\phi)$ by
adopting the TAR for a uniformly rotating star
\citep{LeeSaio1987,Bildsten1996,LeeSaio1997,Townsend2003,Savonije2005,Mathis2009}.
In that case the radial and azimuthal dependences are the same as for
the non-rotating case, while the latitudinal ($\theta$) dependence is
governed by Laplace's tidal equations, whose solutions are written in
terms of the Hough functions \citep{Hough1898} depending only on the
spin parameter $s=2\Omega/\omega$. Each eigensolution of the
oscillation equations in the TAR is connected to the eigenvalue
$\lambda_{l,m,s}$ of the latitudinal equation, instead of $l(l+1)$ as
would occur in the $\theta$-component of the oscillation equations for
the non-rotating case.  The TAR is valid for modes with frequencies
fulfilling $\omega<\!<S_l$, $\omega<\!<N$ and $\omega<2\Omega$, and
whose $\vec{\xi}(\bm{r})$ is dominantly horizontal such that one can
ignore the horizontal component of the rotation vector and assume
$\bm{\Omega}\simeq(\Omega\cos\theta,0,0)$. This is an excellent
approximation for the majority of modes in the $\gamma\,$Dor and SPB
stars. A modern public tool to compute oscillation modes of rotating
{spherical} stars in the TAR is available in \citet{Townsend2018}.
{\citet{LeeBaraffe1995}, on the other hand, describe a method for mode calculations
in the sub-inertial regime taking the full rotation vector
$\bm{\Omega}$ into account, as well as the deformation of the star
up to second order in $\Omega$.}

\begin{figure*}[t!]
\begin{center}
  {\resizebox{9.cm}{!}{\includegraphics{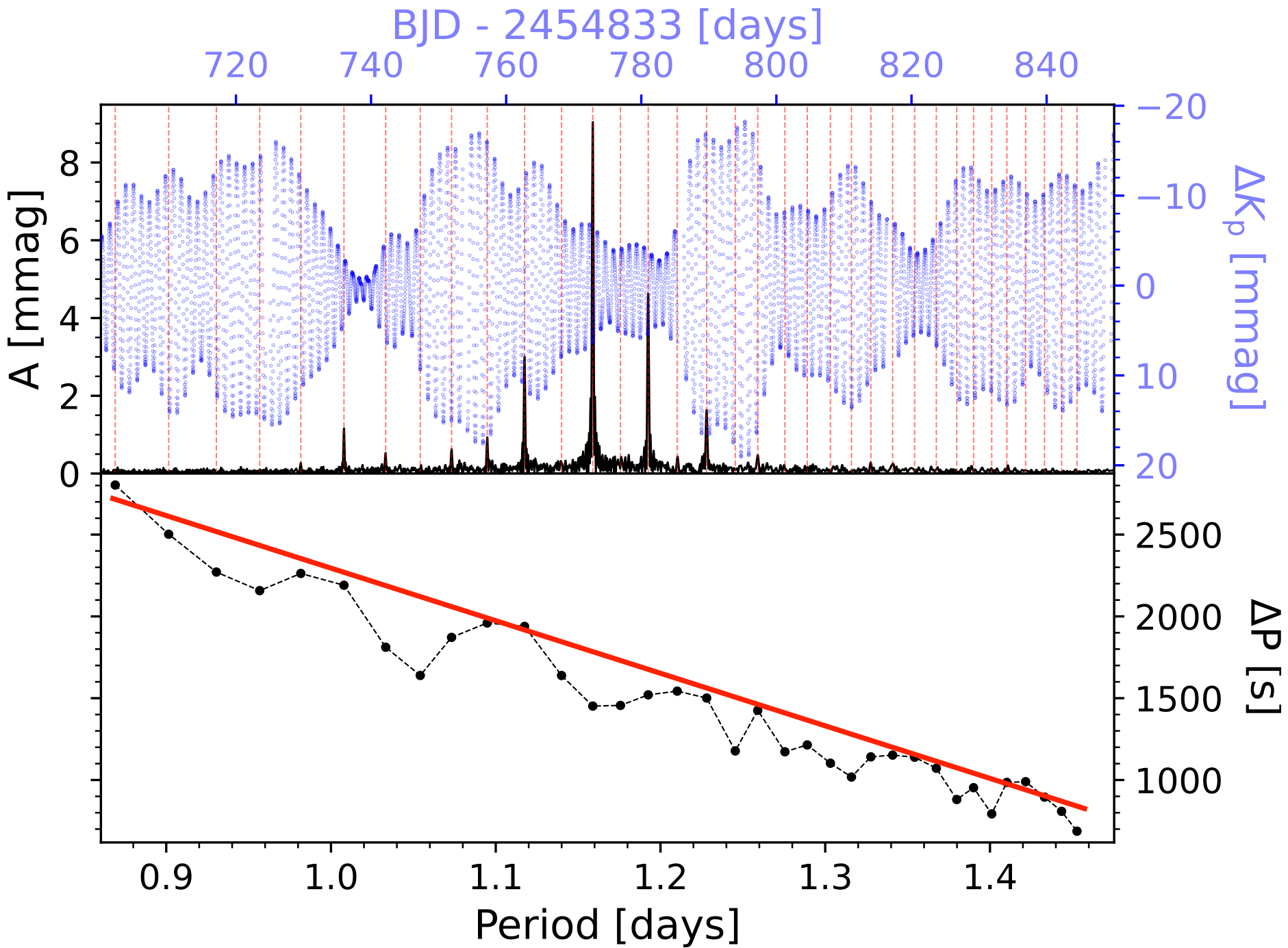}}}
  {\resizebox{9.cm}{!}{\includegraphics{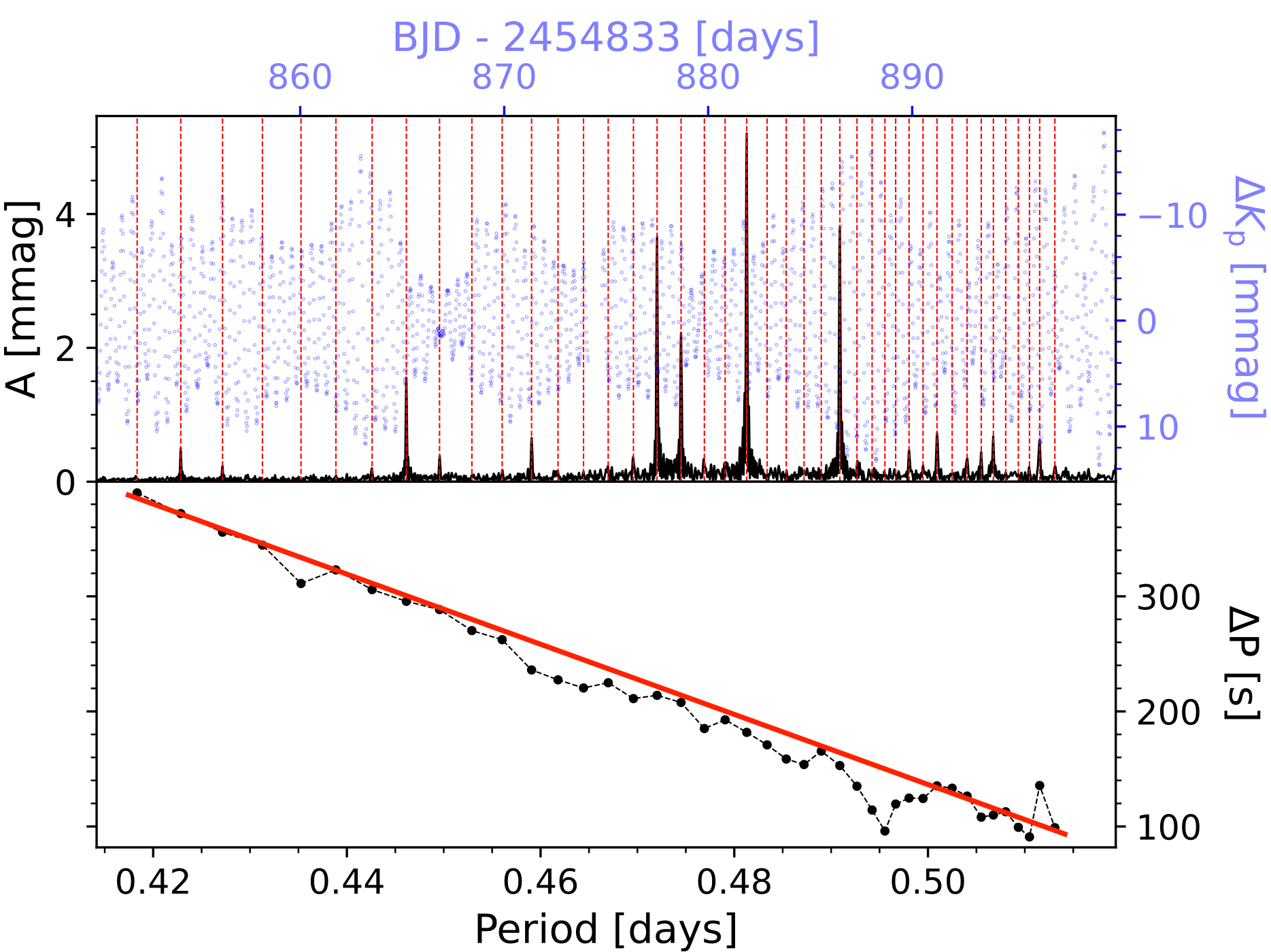}}}  
\end{center}
\caption{\label{GDOR-SPB-DeltaP} Illustration of the method used to deduce
  the near-core rotation frequency of a gravito-inertial mode
  pulsator.  The upper panels show a section of the {\it Kepler\/}
  light curve of the $\gamma\,$Dor star KIC\,8375138 (left) and the
  SPB star KIC\,7760680 (right) as derived respectively by \citet{Papics2015} and
  \citet{VanReeth2015-sample} (blue dots).  The amplitude
  spectra of the 4 yr light curve of these pulsators is overplotted
  (in black) and reveals a regular pattern of high-order dipole g~modes
  of consecutive radial order (red dashed vertical
  lines). The lower panels show the measured period spacings of these
  modes as a function of the mode period (black dots connected by
  dashed black lines to guide the eye). The red solid line is a linear
  fit avoiding the trapped modes causing dips in the pattern. The
  slope of the linear fit is dictated by the internal rotation
  frequency, $\Omega_{\rm rot}$, in the transition zone between the
  convective core and the radiative envelope, as deduced by  
  \citet{VanReeth2015-method}. The figure was reproduced from Chapter\,1 in
  \citet{VanBeeck2023} by Jordan Van Beeck.}
\end{figure*}

The period spacing in the co-rotating frame
among low-frequency g~modes of the same low degree $l$ and azimuthal
order $m$, while being of consecutive radial order $n$
{is denoted as $\left(\Delta P\right)_{\rm co}$. It}
can be approximated by
\begin{equation}\label{DeltaP}
  \displaystyle{
    \left(\Delta P\right)_{\rm co} \simeq \frac{2\pi^2}
             {\sqrt{\lambda_{l,m,s}} \int_{r_1}^{r_2} \frac{N(r)}{r}
                 {\rm d}r\  
\left(1 + \frac{1}{2} \frac{{\rm d}\ln\lambda}{{\rm d}\ln s}\right)}
}\, ,
\end{equation}
where $r_1$ and $r_2$ denote the inner and outer position
of the mode propagation cavity \citep{Ballot2012,Bouabid2013}.
In the limit of zero rotation, the spin parameter is zero and Eq.\,\eqref{DeltaP} reduces to
$\Delta P = \Pi_0/\sqrt{l(l+1)}$, where the 
 quantity
 \begin{equation}
   \label{pi-not}
\Pi_0 \equiv 2\pi^2 \left(\int_{r_1}^{r_2} \frac{N(r)}{r}
                 {\rm d}r\right)^{-1} 
  \end{equation}
has been termed the buoyancy radius of the star
\citep{Miglio2008,Ouazzani2017}. It characterises the size and other properties of the
g-mode cavity. For dwarfs the g~modes have their dominant probing
power just outside the convective core at the deep bottom of the
radiative envelope \citep[see][for illustrations]{Aerts2021-GIW}.
Since $\Pi_0$ has the dimension of a time
and represents the characteristic period for the g~modes of the star, 
it has also been termed the buoyancy travel time
\citep{Aerts2021-RMP}.

{
The mode period spacings observed in an inertial frame ($\Delta P$)}
can be computed from the mode periods expressed in an inertial frame
($P_{\rm in}$) and in the corotating frame ($P_{\rm co}$)
following the transformation formula
\begin{equation}
  \label{transfo}
  \displaystyle{
P_{\rm in} = \frac{P_{\rm co}}{1 - m \frac{P_{\rm co}}{P_{\rm  rot}}} 
  }\, ,
  \end{equation}
with $P_{\rm rot}$ the rotation period of the star.
Equations\,\eqref{DeltaP} and \eqref{transfo}
offer a powerful observable diagnostic tool to achieve the
identification of detected modes from patterns occurring
in the amplitude spectra. Such patterns allow us to estimate $\Omega$
and $\Pi_0$ 
via the involvement of the spin parameter in Eq.\,\eqref{DeltaP}.
{This type of asteroseismic inference involving moderate or fast
  rotation   requires long and uninterrupted data strings in order
  to reach proper precision for the mode periods.}

Major efforts to assemble and interpret period spacing patterns
for ensembles of gravito-inertial pulsators were undertaken from
4 yr {\it Kepler\/} light curves by
\citet{Papics2014}, \citet{Papics2015}, 
\citet{VanReeth2015-sample},
\citet{Papics2017}, \citet{Szewczuk2018},
\citet{GangLi2019,GangLi2020},
\citet{Szewczuk2021}, and \citet{Pedersen2021}.
{Being more than twice as long as the 
CoRoT data strings, the TESS light curves for stars in 
the CVZs also turn out to be
suitable to} hunt for such patterns
\citep{Garcia2022a,Garcia2022b}.
We anticipate the discovery of thousands more
stars with such oscillation
patterns in the future as the TESS light curves become longer.
Overall,
the observed diagrams of $\Delta\,P$ versus mode periods in the inertial
frame of an observer are
already available for more than 700 g-mode pulsators,
covering stars with masses in the range $[1.3,9.0]\,$M$_\odot$ and all
rotation rates from almost zero to critical rotation.

Analysis methods
{based on the TAR with the aim}
of deducing the rotation frequency of $\gamma\,$Dor and
SPB stars were developed by \citet{VanReeth2015-method} and
\citet{Ouazzani2017}.
{The latter authors compared predictions for $\Delta P$ from the
  asymptotic expression in Eq.\,\eqref{DeltaP}, from numerical
  computations adopting the TAR, and from full 2D computations with
  the ACOR code. This led to the important conclusions that $\Delta P$
  values computed from the TAR for zonal and prograde modes differ only by
  a few percent from those obtained with ACOR, while retrograde modes
  may led to differences up to about 10\%.}
 {The method used to derive $\Pi_0$ and $\Omega$
  based on the TAR was} applied to a sample of 37 $\gamma\,$Dor stars
by \citet{VanReeth2016}, with independent validation of the results
by \citet{Christophe2018}. A slightly different
analysis procedure to obtain $\Omega$ was developed by
\citet{Takata2020}, with overall consistent outcomes for $\Omega$
\citep{Ouazzani2019}, but larger diversity (although still consistent) for
the somewhat less precise $\Pi_0$ estimation.
\citet{Pedersen2021} tackled the challenging modelling for a
sample of 26 SPB stars, where mixing in the core boundary layer and in
the envelope is of major importance.

The asteroseismic procedure used to
deduce the internal rotation frequency of $\gamma\,$Dor and SPB stars
as commonly applied now is illustrated for prograde dipole modes in
Fig.\,\ref{GDOR-SPB-DeltaP}.  The higher the rotation, the steeper the
slope in the $\Delta\,P$ versus $P_{\rm in}$ diagrams for
prograde modes. Such modes represent waves travelling along
the rotation of the star, and thus their periodic variation
re-occurs faster in the line of sight for
faster rotation. Hence, the mode periods, and by implication the
difference of mode periods for consecutive overtones, decrease. This
creates the characteristic downward tilt in the bottom panels of
Fig.\,\ref{GDOR-SPB-DeltaP}, as predicted theoretically   by
\citet{Bouabid2013} and put into practice by
\citet{VanReeth2015-method,VanReeth2016,Ouazzani2017}.
Retrograde gravito-inertial modes have the opposite effect
and lead to an upward trend characterised by a positive slope
\citep[see][for numerous examples]{VanReeth2015-sample}.

The detected upward slopes of period spacing patterns in fast rotators
laid the foundation for the discovery of toroidal Rossby modes in
$\gamma\,$Dor stars by \citet{VanReeth2016}, in additon to the
detection of retrograde gravito-inertial modes.
It is remarkable that Rossby modes were
discovered in many $\gamma\,$Dor stars from {\it Kepler\/} data
already in 2016, while they were only detected two years later in the
Sun \citep{Loptien2018}, as discussed in Sect.\,4.1.1.  From their
dedicated description of and search for Rossby modes,
\citet{Saio2018-Rossby} found them to occur in the fastest rotators
among the $\gamma\,$Dor stars, and in pulsating Be stars with
outburst phenomena
\citep{Huat2009,Neiner2012b,Balona2015,Rivinius2016}.  Although
several of the {\it Kepler\/} SPB stars also reveal photometric
outbursts, as found by \citet{VanBeeck2021}, they have smaller
amplitudes than those of the pulsating Be stars and do not necessarily
lead to H$\alpha$ emission in their spectra as is the case for the Be
stars.  Rossby modes have not yet been found in SPB stars, but the
outbursting SPB and pulsating Be stars do share the common property of
non-linear mode couplings \citep{Baade2016,VanBeeck2021}. This detected
behaviour and theoretical developments of non-linear asteroseismology (e.g.\  \citet{Lee2012,Lee2022,VanBeeck2024})
offer the great future potential to also include mode amplitudes in
forward seismic modelling, aside from the frequencies.

A summary of the findings based on 4 yr {\it Kepler\/} light curves
of 611 $\gamma\,$Dor stars in terms of stellar properties is available
in \cite{GangLi2020} and for the internal rotation properties in
Fig.\,6 of \citet[][not repeated here]{Aerts2021-RMP}. The evolutionary
context of the g-mode pulsators summarised in that review paper
reveal overwhelmingly that dwarfs of intermediate mass have nearly
uniform rotation throughout their entire main-sequence phase
\citep[see also][]{VanReeth2018}, irrespective of their mass and
current rotation rate.  Levels of core-to-surface rotation ratios of
$\gamma\,$Dor stars remain below about 10\% \citep{GangLi2020},
irrespective of the value of the rotation rate itself.  The summary of the
internal rotation {frequencies provided} in \citet{Aerts2021-RMP} also coupled the
angular momentum of intermediate-mass dwarfs to that of evolved low-
and intermediate-mass (sub-)giants.
 {This revealed}  that all these
stars transport and lose their angular momentum in such a way that the
angular momentum of their helium-burning cores, by the time they are
red giants, is in agreement with those of sub-dwarfs and white
dwarfs \citep{Aerts2019-ARAA}.  

The ability to estimate $\Omega (r)$ and $\Pi_0$
from $(\Delta\,P)_{\rm  in}$ versus $P_{\rm in}$ diagrams, as illustrated in
Fig.\,\ref{GDOR-SPB-DeltaP}, along with the envelope rotation from p
modes and/or surface rotation from rotational modulation
\citep{VanReeth2018} laid the foundation for the new sub-field of
gravito-inertial asteroseismology for moderate and fast rotators
\citep{Aerts2018}. This was  applied to samples of F and
B pulsators as a way to estimate the  level and type of core
boundary and the envelope mixing. Calibrating the internal mixing is
necessary to properly age-date such stars and to deduce the mass of
the helium core they will produce by the time of their core hydrogen
exhaustion. Gravito-inertial asteroseismology currently delivers
a typical precision of  between 5\% and 20\% for the age and helium core
mass at the TAMS
\citep{Mombarg2019,Mombarg2021,Pedersen2021,Pedersen2022a,Pedersen2022b,Fritzewski2024}.
It also provides overwhelming evidence of extra mixing in the
near-core region of rotating stars, fully in line with the conclusion
drawn from eclipsing binaries \citep{Tkachenko2020,Johnston2021}. It
also shows that the envelope mixing is connected to the internal
rotation, but that rotation is not the only physical cause of
the element transport \citep{Pedersen2021,Pedersen2022a,Mombarg2023}.

Numerically stable tools to compute the eigenvalues of Laplace's
tidal equations are available in \citet{Townsend2020} and make the
estimation of the near-core rotation frequency of $\gamma\,$Dor and
SPB stars a high-precision observational science. Given that the
internal rotation frequency is   also accessible from TESS CVZ
data, period spacing patterns due to low-degree gravito-inertial modes
are within reach for many more pulsators
\citep{Garcia2022a,Garcia2022b}, including stars in young open
clusters \citep{Fritzewski2024,GangLi2024}. This opens
up the territory of applying gravito-inertial asteroseismology to
samples of open clusters with a variety of ages, rotation rates,
metallicities, single and binary populations, and merger products with
the near-future potential to calibrate the internal structure and
angular momentum of young clusters at their birth (see
Fig.\,\ref{Fountain}).

Aside from the near-core rotation frequency deduced from
gravito-inertial modes in the radiative envelope of intermediate-mass
stars, a small fraction of  $\gamma\,$Dor stars also offer a
powerful observational signal that can be used to derive the rotation rate of the
convective core itself. This was first achieved by
\citet{Ouazzani2020} from a pertinent characteristic dip in the period
spacing pattern of the rapidly rotating $\gamma\,$Dor pulsator
KIC\,5608334, pointing to mode coupling between a core inertial
mode and an envelope gravito-inertial mode in the presence of fast
rotation. \citet{Saio2021} subsequently hunted for these inertial
core modes, and found the signal in the period spacing diagram for 16
of the 611 $\gamma\,$Dor stars in \citet{GangLi2020}.  These
observational findings are   readily understood in terms of
theoretical expressions for a mode coupling coefficient between
inertial core modes and gravito-inertial envelope modes for
$\gamma\,$Dor stars derived by \citet{Tokuno2022}. Numerical
computations of these coupling coefficients
by \citet{AertsMathis2023} from the best characterised
$\gamma\,$Dor stars are in agreement with the theory.
Although it is pertinent to only a small fraction of
$\gamma\,$Dor stars, notably the most rapid rotators, the detection of
this mode coupling is of major importance. It opens up the
fast-rotating convective cores of F dwarfs for observational
scrutiny and confirms that their rotation is almost the same
everywhere in their interior, from the core across the entire
radiative envelope.  \citet{AertsMathis2023} showed that the coupling
coefficients for SPB stars are smaller, explaining why such mode
coupling is absent for these hotter pulsators.

Another way to probe the rotating convective core was established for
a few pulsating late-type Be stars whose modelling is based on their
fast rotation and on the observed properties of their stochastically
excited low-frequency gravito-inertial modes, without having
identifications for the individual modes
\citep{Neiner2012a,Neiner2012b,Neiner2020}. These cases are for stars that rotate so 
rapidly   that the Coriolis force cannot be treated either
perturbatively or in the TAR.
{Full pulsation computations in 1D, treating the rotational
  deformation up to second order in
  $\Omega$,   were done for the Be pulsator HD\,49330 by
  \citet{Neiner2020}. The predictions of the oscillation modes were
based on the formalism by \citet{LeeBaraffe1995}, with the aim of explaining
the observed variability spectrum qualitatively, without fitting 
individual mode frequencies.}

All the observational findings on the internal rotation of
gravito-inertial mode pulsators give rise to asteroseismic and
evolutionary calibrations of angular momentum from the stellar birth
to the evolved phases. Such studies are ongoing for single field
stars, and involve the global effect of various rotationally induced
processes \citep{Pedersen2022b,Mombarg2023} and/or core magnetic
dynamo fields \citep{Moyano2023}. The latter are based on the magnetic
Tayler instability developed by \citet{Fuller2019} as an improvement
on the classical
{Tayler-Spruit}
dynamo \citep[see
  also][]{Takahashi2021}.  \citet[][their Fig.\,8]{Aerts2021-GIW}
estimated convective and wave Rossby numbers, core boundary stiffness
values, and internal magnetic field strengths for a sample of 63 well-characterised gravito-inertial mode pulsators. Assuming a stable
dipole field, they predict field strengths in the ranges
$[10^{4.5},10^{5.5}]$\,G and $[10^{5},10^{6.5}]$\,G for SPB and
$\gamma\,$Dor stars, respectively. For the $\gamma\,$Dor stars, this
is fully in line with the predictions by \citet{Cantiello2016} and
\citet{Stello2016} based on dipole mixed mode suppression in about
20\% of the {\it Kepler\/} red giants.  Direct detections of core
magnetic fields from mode splittings or magnetic characteristics in
dipole mode period spacing patterns, as recently found in a small
fraction of red giants (see Sect.\,4.1.3), have not yet been established for
gravito-inertial mode pulsators. However, given the numerous recent
theoretical formalisms developed to hunt for specific signatures
\citep{Prat2019,Prat2020,VanBeeck2020,Dhouib2022,Rui2024}, it {should only be} a matter of time for true direct detections to be achieved.
 {For now, an
  upper limit of the  order of $5\cdot 10^5\,$G was found}
for the rapidly rotating
CoRoT SPB star HD\,43317 from asteroseismic modelling by
\citet{Buysschaert2018}.  {\citet{Lecoanet2022} confirmed this
result by confronting the mode
suppression observed in the star with predictions from 3D MHD
simulations}.

Aside from probing the internal magnetism in or near the core of
intermediate-mass pulsators via their gravito-inertial modes, Rossby
modes and rotational modulation detected in 162 stars known as `hump
and spike' stars gives indirect evidence of magnetic fields in the
envelopes of these stars \citep{Henriksen2023a,Henriksen2023b}. However, it is still
unclear whether the observed variability is due to magnetic spots on
the stellar surface triggered by a dynamo field generated in
sub-surface layers or is rather caused by overstable convective modes
that excite resonant g~modes with frequencies equal to the rotation
frequency of the convective core \citep{LeeSaio2020,Lee2021,Lee2022}.
Further research on the
interaction between time-dependent convection, rotation, and magnetism
in the nonadiabatic outer envelope is required to fully exploit the
detected variability of these fastest rotators.

In conclusion, gravito-inertial asteroseismology of intermediate-mass
stars already made internal rotation and will soon make the internal
magnetism of moderate and fast rotators an observational
science. Exploitation of the detected and identified gravito-inertial
modes is  applied to hundreds of dwarfs of intermediate mass
in the galaxy, with a large potential for asteroseismology of
young open clusters of less than a few billion years in age
\citep{Fritzewski2024,GangLi2024}. Such applications
{of asteroseismology} to clusters observed by TESS and PLATO
will deliver tight constraints on the angular momentum at stellar birth
for various birth masses and metallicity regimes in the not too distant future
(see Fig.\,\ref{Fountain}).

\section{Adding accurate dynamical masses and radii
  to increase asteroseismic precision}

Asteroseismic modelling of rotating pulsators is a high-dimensional
regression problem to solve \citep{Aerts2018};   the aim is to obtain
high-precision estimates of the free parameters occurring in the
stellar models.  Aside from the measured effective temperature and
gravity from high-resolution spectroscopy, and the star's luminosity
from {\it Gaia\/} astrometry, it is beneficial to limit the parameter space
from additional and preferably model-independent input parameters.

Major advances on this aspect may come from (future) interferometric
radii, but this is limited to relatively bright stars. To date, combined
interferometry and asteroseismology has mainly been applied to
bright slow-rotating low-mass Sun-like pulsators
\citep{Cunha2007,Huber2012}. Applications to fast rotators are scarce
and   so far can only rely on uninterrupted space photometric light
curves of short duration, limiting the identification of sufficient modes
\citep[see Rasalhague and Altair
{analysed}  by][respectively]{Monnier2010,LeDizes2021,Rieutord2024}.  It is noteworthy that
the g~modes of the bright hybrid  $\delta\,$Sct--$\gamma\,$Dor pulsator
Rasalhague discovered in its MOST data were already correctly
interpreted as gravito-inertial modes by \citet{Monnier2010}. In our
definition, this pulsator is a moderate rotator with a value of 
$\Omega/\Omega_{\rm crit}^{\rm Kepler}\sim 48\%$ 
(or $\Omega/\Omega_{\rm crit}^{\rm Roche}\sim 88\%$) for
$\Omega\simeq\,1.65\,$d$^{-1}$.  However, the low frequency resolution
of the light curve prevented the detection of clear period spacing
patterns and the derivation of $\Delta P$. Moreover, the methodology
described in Sect.\,4.4 based on the observable diagnostic in
Fig.\,\ref{GDOR-SPB-DeltaP} did not yet exist in 2010.  It is
definitely worthwhile to revisit the few bright fast-rotating
pulsators with a good interferometric radius and to hunt for
g~modes in their space photometry.  Moreover,
the new SPICA instrument
in development at the CHARA array \citep{Mourard2022} holds great
promise to intertwine interferometry and asteroseismology for a large
sample of pulsators in the not-too-distant future.

\begin{figure*}[t!]
   \centering
   \includegraphics[width=18.5cm]{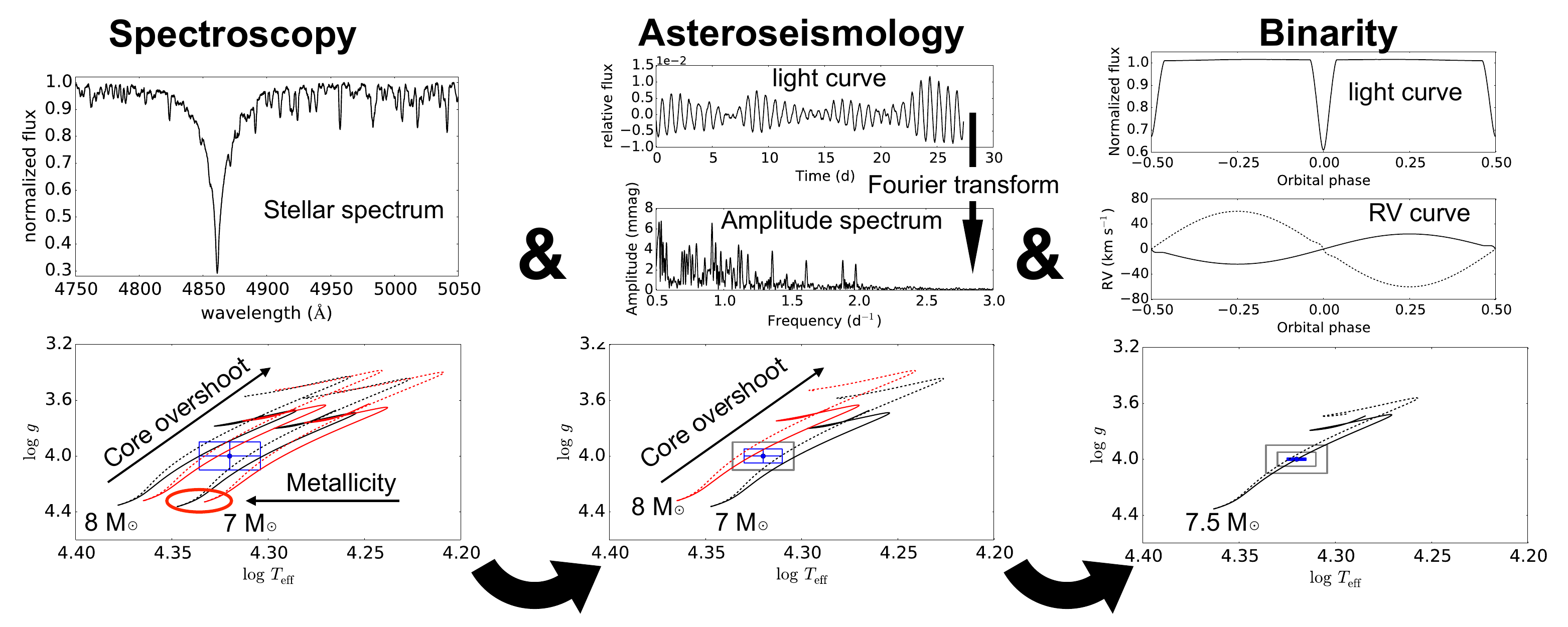}
      \caption{Schematic representation of the complementarity among
        atmospheric, binary, and asteroseismic analyses. The top row
        shows observables for each method, while the bottom row
        highlights the gain in precision and accuracy as the probing
        power of the different methods is combined.
         {The red curves in the lower left panel represent models
          for two masses with 
          $Z=0.014$, while the black curves have $Z=0.006$. The solid lines
have no core boundary mixing, while the dotted lines do, increasing the
mass of the convective core along the evolutionary path. In the middle
lower panel, spectroscopy has allowed us to fix the metallicity, and the
red and black curves now differ only in mass, representing evolutionary
paths with (dotted lines) or without (solid lines)
core boundary mixing. The lower right panel illustrates that
asteroseismology further allows us to pinpoint the mass and limit the
amount of core boundary mixing.}}
    \label{Andrew}
\end{figure*}

A highly successful and already applied extra constraint on
asteroseimic modelling comes from the dynamical masses of detached
eclipsing binaries.  Binary stars are found in diverse configurations
in terms of orbital characteristics and physical properties of their
individual components.  In this section we focus on the classes of
astrometric and eclipsing binaries as ideal test beds to help
asteroseismic modelling.  The key point is that such objects allow us
to infer the masses and radii of the individual components by only relying
on the basic laws of binary dynamics. Hence, the masses and radii
are model-independent and their
precision, and more importantly accuracy, is limited solely by the
quality of the observations.

\citet{Torres2010} presented a sample of 95 binary systems (190 stars in
total) in a detached configuration, that is, systems where both
components are well within their respective Roche equipotential
surfaces and do not interact with each other.  Their masses and radii
are measured to better than $3\%$ accuracy. This sample
covers $M\in[0.2,27]$\,M$_{\odot}$ and $R\in[0.2,25]$\,R$_{\odot}$.
\citet{Serenelli2021}
discussed state-of-the-art methods for the accurate determination of
stellar masses and presented a list of 200 benchmark stars, among which
binary systems, with relative mass accuracy between [0.3,2]\% for the
mass range $M\in[0.1,16]$~M$_{\odot}$. For the sake of brevity, we do
not consider every type of binary system in this review,
but restrict our analysis to cases with the most precise and accurate
determinations of stellar masses and radii.
A recent review of eclipsing binaries
with pulsating components in the era of space-based mission is
provided in \citet{Lampens2021}. 

Having an accurate mass measurement alone does not enable stellar
modelling at high precision due to the multitude of free parameters
and the complex correlation patterns involved in the
process.  However, when combined with asteroseismic modelling, binary
observables can break some of the parameter degeneracies.
Both components of the binary are coeval and were born with
the same initial chemical composition. Moreover, high-precision
inferences of the effective temperatures of the stars from
spectroscopic methods are possible, thanks to the fact that the
components' surface gravities are known to high accuracy from the
dynamical measurements of the stellar masses and radii.

By positioning   the binary components in the HRD or Kiel
diagram and demanding that the two stars were born together from 
the same molecular cloud, one can assess whether and how well
stellar models are able to predict the dynamical masses and
evolutionary paths of both stars.  This way, binarity has been
exploited in numerous studies in the literature; the studies demonstrate that
current stellar models can rarely reproduce the accurately measured
properties of individual binary components and that extra physical
ingredients are needed to reconcile models and observables. This has
led to the need for extra mixing, from relatively small-scale near-core
mixing \citep[often simply referred to as convective core
  overshooting,][]{Schroeder1997,Guinan2000,Lacy2008,Torres2014a,Tkachenko2014,Meng2014,Claret2016,Claret2017,Claret2018,Claret2019,Higl2017,Pavlovski2022,Rosu2020,Rosu2022}
to large-scale mixing caused by rotation and/or internal gravity waves
\citep[e.g.][]{Ryan1995,deMink2009,Garland2017,Pavlovski2009a,Pavlovski2009b,Pavlovski2018,Pavlovski2023}. The
need for extra near-core and/or envelope mixing comes from the need of
more massive convective cores of the binary components
\citep[e.g.][]{Tkachenko2020,Johnston2021,Martinet2021,Rosu2022}.
Interactions between magnetic fields, convection, atomic diffusion,
and stellar spots have also been invoked to bring the radii of the
stars into agreement with model predictions
\citep[e.g.][]{Birkby2012,Torres2014b,Higl2017,Cruz2018,Garrido2019,Morales2022}.

Figure\,\ref{Andrew} depicts the complementarity among the
methods of spectroscopy, asteroseismology, and binarity for stellar
modelling. The top row illustrates  the type of observations for each
method: stellar spectra for atmospheric analysis, periodic brightness
variations caused by stellar oscillations for asteroseismic analysis,
and periodic dimming of the stellar light and radial velocity
variations for binary analysis. The bottom row highlights the gain in
precision and accuracy in the modelling as  the
spectroscopic and asteroseismic methods are combined, at the same time benefiting
from the accurate masses of the star from the
binary dynamics.  If the position of the star in the
\teff-\logg\ Kiel diagram is only constrained by its atmospheric parameters
inferred from spectroscopy, the typical error box (blue error box in the
bottom left panel) is too large to conclude whether the star will follow an
evolutionary track of a 7\,M$_{\odot}$ intermediate-mass star with a
white dwarf as the end-of-life product, or the one of an 8\,M$_{\odot}$ star that
ends its life as a neutron star. The effects of the initial
metallicity (red versus black line) and of core boundary mixing (solid
versus dashed line) on the evolutionary path of the star are too small
to be distinguished with spectroscopic methods alone.

Asteroseismology offers unique sensitivity to the physics in the deep
stellar interior and is largely complementary to binary and
atmospheric analyses in terms of the physical mechanisms these methods
probe.  Adding asteroseismic constraints delivers the smaller blue box
instead of the grey one for a pure spectroscopic solution in the bottom
middle panel. This asteroseismic addition improves the modelling
substantially as it allows us to eliminate some of the highly improbably
solutions, such as those based on the higher- and lower metal content for
the 7\,M$_{\odot}$ and 8\,M$_{\odot}$ models, respectively.  Moreover,
asteroseismic constraints on the amount and functional form of the
near-core mixing is often tighter than suggested by the blue error box
in the bottom middle panel in Fig.\,\ref{Andrew}, such that
the figure represents a pessimistic scenario. Ultimately, the additional
model-independent constraint from binary dynamics decreases the error
box to a fraction of the pure spectroscopic error box or the combined
spectroscopic and asteroseismic error box (blue versus grey boxes in the
bottom right panel). This way, one can characterise the physics of the
binary pulsator with high accuracy.

Finding binaries for which all three analysis methods depicted in
Fig.\,\ref{Andrew} can be combined is a difficult task, and such
studies are therefore still scarce in the literature. The 
following criteria need to be met simultaneously to do the
above-described analysis: (i) the object is an eclipsing or
astrometric binary with high-quality photometric or astrometric data;
(ii) the object is a spectroscopic double-lined binary such that the
radial velocities of both binary components can be measured and their
masses can be inferred; (iii) at least one of the components is a
pulsator with time series photometry of its brightness variations
sufficiently long to achieve high-precision 
frequencies for sufficient modes; and (iv) at least some of the
detected modes can be identified in terms of their
$(l,m,n)$.
 {Even if these four conditions are met, it may still be
  impossible to determine which of the detected oscillation
  frequencies originates in which star, notably in
  binaries with components of almost equal mass.}

Various slow-rotating low-mass solar-like pulsators in eclipsing or astrometric
binaries satisfy all four criteria. Moreover, because the mass and
radius of a solar-like pulsator can be inferred directly from its
oscillations through scaling relations when the effective temperature
of the star is known \citep{Kjeldsen1995}, these asteroseismically
inferred quantities can be confronted with the model-independent
counterparts inferred from binary dynamics
\citep[e.g.][]{Gaulme2013,Appourchaux2015}.  While asteroseismic
measurements of mass and radius are in a good agreement with the
dynamical values for unevolved solar-like pulsators
\citep[e.g.][]{White2017}, the asteroseismic masses and radii of red
giants are systematically overestimated
\citep[e.g.][]{Brogaard2016,Gaulme2016,Themessl2018,Kallinger2018,Benbakoura2021}.

Intermediate-mass p-mode $\delta$\,Sct pulsators in eclipsing
binaries constitute another category of objects for applications,
although mode identification remains a challenge for such moderate to
fast rotators (see Sect.\,4.3).  For this reason, validation of
stellar models with $\delta$\,Sct pulsators in eclipsing binaries
is currently primarily limited to the use of empirical relations between
their oscillation properties and mean densities obtained from binary
dynamics
\citep[e.g.][]{Garcia-Hernandez2015,Garcia-Hernandez2017}. Because
many $\delta$\,Sct pulsators are found in close Algol-type binary
 systems \citep[also known as oEA stars;][]{MKrtichian2004,Mkrtichian2022} and
p-mode oscillations reach their highest amplitudes in the outer layers
of stars, oEA stars are often exploited to study binary interactions
(see Sect.\,6). As for high-mass p-mode pulsators, the
{discovery}
of 78 $\beta\,$Cep pulsators in eclipsing binaries by
\citet{Eze2024}
in the TESS data is extremely promising to
calibrate binary supernova progenitors, of which some may be on their
way to becoming gravitational wave emitters.

The group of g-mode pulsators in eclipsing binaries is among the most
promising in the context of validation and calibration of stellar
models. The main reason lies in the effective probing power of the
deep stellar interior facilitated by g~modes, in particular the
near-core regions (see Sect.\,4.4).  Binary and asteroseismic analyses
become truly complementary in this particular regime and make
strategies shown in Fig.\,\ref{Andrew} optimally suitable.
Currently, only a limited number of $\gamma$~Dor,
SPB stars, or white dwarf g-mode pulsators in eclipsing binaries
with mode identification are known. 
\citet{Gaulme2019} presented a compilation of  300 
pulsating stars in eclipsing binaries observed with {\it Kepler\/} of
which about  one-third have $\gamma$\,Dor pulsators. Among such pulsators
found by \citet{Li2020} is the remarkable
quadruply eclipsing heptuple system
KIC\,4150611 \citep{Kemp2024}, whose primary exhibits a long series of prograde
dipole gravito-inertial modes yet to be modelled asteroseismically.
Further, \citet{Southworth2022a} and \citet{Southworth2022b} reported the detection of
22 $\gamma$\,Dor, SPB, and $\beta$\,Cep pulsators in
eclipsing binaries observed with TESS. For now, however, none of these
systems have identified modes.  

\citet{Sekaran2020} undertook a systematic search for unevolved g-mode
pulsators in eclipsing binaries with period spacing patterns. Among
the 93 detected eclipsing systems with a g-mode pulsating component,
seven were found to exhibit period spacing patterns. In a follow-up
study, \citet{Sekaran2021} investigated one of the detected systems in
great detail. This deals with the slow rotator KIC\,9850387, whose 
dynamical parameters are found to be in agreement with the parameters extracted
from asteroseismic modelling provided that a high level of internal mixing for
the pulsating component is used for its internal structure. The
authors proposed that this is a result of intrinsic
non-tidal mixing mechanisms.
The findings in \citet{Sekaran2021}
reinforce the common conclusion in the literature from large samples
of single-star pulsators and non-pulsating eclipsing binaries: 
stellar models require extra internal mixing compared
to standard models to achieve a meaninful calibration. Similar studies
of moderate to fast rotators with g~modes in detached eclipsing binaries
are not yet available, given that almost all such objects reveal tidal
interactions, as discussed in the next section.

\section{Tidal asteroseismology: Interplay between rotation, tides, and
  oscillations in binaries}\label{Sec:BinPulsInterplay}

The space photometry revolution gave rise to a `zoo' of observed
phenomena in {\it Kepler\/} and TESS data caused by tidal interactions
in close binaries. Systems with free oscillation modes perturbed by
tides, with tidally excited modes, with tidally tilted pulsation axes,
with tidal deformation of the mode cavities, and with non-linear mode
excitation due to tides have all been found \citep[][for a
  review]{Guo2021}.  In order to study the effect of linear tides in
binaries with rotating components we need to solve
Eq.\,\eqref{SL}, while additional terms (as in Eq.\,\eqref{nonlinear}) have
to be included for non-linear tides.

Tidally excited oscillations occur at exact multiples of the orbital
frequency. They come in various flavours and may be caused by an
equilibrium tide in case of a circular orbit and/or dynamical tides in
eccentric orbits. An equilibrium tide is a static phenomenon occurring
due to the balance between the pressure force and gravity and
manifests itself as a geometrical deformation of one star due to the
force exerted by the companion. Dynamical tides, on the other hand,
cause a periodic acceleration due to $\bm{a}_{\rm extra}^{\rm
  external}$ in Eq.\,\eqref{SL} as the components revolve around each other
in their orbit \citep[see the pedagogical paper by][]{Fuller2017b}.
Tidal oscillations give rise to slow waves at low frequencies, which
naturally occur in close binaries because the orbital, rotation, and
g-mode periods are all  of similar order and reside in
 {the}
left
part of Fig.\,\ref{Frequency-Axis}.  On the other hand, self-excited
oscillations may get perturbed due to both the equilibrium and
dynamical tides, and such deformed oscillations are referred to as
tidally perturbed oscillations.  They may occur across the whole
frequency axis in Fig.\,\ref{Frequency-Axis}, including short-period
p~modes, when tidal forces are significant compared to the other forces
at play in the system.

Tidally excited oscillations represent the main mechanism of tidal
dissipation in binaries composed of intermediate- or high-mass stars
with a convective core and a radiative envelope
\citep{Zahn2013,Alvan2013,Ogilvie2014}.  In a nutshell, the mechanism
acts as follows: (i) the tidal potential excites internal gravity
waves in the deep stellar interior at the interface of the convective
core and the radiative envelope of the star; (ii) these low-frequency
travelling waves experience strong damping due to radiative diffusion
\citep{Zahn1997}; and (iii) the waves break in the low-density
near-surface layers of the star, depositing their energy and angular
momentum \citep{Rogers2013}, causing tidal heating and
(pseudo-)synchronisation. Such tidal synchronisation is first
initiated in the outer layers of the star and gradually proceeds
inwards as more waves are being generated through tidal forcing and
continue to dissipate. The dissipation is particularly efficient at
the critical layers deep inside the star characterised by
resonant waves having frequencies equal to the local co-rotation rate
\citep{Alvan2013}.  

Tidally excited oscillations corresponding to standing waves occur at
exact multiples of the orbital frequency. They do not suffer from
strong damping, and can therefore be observed in stars. Because the
dominant component of the tide-generating potential is a spherical
harmonic of $l=2$, linear dynamical tides are expected to give rise to
quadrupole modes.  Some tidally excited oscillations can reach
amplitudes exceeding a parametric instability threshold, in which case
their mode energy will be transferred to daughter modes through
non-linear mode coupling. The observational manifestation of such
non-linear mode coupling is the occurrence of pairs or multiplets of
daughter modes whose linear combination(s) satisfy resonance
conditions
\citep[e.g.][]{Oleary2014,Manzoori2020,Guo2020a,Guo2021,Guo2022}. Moreover,
some of the tidally excited oscillations can enter into resonance,
locking with the orbital frequency of the binary. This happens when
the two frequencies evolve in concert. Such orbital resonant mode locking
substantially increases the efficiency of tidal dissipation and
orbital circularisation
\citep[e.g.][]{Witte1999,Witte2001,Willems2003a,Willems2003b,Burkart2012,Fuller2012,Fuller2017a,Fuller2017b,Zanazzi2021,Kolaczek2023}.

It was recently also shown by \citet{Li2020} and \citet{Fuller2021}
that `inverse tides' may occur in close binary systems hosting a
g-mode pulsator. In this process, which occurs when the tidally forced
mode amplitude approaches the mode's saturation amplitude, a
self-excited oscillation mode transfers (some of) its energy to the
binary orbit. This may pump up the orbital eccentricity, create
spin-orbit misalignment, and cause asynchronous rotation. This action
of tidal dissipation is thus opposite to the standard view and rather
pushes the state of the binary away from synchronised circular orbits.
Another non-standard effect with respect to tidal theory in a two-body
system was presented by \citet{Felce2023}. They describe how the
presence of a tertiary component may trap close binaries into a
spin-orbit equilibrium known as a Cassini state. In this case, orbital
precession due to the tertiary lies at the origin of the stable
spin-orbit configuration.  This mechanism has been invoked as an
explanation for a minority of close binaries with stars of
intermediate mass having extremely slow sub-synchronous rotation. Such
a state can be understood by alterations of the spin-orbit
misalignment angle and of the internal rotation rate caused by tidal
dissipation due to inertial waves \citep{Fuller2021}.

Although the first (ground-based) observations of tidally excited
oscillations in close binaries date from  two decades ago
\citep[][]{Willems2002,Handler2002,Quaintrell2003}, numerous
detections had to await uninterrupted space-based photometric
observations and gave rise to the great variety of observed behaviour
we know today. \citet{Maceroni2009}
{{reported}
  the detection of
tidally excited  oscillations in the eccentric binary HD\,174884 from
CoRoT data. A few years later, \citet{Welsh2011} presented
spectacular tidally excited oscillations at the 90${\rm th}$ and
91${\rm st}$ harmonic of the orbital frequency in
{HD\,187091, also known as KOI-54.}
The light
curve of KOI-54 is characterised by periodic brightness variations
near periastron passage caused by geometrical deformation of the star
as a response to the instantaneous tidal force. KOI-54 became the
prototype of the class of eccentric binaries exhibiting periodic
variability in its light curve similar to cardiograms of human
heartbeats. These eccentric binaries are therefore also dubbed
`heartbeat stars' \citep{Thompson2012}.  Just like KOI-54 itself,
the majority of heartbeat stars exhibit tidally excited oscillations
\citep[e.g.][]{Hambleton2013,Hambleton2016,Kjurkchieva2016,Pablo2017,Hambleton2018,Guo2017,Guo2019,Cheng2020,Kolaczek2021,Jayasinghe2021,Ou2021a,Ou2021b,Wrona2022,Sharma2022,Wang2023}. This
makes them unique astrophysical laboratories to validate and improve
tidal excitation theory
\citep[e.g.][]{Burkart2012,Fuller2012,Fuller2017b,Oleary2014,Guo2022,MengSun2023}. We
refer to \citet{Ogilvie2014} and \citet[][and
  references therein]{Guo2021} for comprehensive discussions
on the theoretical and observational aspects of tidally excited
oscillations in close binaries.

Another distinct group of pulsators in close binary stars is the
class of stars with tidally perturbed oscillations, which means that
the star's free oscillations are affected by the presence of a
companion.
Tidal deformation in a close binary system affects the
mode propagation cavities inside the components, as well as the
alignment of the pulsation axes. While the effect of the tidal force
on the eigenfrequencies of the star has been investigated to a
reasonable level of detail with polytropic models
\citep[e.g.][]{Saio1981,Smeyers1983,Martens1986,Reyniers2003a,Reyniers2003b,Smeyers2005},
the consequences of the equilibrium tide on the mode eigenfunctions
has been studied only recently \citep{Fuller2020}. While the former
effect results in tidal splitting of a self-excited oscillation mode
of a given spherical degree $l$, the \citet{Fuller2020} formalism
allows   tidal coupling between tidally split multiplets with
different angular degree $l$. The effect of such tidal mode coupling
can be three-fold: (i) tidal alignment, where the axis of the
oscillations gets aligned with the tidal axis rather than with the
rotation axis of the star; (ii) tidal trapping, such that oscillation modes
are confined to regions corresponding to the tidal pole or equator; and
(iii) tidal amplification, increasing flux perturbations in
the tidal polar regions due to the propagation of acoustic waves close
to the stellar surface.

Observationally, tidally perturbed oscillations (which include both
tidally split and trapped oscillations) manifest themselves as (a
series of) modes that are offset from their adjacent orbital harmonics
and are (quasi-)equidistantly spaced with the orbital frequency of the
star
\citep[e.g.][]{Balona2018,JWLee2021,Steindl2021,VanReeth2022,Jennings2023}.
Some of the class members exhibit readily interpretable observed
frequency patterns due to one of the above theoretical
scenarios \citep[e.g.\ some of the tidally tilted
  pulsators;][]{Handler2020,Kurtz2020,Rappaport2021,Jayaraman2022,Kahraman2022,Kahraman2023}. However,
others show patterns that are  either very complex or too sparse  to interpret
\citep[e.g.][]{Bowman2019,Southworth2020,Jerzykiewicz2020,Southworth2021a,Southworth2021b,Kalman2022,VanReeth2023,Johnston2023,Kalman2023}. Furthermore,
in the specific case of eclipsing binaries, the geometrical effect of
eclipse mapping due to particular obscuration of the visible disk of
the pulsator by its companion \citep[][]{Reed2001,Reed2005} causes
extra orbital phase-dependent variability of the mode
amplitudes. Depending on the mode geometry, on the misalignment of the
axis with respect to the rotation axis of the star, and on the
geometry of the eclipse, the amplitude modulation can manifest itself
as a complex frequency pattern in the Fourier transform of the light
curve, as in the case of U\,Gru \citep{Johnston2023}.

Recent advances in asteroseismology of (close) binary stars can hardly
be overestimated, both from the theoretical and observational point of
view. Major efforts are undertaken to develop a general enough tidal
oscillation theory as a pertinent need to explain the zoo of observed
phenomena resulting from the interplay between tides and oscillations
brought by the high-precision space photometry.   The novel theories
mentioned above are powerful in terms of the effects due to the tidal,
Lorentz, Coriolis, and centrifugal forces acting on a rotating
pulsating star in a close binary or triple system, taking into account
misalignments between the axes of apsides and rotation
\citep{Fuller2020,Fuller2021,MengSun2023}.

The classes of observed stars exhibiting tidally excited and tidally
perturbed oscillations open up the new window of tidal
asteroseismology, irrespective of the rotation rate, from ultra-slow
to very fast. Progress in the forward modelling of the internal
physical properties from carefully identified modes is the next 
extremely challenging step, given the numerous orbital parameters
involved, as well as the numerical aspects involved in modal
decompositions and the computation of damping coefficients, and the
complex physical phenomenon of pseudo-synchronisation
\citep{TownsendSun2023}. While we are still waiting for a concerted plan of attack for tidal
asteroseismic modelling, applications to large
samples of close binaries are the best way to understand and calibrate
tidal dissipation and close binary evolution. Achieving this would be
a major step ahead as input for proper binary population synthesis and
gravitational-wave progenitor modelling (see Fig.\,\ref{Fountain}).
The modern public tool to model stellar tides developed by
\citet{MengSun2023} is highly relevant in this context.

\section{The special cases of sub-dwarf and merger seismology}

Sub-dwarfs play a special role in terms of close binarity as their
bare existence demands binary interaction at evolved stages of stellar
evolution, including mergers \citep{Han2003}. Either they
occur in a binary or they are single sub-dwarfs as the result of a
merger \citep{Heber2009}. Just as is the case for white dwarfs,
the asteroseismology of sub-dwarfs has advanced appreciably thanks to
short-cadence (1\,min,  2\,min, or 20\,sec) light curves assembled
with {\it Kepler}, K2, and TESS
\citep{Hermes2017,Bell2019,Reed2021,Uzundag2021,Corsico2022,Uzundag2022,Romero2022},
including post-merger sub-dwarfs \citep{Vos2021}.  Asteroseismic
probing of almost all sub-dwarfs and all white dwarfs is done from
the modelling of their identified g~modes via period spacing patterns
and/or rotationally split triplets and quintuplets, where the Coriolis
and tidal forces can be treated as a small perturbation for the
computation of the modes because the periods of all detected modes
(minutes to hours) are much shorter than the rotation or orbital
periods (days).

A review of white dwarf asteroseismology was presented by
\citet{Corsico2019}, but no such comprehensive summary is available
for sub-dwarfs. This is far beyond the scope of this paper, and hence we
just highlight some recent findings for sub-dwarf binaries.
Two major conclusions connected with transport processes can be drawn
from published sub-dwarf asteroseismology:
\begin{enumerate}
  \item
The angular momentum of the convective core of red giants and of sub-dwarfs
is in agreement with the angular momentum of white dwarfs
\citep{Aerts2019-ARAA};
\item
  Sub-dwarfs and white dwarfs have respectively
  more massive helium and carbon--oxygen cores compared to those
  predicted by 
standard stellar evolution models without extra mixing beyond the
Schwarzschild boundary. Hence, core boundary mixing is prominent
throughout the evolution of low- and intermediate-mass stars
\citep[][where the latter
  review paper places both types of compact pulsators and the
  properties of their carbon-oxygen cores into an evolutionary
  context]{VanGrootel2010a,VanGrootel2010b,Charpinet2011,Hermes2017,Charpinet2019a,Charpinet2019b,Giammichele2018,Uzundag2021,Giammichele2022}.
\end{enumerate}

Future progress in the calibration of close binary evolution at
evolved phases can come from several aspects. First of all, many of
the modes of sub-dwarfs and white dwarfs exhibit amplitude and
frequency modulations on timescales of weeks to months, pointing to
non-linear resonant mode interactions. The exploitation of such
interactions is easier than for the rapidly rotating SPB and
$\gamma\,$Dor stars discussed in Sect.\,4.4, and is therefore somewhat
more advanced \citep{Zong2016a,Zong2016b}. However, much remains to be
done in the development and application of non-linear
asteroseismology, also for the compact pulsators. The recent finding
of radial-differential rotation between the core and surface
at the level of about 1.2 in the sub-dwarf EPIC\,220422705 by
\citet{XiaoYuMa2022} is an important step towards the future modelling of
its non-linear modes.

As for the calibration of binary evolution theory, several recent key
findings were delivered by asteroseismology. First of all,
short-period sub-dwarf--M\,dwarf binaries are not necessarily
synchronised. A key laboratory for further study is TIC\,137608661,
whose core rotates slower than its surface, with a period of 4.6\,d
deduced from g-mode asteroseismology. This core rotation period is
much longer than the short orbital period of only about 0.3\,d. The
asteroseismic study of this sub-dwarf binary was placed into the
context of sub-dwarf rotation and binary properties by
\citet{Silvotti2022}.  Their Table\,4 and Fig.\,16 suggests close to
rigid rotation for most of the other sub-dwarfs in short-period
binaries with either an M-dwarf {or} a white dwarf companion. Another
striking result comes from the tidally tilted oscillations \citep[see
][as discussed in the previous section]{Fuller2020} of the binary
sub-dwarf HD\,265435 discovered and analysed by \citet{Jayaraman2022}.
Its white dwarf companion revolves around it in only 1.65\,hr. The
sub-dwarf was found to be near or just beyond the end of its central
helium burning, and hence this object is a Type Ia supernova
progenitor.  Most, but not all, of the sub-dwarf's oscillations have
frequencies that are exact multiples of the orbital frequency.  This
star was diagnosed with acoustic modes having periods of around
250\,s. To date, these tidally tilted oscillations have been modelled,
while the Coriolis and centrifugal forces have been ignored.
{However,} the tidal distortion for a binary with such a tight orbit
shifts the mode frequencies appreciably compared to the case of a
spherical star as the authors have assumed.  Future measurements of
the internal and surface rotation for more short-period binaries are
key to guiding theories of tidal synchronisation.

Backtracking to earlier phases, \citet{Deheuvels2022},
\citet{YaguangLi2022}, and \citet{RuiFuller2024} used asteroseismology
of red clump red giants to identify groups of stripped stars due to
binary interactions.  One group comprises several underluminous red
giants with smaller than usual helium burning cores, which are
interpreted as resulting from heavy mass loss due to envelope
stripping by a companion while they ascended the red giant
branch. Another group consists of tens of red giants whose ages would
exceed the age of the Universe if interpreted from single-star
evolution with no or nominal mass loss on the red giant branch.  However, 
another mechanism leads to core-helium burning red giants with higher
than usual luminosity as a result of the consumption of a white
dwarf during the earlier main-sequence phase.  The small sample sizes
of these observed anomalous groups are consistent with binary
fractions as in Fig.\,\ref{BinaryFraction}. Coupling the properties of
these red giants to models of post-mass-transfer binary systems will
be invaluable to calibrate binary evolution theories of low- to
intermediate-mass stars.

\begin{figure*}[t!]
\begin{center}
  {\resizebox{9.cm}{!}{\includegraphics{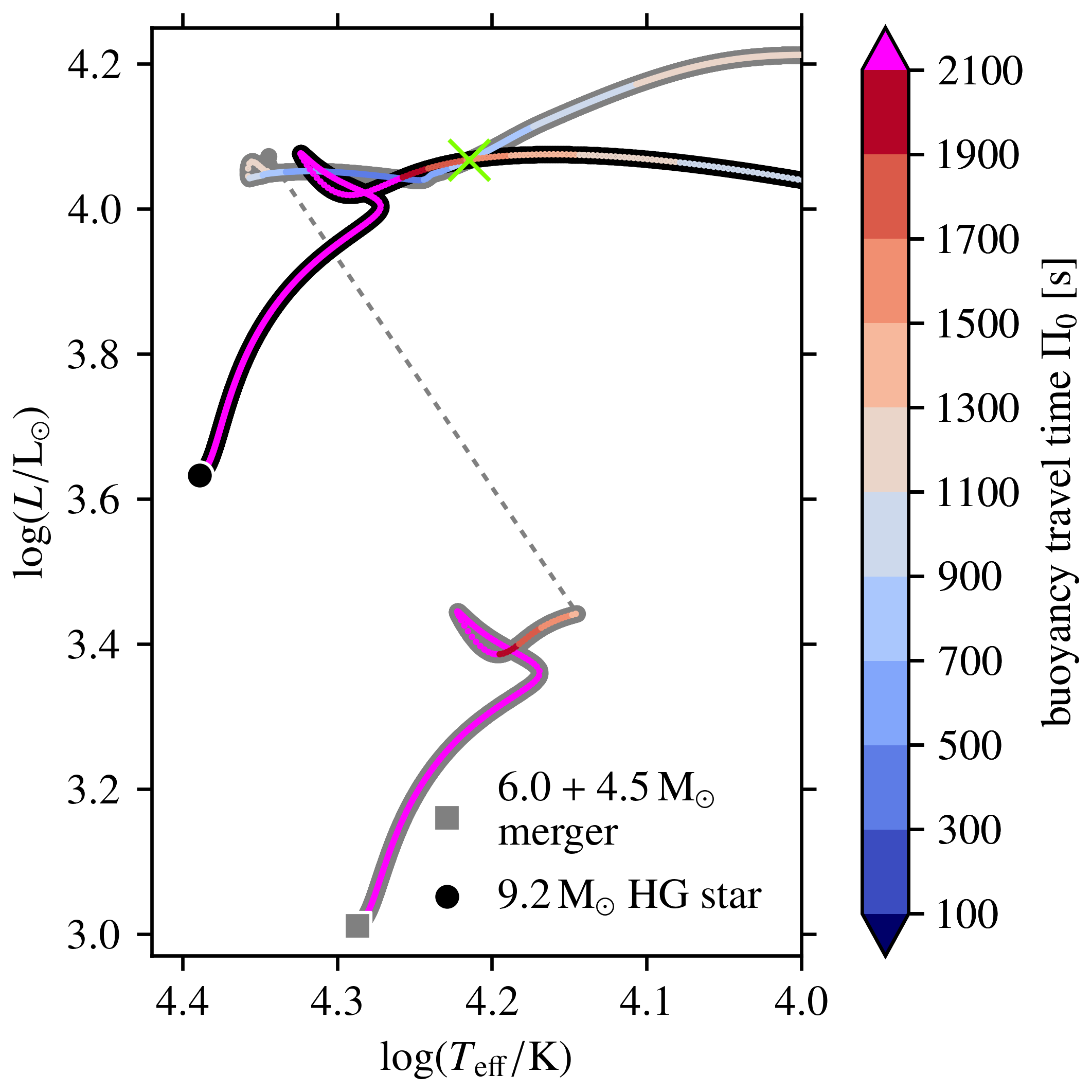}}}
  {\resizebox{9.cm}{!}{\includegraphics{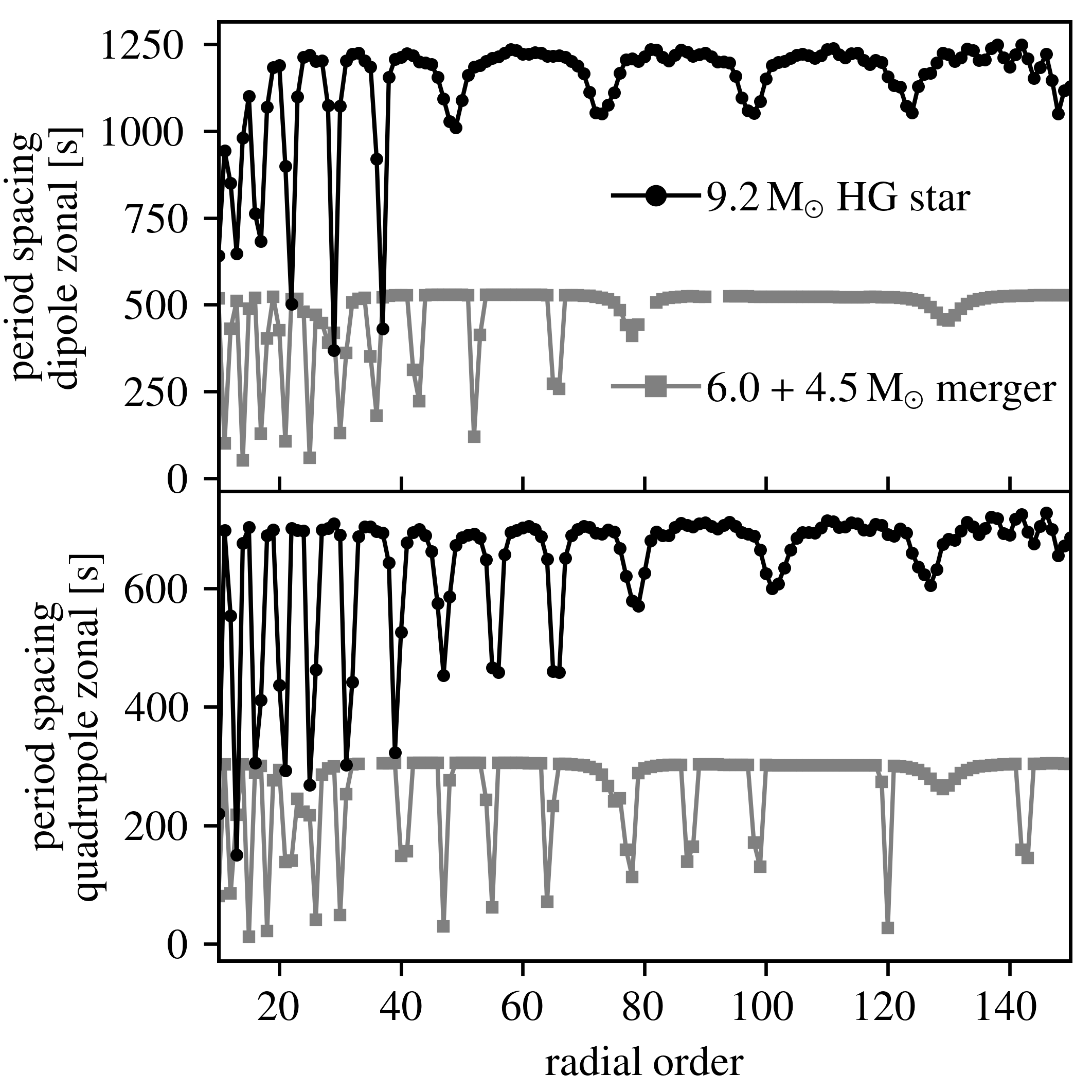}}}  
\end{center}\vspace{-0.5cm}
\caption{\label{Jan} Illustration of the potential of merger
  seismology.  The left panel shows two evolutionary tracks for
  non-rotating stellar models, {colour-coded by the value of the
    buoyancy travel time $\Pi_0$ defined in Eq.\,\eqref{pi-not}. The
    black dot indicates the position of a single star born with
    9.2\,M$_\odot$ at the ZAMS. The grey square is the position of a
    6.0\,M$_\odot$ primary at the ZAMS, accompanied by a
    4.5\,M$_\odot$ close companion. The binary components merge via a
    Case\,B mass transfer following the description by
    \citet{Schneider2021}. This happens when the primary evolves
    beyond the TAMS, as indicated by the dotted line. The single star
    and merger product reside in a similar position while crossing the
    Hertzsprung gap (HG).  The  right panels show the predicted period
    spacing patterns for dipole (upper panel) and quadrupole (lower
    panel) high-order g~modes of the single star undergoing hydrogen
    shell burning (black circles) and of the merger product (grey
    squares), at the evolutionary stage indicated by the green cross
    in the left panel.  At this time the merger burns helium in its
    convective core and has a convective region due to hydrogen shell
    burning, while the inner structure of the single star is fully
    radiative, including its shell burning region. The figure
    was produced by Jan Henneco, from models in \citet{Henneco2024}.}}
\end{figure*}

Moving on to the even earlier phases and to more massive stars,
Fig.\,\ref{BinaryFraction} and binary evolution calculations by
\citet{Schneider2020} tell us that 20\% of OB stars come into contact
with their companion.  Hence, a large population of binary mergers must
exist, either formed from collision or from more gentle coalescence
\citep{Schneider2019}. The properties of such mergers are highly
uncertain and searching for such products is
difficult. Asteroseismology could play a major role here, both in
terms of establishing a proper calibration for the statistics of the
sample of high-mass mergers and of probing their internal
structure. This would allow us to check whether mergers indeed have a strong
large-scale magnetic field, as the MHD simulations by
\citet{Schneider2019} predict.  The first calculations of seismic
diagnostics for stellar mergers were compared with those of single
blue supergiants in the hydrogen shell burning phase (called the
Hertzsprung gap (HG) in the HRD), of main-sequence stars with an
exceptionally high level of core overshooting, and of evolved single
blue giants undergoing a blue loop during their core-helium burning by
\citet{Bellinger2024}. These authors computed grids of non-rotating
equilibrium models and their oscillation frequencies for these four
different populations, considering the mass range
$[10,20]$\,M$_\odot$. This theoretical study shows that
asteroseismology in principle offers the potential to distinguish
among these four cases from the large frequency separation of
identified p~modes and the average period spacing of identified
g~modes.
A major point to be improved is to include internal rotation in
such models.

To illustrate the potential of `merger seismology' (a term suggested
by Dr.\ Fabian Schneider from the Heidelberg Institute for Theoretical
Studies) from a practical point of view, we show in Fig.\,\ref{Jan}
the predicted seismic g-mode period spacings of a merger product in the HG
and compare them to those of a single blue giant residing in the same
part of the HRD, following the study by \citet{Henneco2024}. It can be seen that the
period spacing values of these two models are {clearly} different
in the sense that the average period spacing of the merger product is
lower than that of a model resulting from nominal single-star
evolution. This is in agreement with the results found by
\citet{Bellinger2024}.  {Moreover, the oscillations of the
  merger shown in Fig.\,\ref{Jan} lead to different mode periodicities
  and dip structures in the period spacing patterns compared with
  those of the single HG star.  The shorter mode periods and the
  sharper dips at high radial order for the merger are due to
  particular structures in its Brunt-V\"ais\"al\"a frequency,
  originating from the chemical properties near its core reflected in
  its $\mu$-gradient.}  {The occurrence or absence of intermediate
  convection zones, often caused by semi-convection,
  also plays a role.  The profile of $N(r)$ for the two models with
  similar positions in the HRD are markedly different, resulting in
  different mode trapping and mode coupling according to the
  occurrence of various mode cavities.  Signatures such as those in Fig.\,\ref{Jan}, when
  computed for a representative population of mergers with proper
  inclusion of internal rotation, will provide a powerful tool to
  search for merger products in a machine-learning context when
  applied to {\it Kepler\/} and TESS light curves. Once found,
  modelling the morphology of the period spacing pattern(s) would
  allow us to assess the profiles of the internal chemical
  composition, rotation, and magnetic field of mergers.}

TESS is currently monitoring about 1500 massive OB
dwarfs, giants, and supergiants across the sky at a cadence of 2\,min.
Figure\,\ref{BinaryFraction} makes it clear that dedicated
searches for signals of merger products among these stars based on
the predictions  in Fig.\,\ref{Jan} offer an interesting route for
massive star asteroseismology. Future searches in the TESS data are highly relevant to
calibrating populations of gravitational wave progenitors seismically
because mergers should also occur in what were originally double,
triple, or multiple systems in general.  The discovery of 78
$\beta\,$Cep stars in eclipsing binaries, including 59 new pulsators,
by \citet{Eze2024} offers a fantastic new data set
to apply the principles in Fig.\,\ref{Andrew} and to predict how many
of these targets will evolve into a merger before or after the
supernova explosion(s) that will be happening during their future
evolution.  Discovering mergers from seismology, unravelling their
internal structure, and comparing it to the structure of massive SPB
or $\beta\,$Cep detached or semi-detached binaries offers a great way
to calibrate tidal dissipation and close binary evolution.  It would
also provide an elegant and independent way to evaluate the measured
mass distribution of the black hole and neutron star mergers that 
lead to gravitational wave signals (see Fig.\,\ref{Fountain}).

\section{Onward to a bright future for asteroseismology, for slow and
  fast rotators}

The TESS mission keeps on going, and currently provides ever longer time series
photometry covering many years  (albeit with interruptions of a
year for each of its covered sectors). Moreover, the launch of the
PLATO mission is on the horizon (late 2026, at the time of
writing). With its core programme focused on exoplanet host candidates
and its open Guest Observer programme, it will cover slow to fast
rotators in all evolutionary phases, from single stars through binaries
and multiples to clusters of various ages and metallicities
\citep{Rauer2024}. Even without the additional candidate space missions based on
a similar principle of high-cadence high-precision uninterrupted space
photometry, such as the
High-precision AsteroseismologY of DeNse stellar fields mission
\citep[HAYDN,][]{Miglio2021} and the mission
ExtraTerrestrial: To Find the First Earth\,2.0
\citep[ET,][]{JianGe2022},
TESS and PLATO guarantee a bright future for
asteroseismology for many years to come.  A wealth of suitable light
curves covering many years with high duty cycle will become available
for millions of stars.
{These}
data feed the fountain of opportunities
shown in Fig.\,\ref{Fountain} and will allow us to push asteroseismology
to high-mass stars (unavoidably in close binary systems;
see Fig.\,\ref{BinaryFraction}), covering from early to evolved
evolutionary phases, with some on their way to becoming gravitational wave
sources.
  
As we have tried to emphasise in this review, asteroseismology based on
numerous low-frequency modes in moderate to fast rotators requires
long and uninterrupted data strings to reach sufficient frequency
precision and to unravel long and complex beating patterns
(see Fig.\,\ref{GDOR-SPB-DeltaP}).  Even the 4 yr uninterrupted
{\it Kepler\/} light curves are insufficient to resolve some of the
individual modes of $\gamma\,$Dor, SPB, and Be pulsators.  Years-long
time strings are also necessary to detect and model the
internal rotation and magnetic field profiles inside stars.

\subsection{Pushing {\it Gaia\/} beyond its limits to assist in the asteroseismology of
  intermediate- and high-mass stars}

The {\it Gaia\/} space mission of the European Space Agency \citep{Prusti2016}
was not designed to do asteroseismology.
High sampling rates and continuous monitoring with duty cycles above 90\%
are required to perform asteroseismology. This is needed to
push the Nyquist frequency of the light curves to high enough values,
and to detect and identify a sufficient number of oscillation modes.  {\it Gaia\/} cannot
deliver that. In particular it cannot offer mode detections for amplitudes below
the millimagnitude range
as required to detect solar-like oscillations.  Nevertheless,
{\it Gaia\/} has a significant role to play, particularly for the asteroseismology
of moderate and fast rotators.

All asteroseismic modelling efforts
currently benefit from {\it Gaia\/}'s data, notably the high-precision
parallaxes \citep{Brown2018}, leading to luminosities that guide
the computation of grids of equilibrium models required for the
modelling.  For slow-rotating low-mass stars, the {\it Gaia\/} effective
temperatures additionally lead to fairly precise radius estimates,
which can be confronted with the asteroseismic values. Conversely,
combined asteroseismic radii and {\it Gaia\/} effective temperatures lead to
asteroseismic distance estimates of remarkable precision, up to
distances of kiloparsecs \citep{Huber2017,Davies2017}.

\begin{figure*}
\sidecaption
\includegraphics[width = 12cm]{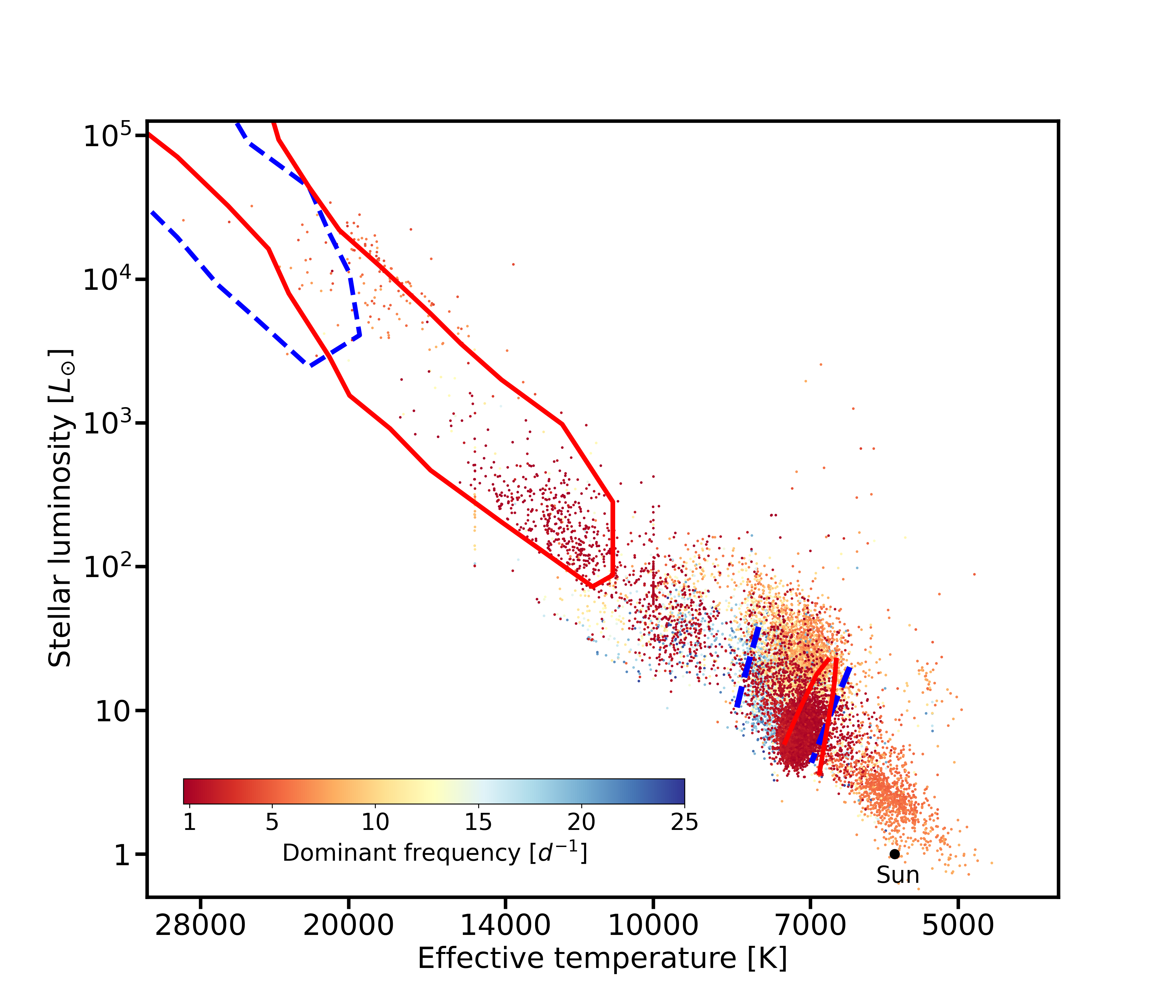}
  \caption{\label{GaiaDR3-HRD} HRD showing the position of more than 100,000
    candidate (non-)radial oscillators discovered from the {\it Gaia\/} DR3
    G-band light curves, colour-coded according to their dominant
    frequency. The thick lines indicate the borders of four
    instability strips for solar metallicity computed by
    \citet{Burssens2020} and \citet{Dupret2005a}, enclosing areas where
    theoretically predicted g~modes (solid red lines) and p~modes
    (dashed blue lines) are expected for the chosen input physics of
    the models.  The figure was produced by Dr.\ Joris De Ridder from data in
    \citet{DeRidder2023}.}
\end{figure*}

The continued improvement of the {\it Gaia}, {\it Kepler\/} and TESS data
reductions provide excellent means for more accurate instrument
calibrations, with ever better future capacity for large and
homogeneous population studies.  Along with the {\it Gaia\/} distances and
proper motions, precise asteroseismic ages are particularly valuable
for galactic {archaeology} studies (see Fig.\,\ref{Fountain}).

We also point to additional roles that {\it Gaia\/} can play by pushing
beyond the nominal use of its data.  The three light curves made available
in {\it Gaia\/}'s Data Release 3 (DR3) in the G, RP, and BP photometric bands
delivered millions of new variable stars \citep{Eyer2023}, among which
thousands of candidates with non-radial oscillations having amplitudes
above a few mmag \citep{DeRidder2023}. Figure\,\ref{GaiaDR3-HRD} shows
the position of more than 100,000 new (non-)radial pulsators in the
upper HRD discovered from {\it Gaia\/} DR3. This figure includes a comparison
with the positions of four instability strips where non-radial modes
are expected to be excited based on one particular choice of input
physics for each of the $\beta\,$Cep, SPB, $\delta\,$Sct, and
$\gamma\,$Dor pulsator classes.  We know  from
asteroseismology that such comparisons with one strip per class based
on only one choice of input physics is too limited, in view of the
large variety of internal rotation rates and mixing levels in the
interior of single and binary stars in that part of the HRD
\citep[e.g.][]{Moravveji2015,Moravveji2016,Szewczuk2018,TaoWu2019,TaoWu2020,Mombarg2020,Pedersen2021,Pedersen2022a,Szewczuk2022}.
This can explain why many stars occur outside of the particular strips
chosen to illustrate the comparison in Fig.\,\ref{GaiaDR3-HRD}. A
similar conclusion was reached from the summary by \citet{Balona2024}
based on assembled space photometry over the past decade.  Proper
comparisons between the observed and theoretically predicted position
in the HRD can help to improve the input physics and delineate free
parameters used in stellar models. This is a question of various choices for
opacity tables and chemical mixtures, for the mixing
length parameter of (time-dependent) convection calculations, and for
parameters used to describe transport processes caused by rotation,
magnetic fields, and atomic diffusion, including radiative levitation
\citep[][to list just a few papers treating some of these
  phenomena]{Townsend2005,Dupret2005a,Dupret2005b,Dupret2005c,Moravveji2016-OP,Szewczuk2017,Rehm2024}.

The more than 15,000 new candidate g-mode pulsators discovered from
{\it Gaia\/} DR3 shown in Fig.\,\ref{GaiaDR3-HRD} have global properties fully
in line with the {\it Kepler\/} pulsators in these classes
\citep{Aerts2023}, including their chemical composition \citep{Gebruers2021,Laverny2025}.  This offers great
potential for `industrialised' ensemble asteroseismology of
thousands of intermediate- and high-mass stars. The requirement to
embark upon this is that sufficient non-radial modes can be identified
per star, for example from dedicated space photometry by {\it Kepler}, TESS,
or PLATO, in addition to the identification of their dominant mode(s)
discovered in the {\it Gaia\/} photometry. For now, identifications of a
sufficient number of modes are only available for about 700
moderate and fast rotators observed by {\it Kepler\/}
\citep{Pedersen2019,Antoci2019,Cunha2019-TESS,Burssens2020,Skarka2022,Garcia2022a}.

\citet{HeyAerts2024} provided a path forward to extend this {\it Kepler\/}
sample by scrutinising the light curves extracted from TESS FFIs for the
60,000 candidate pulsators in
Fig.\,\ref{GaiaDR3-HRD} with such data available. They confirmed the large
majority of these 60,000 stars to be multi-periodic non-radial
pulsators, and found the same dominant frequency in the TESS FFI light curves
as found in the sparse {\it Gaia\/} DR3 light curves, highlighting
{\it Gaia}'s capabilities to detect good asteroseismology targets.

\citet{Mombarg2025} followed up on \citet{HeyAerts2024} and
determined the (convective core) masses, radii, and evolutionary stages for more than
10,000 g-mode pulsators from their {\it Gaia\/} DR3 data, after having
confirmed their dominant oscillation mode from TESS FFI data. To
achieve this they relied on grids of models calibrated by {\it
  Kepler\/} asteroseismology in terms of element and angular momentum
transport computed by \citet{Mombarg2024}. These {\it Gaia\/}-based results
deliver many non-radial pulsators with a mass between 2 and
3\,M$_\odot$, while none are expected from current excitation theory.
This discrepancy between theory and observations will trigger further
research, given the poorly known uncertainties of the {\it Gaia\/} DR3 observables of hot
stars. Moreover, following
Fig.\,\ref{BinaryFraction}, a large fraction of the stars inside and
  between the instability strips shown in Fig.\,\ref{GaiaDR3-HRD}
could be as-yet-unknown binaries or mergers. Their deduced luminosity and effective
temperature may thus be particularly prone to systematic uncertainty
as the {\it Gaia\/} processing ignored their potential (former) multiplicity.

\begin{figure*}[t!]
\begin{center}
\hspace{-.8cm}\rotatebox{270}{\resizebox{7.65cm}{!}{\includegraphics{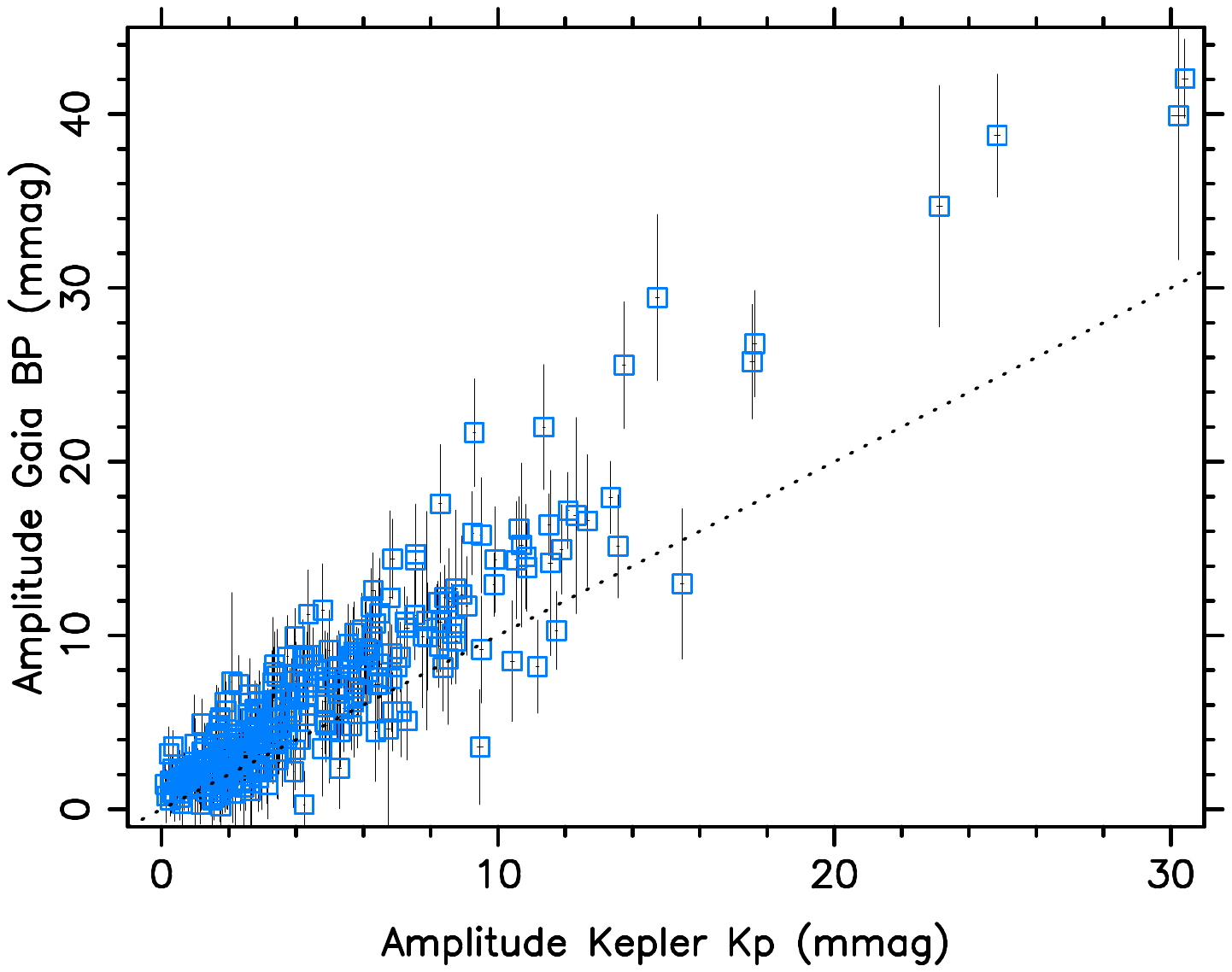}}}
\hspace{-.8cm}\rotatebox{270}{\resizebox{7.65cm}{!}{\includegraphics{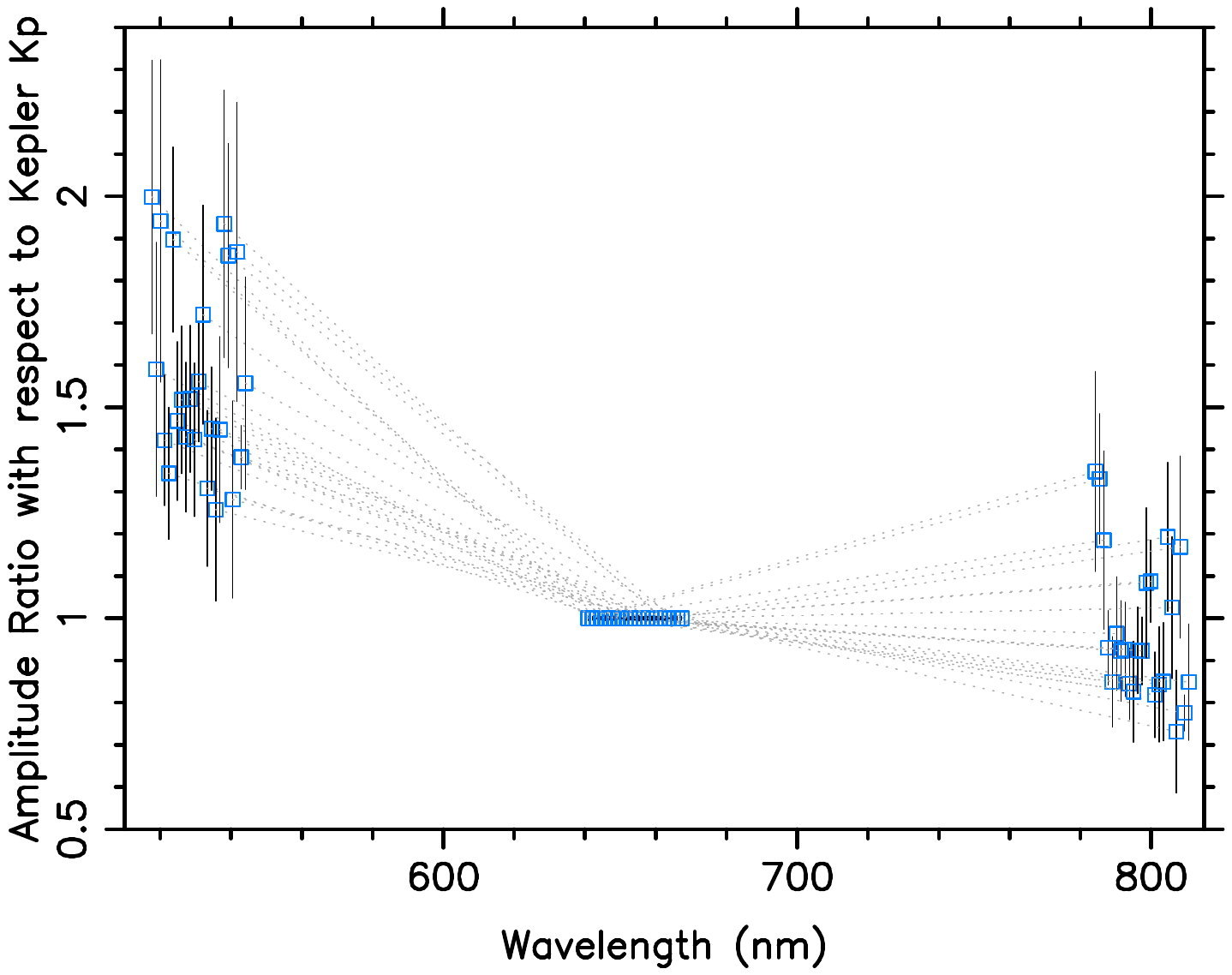}}}
\vspace{-0.5cm}
\end{center}
\caption{\label{amplratios} Comparison between the
  amplitudes of the dominant g-mode frequency in the {\it Gaia\/} BP and {\it
    Kepler\/} $K_p$ bands, whose maxima in the response functions
  occur at 517\,nm and 640\,nm, respectively. The left panel shows the
  results for the 309
  $\gamma\,$Dor stars from \citet{GangLi2020} with time series
  photometry available in the public domain of both surveys. The
  frequency deduced from the {\it Kepler\/} data was imposed on the
  {\it Gaia\/} DR3 data to produce this plot.  The right panel shows
  the {\it Gaia\/} BP and RP
  (response curve peaking at 783\,nm) amplitude ratios with respect to
  the one in $K_p$, plotted as a function of wavelength for the 23
  $\gamma\,$Dor stars in the sample with an identified dominant prograde dipole
  mode explaining more than 50\% of the total variance
  in  the {\it Gaia\/} BP and RP light curves.  The three values for each star are
  connected by a dotted line to guide the eye.  The results for each
  star have been shifted slightly in wavelength for visibility
  reasons. The errors for $K_p$ are smaller than the
  symbol size.}
\end{figure*}

Pushing even more beyond the  limits, we can now fully integrate the
knowledge from {\it Gaia\/}'s three-band and {\it Kepler}'s and/or TESS's (or in the
future PLATO's) single-band light curves to improve asteroseismic
modelling.  Such a synergistic integration of data is
illustrated in Fig.\,\ref{amplratios}, which compares the amplitudes
of the dominant dipole prograde mode in the {\it Gaia\/} BP and {\it Kepler\/}
$K_{\rm p}$ bands for 309 $\gamma\,$Dor stars.
Theory predicts the amplitudes of dipole modes in $\gamma\,$Dor
pulsators to be higher at the BP than at the $K_{\rm p}$ wavelengths
for $\gamma\,$Dor pulsators \citep{Aerts2004}.  This is in agreement
with the results for g-mode pulsators shown in the left panel in
Fig.\,\ref{amplratios}. This panel shows the amplitude ratios of the
23 $\gamma\,$Dor stars among the sample of 309, for which a harmonic
fit with the dominant {\it Kepler\/} frequency explains at least 50\%
of the overall variance in the {\it Gaia\/} BP and RP time series. {Measured
  amplitudes depend on the inclination angle of the star's symmetry
  axis of the oscillations with respect to the line of sight and on
  the position of the surface nodal lines. However, for slow rotators,
  both dependencies} drop out of the theoretical expression of
amplitude ratios with respect to a reference wavelength \citep[see
  Chapter\,6,][for details]{Aerts2010}.  Theoretical expressions for
the ratios in the approximation of the TAR are also available
\citep{Townsend2003-AR}.  This implies that measured amplitude ratios
contain integrated astrophysical information about the properties of
the stellar atmosphere, notably its aspect-angle dependent effective
temperature, gravity, and limb darkening, all of which occur in the
theoretical expressions of the ratios. While this was used as an attempt
to   identify mode degrees from models with fixed input physics
prior to space asteroseismology, {\it Gaia\/} now also offers the reverse for
stars whose frequencies and amplitudes of already identified modes are
detected in the {\it Gaia\/} BP and RP photometry.
  
In addition, \citet{Fritzewski2025} showed that the {\it
  Gaia\/} BP and RP photometry may lead to the identification of the
degree of low-order modes in $\beta\,$Cep stars.  They designed a
novel probabilistic identification method and applied it to the
dominant mode detected consistently in the {\it Gaia\/} BP, RP, and
TESS data of 164 $\beta\,$Cep stars included in Fig.\,\ref{GaiaDR3-HRD},
among which 121 new pulsators discovered from {\it Gaia}.
This mode identification and the detection of rotationally split modes
in the TESS FFI data led to the first forward modelling of a
population of high-mass pulsators, resulting in masses between
9\,M$_\odot$ and about 20\,M$_\odot$ for these 164 stars. For 48
of them, limits on the core-to-surface rotation ratios were derived
from the combined {\it Gaia\/} and TESS data;   most of these stars have
values below 2, but a few objects reveal a ratio   up to $\sim$ 4. Such internal
rotation estimates were only available for five class members before {\it
  Gaia\/} DR3.

These examples illustrate that {\it Gaia\/} already has an important role to
play in asteroseismic modelling of intermediate- and high-mass
pulsators. Much more is to come in the near future.
Since the 23 $\gamma\,$Dor stars in
Fig.\,\ref{amplratios} have similar metallicity and the same
mode identification in terms of $(l,m)$, future exploitation of their
amplitude ratios may reveal their atmospheric properties, which  are
currently unknown.  This may even help us to find optimal stars with a
variety of surface phenomena, such as spots and
lattitudinal-differential rotation, for calibration purposes of new high-resolution
spectrographs to study exoplanet hosts (see Fig.\,\ref{Fountain}).
  Figure\,\ref{amplratios} shows that the uncertainties in the
{\it Gaia\/} BP and RP amplitude ratios are currently still too large to
exploit this potential, but they should decrease appreciably for the
longer light curves to be expected from {\it Gaia\/} Data Releases 4 (2026)
and 5.  {These will occur at the time when the PLATO mission should be
  operational.  PLATO itself offers two-colour photometry from its two
  fast cameras, and is even more suited to push this research beyond
  current instrument barriers than {\it Kepler\/} or TESS combined
  with {\it Gaia}.  In addition, ground-based survey facilities with
  the capacity to deliver multi-colour photometry at the millimagnitude level are operational or
  will become operational soon, such as BlackGEM \citep{Groot2024} and
  the {\it Vera Rubin\/} Observatory \citep[formerly known as the
    Large Synoptic Survey Telescope (LSST),][]{DiCriscienzo2023}.
  The basics of
  modelling tools used to interpret the amplitude ratios in
  moderate and fast rotators had already been developed long before the
  era of space asteroseismology
  \citep[e.g.][]{Daszynska2002,Townsend2003-AR}. Now, about two
  decades later, modern tools such as  those designed by
  \citet{Fritzewski2025} 
  can be applied massively from combined
  {\it Gaia}, {\it Kepler}, TESS, PLATO, BlackGEM, and {\it Rubin\/} multi-colour
survey photometry for all the pulsators with proper identification of
their mode(s). Exploitation of the amplitude ratios in terms of
  internal and atmosphere physics from modern multi-colour photometry
  should be within reach for modes with amplitudes of several mmag.

\subsection{The  need for better theory and modelling tools}

Optimal exploitation of the wealth of available and future
observations of stellar oscillations requires concerted
efforts to develop better theory. This need is particularly prominent
for the theory and modelling of non-radial oscillations in fast-rotating
single stars and close binaries. In order to treat moderate and fast rotators up to
the level of the precision of modern data, we must go beyond
{the current modelling tools and develop 
  non-linear pulsation theory involving mode coupling and non-linear
  inversion tools
  \citep{Vanlaer2023,Farrell2024}.} 
 {Ample} detections of amplitude modulation
\citep[e.g.][]{Bowman2016-Springer} and non-linear mode coupling
\citep[e.g.][]{Huat2009,Breger2012,Kurtz2015,Baade2016,VanBeeck2021}
in fast rotators are available, while most asteroseismic modelling currently only
relies on measured frequencies assuming linear and uncoupled modes.
Non-linear asteroseismology would allow us to exploit mode amplitudes from
coupled mode equations.  Following the initial
pioneering steps {in non-linear space asteroseismology
for slow-rotating evolved sub-dwarfs and white
dwarfs by \citet{Zong2016a,Zong2016b}, similar applications to}
moderate and fast rotators along the main sequence
have yet to be tried out.  This can be done by extending the principles
laid out by \citet{Buchler1997,Goupil1998,Mourabit2023}, among others, by means of a
proper treatment of the rotation.  \citet{Lee2012,Lee2022} and
\citet{VanBeeck2024} include the Coriolis force in the sub-inertial
regime (see Fig.\,\ref{Frequency-Axis}), providing enhanced theory to
model the numerous gravito-inertial non-linear pulsators with
identified multi-mode couplings.  Asteroseismic modelling based on
these non-linear formalisms would allow us to scrutinise transport
processes from early phases of evolution onwards. This would
constitute a critical step ahead because the effects of angular
momentum transport and chemical mixing are cumulative, and hence have
major implications for the entire life of the star, from birth to death.

Finally, moderate and fast rotators require attention at an even more
fundamental level than developing and applying non-linear
asteroseismology.  If we truly want to maximise the exploitation
of the asteroseismic data of such pulsators, we must abandon the
premise that stars can be approximated as spheres, both at the level
of the stellar structure equations and in solving the oscillation
equations to compute the modes. Rotating stars are oblate
spheroids, requiring at least 2D (for rotating single stars) and
optimally 3D (for magnetic spheroids and/or binary systems)
treatments. Realistic evolution models including the
chemical evolution of   stars are being constructed
\citep{Deupree1990,Deupree1995,Rieutord2016,Mombarg2023} and so are
numerical computations of their oscillation modes
\citep{Ouazzani2015,Reese2021}. Major long-term mathematical and
computational efforts to achieve this in a systematic approach
are on the horizon\footnote{\tt
www.4D-STAR.org} and offer a fantastic opportunity to bridge
the fields of asteroseismology, close binary evolution, binary
population synthesis, and gravitational wave astrophysics, as
highlighted in the fountain in Fig.\,\ref{Fountain}.

\onecolumn
\color{purple}
\kader{6.8in}{{\bf Takeaway messages from the asteroseismology of fast
    rotators in the literature include:}
\begin{itemize}
\item
  Asteroseismology of stars rotating up to
  70\% of their critical Keplerian rotation rate is a mature field of
  research, delivering their mass, radius, core mass, and core radius with
  a precision of a few percent for the best cases.
\item
  About 80\% of the detected
 {g-mode}
  oscillations in intermediate-mass
dwarfs are prograde dipole gravito-inertial modes. These modes deliver
the near-core rotation frequency to better than
5\% accuracy. Values range from zero to $30\,\mu$Hz.
\item
For a small fraction of the fastest rotators in the mass range
$[1.3,1.9]\,$M$_\odot$, inertial modes in the convective core couple
to gravito-inertial modes in the envelope, offering a direct
measurement of the rotating core properties.
\item
  Single intermediate-mass dwarfs have modest radial-differential rotation, with their
  core and surface rotation frequencies differing by less than 10\%.  The
  asteroseismic results imply rapid transport of angular
  momentum from the core to the surface during the main sequence.
\item
  About 15\% of the fast-rotating dwarfs in the mass range
  $[1.3,1.9]\,$M$_\odot$, as well as some of the outbursting Be
  pulsators, reveal global Rossby modes.
\item
  Non-linear resonant mode coupling is omnipresent in rotating
  intermediate-mass pulsators and in their sub-dwarf and white dwarf
  successors. This offers a golden
  opportunity to exploit mode amplitudes in addition to frequencies.  
\item
  The fastest rotating B and Be pulsators in the mass range
  $[3,9]\,$M$_\odot$ show outbursts in their photometric light
  curves, often accompanied by non-linear mode coupling.
\item
  Asteroseismic core-to-surface rotation
 {ratios} have been determined for about 50 
 high-mass dwarfs in the mass range
$[8,20]\,$M$_\odot$, revealing values below 2 for most of them, but a value up
 to 5 for a few stars.
\item
  The cumulative effect of the measured near-core mixing
  in dwarfs with a convective core extends the duration of their
  main sequence  and induces large spreads at the
  turnoff. This mixing impacts all phases of evolution, notably
  age-dating and chemical yield production. Up to twice as
  much helium can be produced by the time of the turnoff compared
  to the case without any near-core mixing.
\item
Evidence of enhanced internal mixing during the  core-helium burning
phase is uncovered from asteroseismology of white dwarfs and
sub-dwarfs, but has yet to be calibrated with high precision for large samples.
\item 
  Measured vertical diffusive envelope mixing levels in single
  intermediate- and high-mass dwarfs range from 10\,cm$^2$\,s$^{-1}$
  to 10$^7$\,cm$^2$\,s$^{-1}$. The near-core and envelope mixing profiles are
  diverse and change over time.
\item
  Internal magnetic fields are detected in a fraction of pulsators
  across the entire evolution and mass ranges accessible by
  asteroseismology, revealing field strengths between $10^4$\,G and
  about $10^6$\,G.
\item
  Model-independent dynamical masses of detached eclipsing binaries
  offer tight constraints on asteroseismic modelling, confirming
  that stellar models require near-core boundary mixing to bring
  (core) masses from theory and observations into agreement.
\item
  Tidally excited, tidally perturbed, and tidally tilted oscillations
  occur across a wide range of binaries, including supernova
  progenitors and members of young open clusters.  This makes
  asteroseismic calibration of tidal dissipation processes and of
  close binary evolution models possible, including binaries en route
  to gravitational wave emission.
\end{itemize}
}

\color{black}


\begin{acknowledgements}
  The authors thank Ga\"el Buldgen and Jan Henneco for sharing their
  work that led to Figures 5 and 9; they are also grateful
  to Clio Gielen, Siemen Burssens, Jordan Van Beeck, Jan Henneco, and
  Joris De Ridder for providing Figures 1 and 3, 6, 7, 9, and 10,
  respectively. Jordan Van Beeck and Vincent Vanlaer are acknowledged
  for their comments on a draft version of the text prior to
  submission. {We gratefully acknowledge the positive feedback and
    detailed comments received from two anonymous referees. We also
    express our appreciation to those colleagues from the community
    having done the effort to answer our call for comments by offering
    feedback and suggestions, which helped us to improve the text.}

  The authors acknowledge funding from the
  KU\,Leuven Research Council (grant C16/18/005: PARADISE) and from
  the BELgian federal Science Policy Office (BELSPO) through PRODEX
  grant PLATO.  CA additionally received financial support from the
  Research Foundation Flanders (FWO) under grant K802922N (Sabbatical
  leave) and from the European Research Council (ERC) under the
  Horizon Europe programme (Synergy Grant agreement
  N$^\circ$101071505: 4D-STAR). {Work for this review was partially
  funded by the European Union. Views
  and opinions expressed are however those of the author(s) only and
  do not necessarily reflect those of the European Union or the
  European Research Council. Neither the European Union nor the
  granting authority can be held responsible for them.}

  The content of this review was fed by
  inspiring discussions held at the Munich Institute for Astro-,
  Particle and BioPhysics (MIAPbP), which is funded by the Deutsche
  Forschungsgemeinschaft (DFG, German Research Foundation) under
  Germany's Excellence Strategy -- EXC-2094 -- 390783311.  CA thanks
  the organisers and local staff for the invitation and for the good
  organisation.  She is also grateful for the kind hospitality offered
  by the Center for Computational Astrophysics of the Flatiron
  Institute of the Simons Foundation in New York, by the Max Planck
  Institute of Astronomy in Heidelberg, and by the Centre d'\'Etude
  Atomaire in Saclay during her sabbatical work visits in academic
  year 2022 -- 2023.
\end{acknowledgements}

\bibliographystyle{aa}
\bibliography{AertsTkachenko}

\end{document}